\documentclass{article}
\usepackage{import}
\usepackage{preambles}
\usepackage{cite}
\numberwithin{equation}{section}
\allowdisplaybreaks

\begin{document}
\hfuzz=100pt
\title{Quantum phase transition and Resurgence:\\
Lessons from 3d $\mathcal{N}=4\:$ SQED}

\author{
\Large Toshiaki Fujimori$^{1}$\footnote{toshiaki.fujimori018(at)gmail.com},\quad
Masazumi Honda$^{2}$\footnote{masazumi.honda(at)yukawa.kyoto-u.ac.jp}, \quad
Syo Kamata$^{3}$\footnote{skamata11phys(at)gmail.com},\\ 
\Large Tatsuhiro Misumi$^{1,4}$\footnote{tatsuhiromisumi(at)gmail.com},\quad
Norisuke Sakai$^{1}$\footnote{norisuke.sakai(at)gmail.com},\quad
Takuya Yoda$^{5}$\footnote{t.yoda(at)gauge.scphys.kyoto-u.ac.jp}
\vspace{1em} \\
\\
{$^1$ \small{\it Department of Physics, and Research and 
Education Center for Natural Sciences,}}\\ 
{\small{ \it Keio University, 4-1-1 Hiyoshi, Yokohama, Kanagawa 223-8521, Japan}}\\
{$^{2}$ \small{\it Center for Gravitational Physics, Yukawa Institute for Theoretical Physics}}\\ 
{\small{\it Kyoto University, Sakyo-ku, Kyoto 606-8502, Japan}} \\
{$^{3}$ \small{\it National  Centre  for  Nuclear  Research,  02-093  Warsaw,  Poland}}\\ 
{$^{4}$ \small{\it Department of Mathematical Science, Akita University,  Akita 010-8502, Japan}}\\ 
{$^{5}$ \small{\it Department of Physics, Kyoto University, Kyoto 606-8502, Japan}}
}
\date{\small{March 2021}}
\maketitle
\thispagestyle{empty}
\centerline{}

\begin{abstract}
We study a resurgence structure of a quantum field theory with a phase transition
to uncover relations between resurgence and phase transitions.
In particular, we focus on three-dimensional $\mathcal{N}=4$ supersymmetric quantum electrodynamics (SQED) with multiple hypermultiplets,
where a second-order quantum phase transition has been recently proposed in the large-flavor limit.
We provide interpretations of the phase transition from the viewpoints of Lefschetz thimbles and resurgence.
For this purpose,
we study 
the Lefschetz thimble structure and properties of the large-flavor expansion
for the partition function obtained by the supersymmetric localization.
We show that
the second-order phase transition is understood as a phenomenon
where a Stokes and anti-Stokes phenomenon occurs simultaneously.
The order of the phase transition is determined by how saddles collide at the critical point.
In addition, the phase transition accompanies an infinite number of Stokes phenomena
due to the supersymmetry.
These features are appropriately mapped to the Borel plane structures
as the resurgence theory expects.
Given the lessons from the SQED,
we provide a more general discussion on the relationship between the resurgence and phase transitions.
In particular, 
we show how the information on the phase transition is decoded from the Borel resummation technique.
\end{abstract}

\vfill
\noindent

YITP-21-13, KUNS-2859

\renewcommand{\thefootnote}{\arabic{footnote}}
\setcounter{footnote}{0}

\newpage
\pagenumbering{arabic}
\setcounter{page}{1}
\tableofcontents

\clearpage
\section{Introduction}
One of the most important problems in quantum field theory (QFT) is to determine phase structures in the space of parameters.
It is connected to significant information
such as symmetries, energy gap, critical phenomena, topological order, etc.
In particular, second-order phase transitions are essential
as they often describe the starts and goals of renormalization group flows.
It is natural to expect that
phase transitions are technically related to (anti-)Stokes phenomena
as they both describe some discontinuous behaviors of physical quantities
in a certain limit of parameters.
A well-known example of this is the connection between first-order phase transitions and anti-Stokes phenomena
as there occur switches of dominant saddle points across anti-Stokes lines.
However, the connections for higher order cases are less clear, and
it seems necessary to study in a systematic framework.
One such approach to describe Stokes phenomena is 
{\it the resurgence theory} \cite{Ecalle},
which recently has attracted much attention in the contexts of QFTs\footnote{
See e.g. reviews \cite{Costin:1999798,Marino:2012zq,Dorigoni:2014hea,Aniceto:2018bis,2014arXiv1405.0356S} for details.
}. 
This paper aims to study relations between phase transitions and resurgence in QFTs.

The resurgence theory has a long history in applications to quantum mechanics and differential equations.
It has often been used to cure situations that perturbative expansions are not convergent.
It typically gives relations between non-perturbative effects and 
large order behaviors of perturbative series. 
There have been applications of resurgence 
to many physical systems
including quantum mechanics \cite{Bender:1969si,Bender:1973rz,Balian:1978et,AIHPA_1983__39_3_211_0,ZinnJustin:2004ib,ZinnJustin:2004cg,Jentschura:2010zza,Jentschura:2011zza,Dunne:2013ada,Basar:2013eka,Dunne:2014bca,Escobar-Ruiz:2015nsa,Escobar-Ruiz:2015rfa,Misumi:2015dua,Behtash:2015zha,Behtash:2015loa,Gahramanov:2015yxk,Dunne:2016qix,Kozcaz:2016wvy,Fujimori:2016ljw,Dunne:2016jsr,Serone:2016qog,Basar:2017hpr,Alvarez:2017sza,Behtash:2018voa,Duan:2018dvj,Raman:2020sgw,Sueishi:2019xcj,Sueishi:2020rug,Sueishi:2021xti}, 
hydrodynamics \cite{Aniceto:2015mto,Basar:2015ava,Casalderrey-Solana:2017zyh,Behtash:2017wqg,Heller:2018qvh,Heller:2020uuy,Aniceto:2018uik,Behtash:2020vqk}, 
integrable systems \cite{Ito:2018eon,Schepers:2020ehn,Marino:2019wra,Marino:2019eym,Marino:2019fvu,Marino:2020dgc,Marino:2020ggm,Marino:2021six},
non-critical string \cite{Marino:2008vx,Garoufalidis:2010ya,Chan:2010rw,Chan:2011dx,Schiappa:2013opa} and 
string theory \cite{Marino:2006hs,Marino:2007te,Marino:2008ya,Pasquetti:2009jg,Aniceto:2011nu,Santamaria:2013rua,Couso-Santamaria:2014iia,Grassi:2014cla,Couso-Santamaria:2015wga,Couso-Santamaria:2016vcc,Couso-Santamaria:2016vwq,Kuroki:2019ets,Kuroki:2020rgg,Dorigoni:2020oon}
as well as QFTs.
Recently there are also various applications to QFTs
such as 2d QFTs
\cite{Dunne:2012ae,Dunne:2012zk,Cherman:2013yfa,Cherman:2014ofa,Misumi:2014jua,Nitta:2014vpa,Nitta:2015tua,Behtash:2015kna,Dunne:2015ywa,Buividovich:2015oju,Demulder:2016mja,Sulejmanpasic:2016llc,Okuyama:2018clk,Abbott:2020qnl,Abbott:2020mba,Ishikawa:2019tnw,Ishikawa:2020eht},
the 3d Chern-Simons theory 
\cite{Gukov:2016njj,Gang:2017hbs,Gang:2017hbs,Wu:2020dhl,Garoufalidis:2020nut,Garoufalidis:2020xec,Fuji:2020ltq,Ferrari:2020avq,Gukov:2019mnk,Garoufalidis:2020nut} 
and Skyrme model \cite{Barsanti:2020qxm},
4d non-supersymmetric QFTs \cite{Argyres:2012vv,Dunne:2015eoa,Yamazaki:2017ulc,Mera:2018qte,Itou:2018wkm,Canfora:2018clt,Ashie:2019cmy,Ishikawa:2019oga,Unsal:2020yeh,Ashie:2020bvw,Morikawa:2020agf},
and supersymmetric (SUSY) gauge theories in various dimensions \cite{Russo:2012kj,Aniceto:2014hoa,Grassi:2014cla,Aniceto:2015rua,Honda:2016mvg,Honda:2016vmv,Gukov:2016tnp,Honda:2017qdb,Gukov:2017kmk,Dorigoni:2017smz,Honda:2017cnz,Fujimori:2018nvz,Grassi:2019coc,Dorigoni:2019kux,Dorigoni:2021guq}.\footnote{
There are also studies on relations to
renormalization \cite{Klaczynski:2016mbx,Bersini:2019axn,Bellon:2020uzi,Borinsky:2020vae,Marino:2020ggm,Bellon:2020qlx},
thermalization \cite{Engelsoy:2020tsp} and
large charge expansions \cite{Dondi:2021buw}.
}
However, most of the works have focused on systems without phase transitions
while we will mention some works related to phase transitions.

In this paper,
we study the resurgence structure of a QFT model with a phase transition
to study relations between resurgence and phase transitions.
We take two approaches to address this problem.
The first approach is the Lefschetz thimble (steepest descents) analysis \cite{Balian:1978et,AIHPA_1983__39_3_211_0,Witten:2010cx}\footnote{
See also \cite{Witten:2010zr,Harlow:2011ny,Cristoforetti:2012su,Cristoforetti:2013wha,Fujii:2013sra,Mukherjee:2013aga,Aarts:2013fpa,Cristoforetti:2014gsa,Tanizaki:2014xba,Tanizaki:2015pua,Alexandru:2016ejd,Alexandru:2016san}.
}.
For a Lagrangian QFT,
physical observables admit path integral representations and
we can decompose them in terms of Lefschetz thimbles associated with saddle points
in the field configuration space.  
In general,
the structure of such a thimble decomposition can change discontinuously
as varying parameters continuously in the theory under consideration.
When it happens,
asymptotic expansions around saddle points exhibit discontinuous changes of their forms,
called the Stokes phenomena.
Other interesting phenomena related to thimble decomposition are anti-Stokes phenomena,
where dominantly contributing saddles are switched as varying parameters
while thimble structures themselves are unchanged.
As mentioned above, 
anti-Stokes phenomena typically induce a first-order phase transition and
it is often discussed in theoretical physics simply by comparing values of actions at saddle points.
phase transitions.
Here we mainly focus on relations between second-order phase transitions and Stokes phenomena.
The other approach is to interpret it from the viewpoint of Borel resummation.
The information of such phenomena is expected to be encoded in a perturbative series.
The Borel resummation technique enables us to decode such information.
Therefore, if the resurgence theory works,
then we can approach phase transitions (typically by non-perturbative effects) from perturbative expansions.

Here we mainly study 
the three dimensional $\mathcal{N}=4$ supersymmetric quantum electrodynamics with multiple hypermultiplets (SQED)
to obtain lessons on relations between resurgence and phase transitions.
Recently it has been proposed by Russo and Tierz \cite{Russo:2016ueu} that
there is a quantum second-order phase transition in the SQED 
based on saddle point analysis in the large flavor limit.
They argued that
the number of dominant saddle points changes across a particular value of the Fayet-Illiopoulos (FI) parameter and
assuming that they all contribute to the path integral induces the phase transition.
We provide interpretations of the phase transition from the viewpoints of Lefschetz thimbles and resurgence.
We first justify the assumption in \cite{Russo:2016ueu} that
all the dominant complex saddles contribute to the path integral by the Lefschetz thimble analysis.
Then we interpret the second-order phase transition as a simultaneous Stokes and anti-Stokes phenomena.
Our results show that
the resurgence theory works for describing the second-order phase transition of the SQED.
Given the lessons from the SQED,
we finally provide a more generic discussion on relations between the resurgence and phase transitions.
In particular,
we generally show that the orders of phase transitions are determined by 
how saddle points collide and scatter as varying a parameter through a critical point.
We also show how the information of the phase transition is decoded from the Borel resummation technique.
We believe that our results open up potential applications of resurgence to quantum field theories.

Let us finally comment on previous works closely related to this paper\footnote{
See also \cite{Mukherjee:2014hsa,Tanizaki:2015rda,Ahmed:2018tcs,Marino:2019wra,Marino:2019eym,Marino:2019fvu,Marino:2020dgc,Marino:2020ggm,Marino:2021six}
for works indirectly related to this context.
}.
The work \cite{Kanazawa:2014qma} studied thimble structures of simple fermionic systems such as zero-dimensional versions of the Gross-Neveu model and Nambu-Jona-Lasinio model, and one-dimensional gauge theory coupled to a massive fermion with a Chern-Simons term.
In particular, it was found in the zero-dimensional Gross-Neveu model that
there was a jump in the number of contributing thimbles at the second-order chiral phase transition point in the massless case.
It was also demonstrated that there was an interesting link between anti-Stokes lines and Lee-Yang zeros.
There are also interesting works on the two-dimensional pure $U(N)$ Yang-Mills theory on lattice \cite{Buividovich:2015oju,Ahmed:2017lhl,Ahmed:2018gbt}, which is technically reduced to a unitary matrix model called the Gross-Witten-Wadia model \cite{Gross:1980he,Wadia:1980cp}\footnote{
The Painleve equations were further studied in \cite{Dunne:2019aqp,Costin:2019xql}
}.
It was found that 
there occurred a condensation of complex saddle points at the third order phase transition point in the large-$N$ limit.
Historically, physicists have studied simple models to draw lessons and to uncover general laws.
The field of resurgence and its relation to QFT phase transitions is not an exception either.
Now it is a good time to broaden the reach of resurgence toward more realistic QFTs.
The SQED studied in this paper should be a nice first step along this direction
since it is more realistic and nevertheless its partition function is expressed in a simple manner thanks to the supersymmetry.

This paper is organized as follows.
In Sec.~\ref{sec:transition}, we review the work by Russo and Tierz \cite{Russo:2016ueu}.
In Sec.~\ref{sec:thimble},
we provide interpretations of the phase transition from the viewpoints of Lefschetz thimble.
In Sec.~\ref{sec:Borel},
we discuss relations between the phase transition and resurgence structures.
In Sec.~\ref{sec:phase_trans_stokes_pheno},
given the lessons from the SQED example,
we give a more generic point of view on relations between the resurgence and phase transitions.
Sec.~\ref{sec:discussions} is devoted to conclusion and discussions.
In App.~\ref{sec:app_expansion},
we explain details on calculations of $1/N_f$ flavor expansion,
where $2N_f$ is the number of the hypermultiplets in the SQED.
In App.~\ref{sec:Lefschetz_thimble_larger_arg},
we present thimble structures for larger ${\rm arg}(N_f )$
while the main text focuses around ${\rm arg}(N_f )=0$.
In App.~\ref{app:Borel_comments},
we make some comments on the Pad\'{e}-Uniformized approximation
based on comparisons with the standard Pad\'{e} approximation in some simple examples.
In App.~\ref{sec:trans_finite_eta} and App.~\ref{sec:trans_finite_lam},
we study resurgence structures of the $1/N_f$ expansion from the viewpoint of a difference equation
for finite values of the FI parameter $\eta$ and the rescaled parameter $\lambda =\eta /N_f$ respectively. 
In App.~\ref{app:CSS},
we point out a possible relation between the Borel singularities and 
complex supersymmetric solutions found in \cite{Honda:2017qdb}.

\section{Quantum phase transition in the 3d $\mathcal{N}=4$ SQED}
\label{sec:transition}
In this section, we review the arguments of \cite{Russo:2016ueu}
to find the quantum phase transition in the 3d $\mathcal{N}=4$ SQED with a large number of hypermultiplets.
Let us consider a 3d $\mathcal{N}=4$ SUSY $U(1)$ gauge theory 
coupled to $2N_f$ hypermultiplets with charge 1.
We turn on a Fayet-Illiopoulos (FI) term and
a real mass associated with a $U(1)$ subgroup of the $SU(2N_f)$ flavor symmetry\footnote{
The $U(1)$ subgroup rotates $N_f$ hypers with charge $+1$ and the other $N_f$ hypers with charge $-1$.
}.

Applying the SUSY localization \cite{Pestun:2007rz},
the path integral is dominated by saddle points and
one can exactly compute the $S^3$ partition function of this theory as \cite{Kapustin:2009kz,Hama:2010av,Jafferis:2010un}
\begin{\eq}
Z=\int_{-\infty}^\infty d\sigma 
\frac{e^{i\eta \sigma}} {\bigl[ 2\cosh{\frac{\sigma +m}{2}} \cdot 2\cosh{\frac{\sigma -m}{2}} \bigr]^{N_f} } ,
\label{eq:localization}
\end{\eq}
where $\eta$ is the FI parameter and $m$ is the real mass.
The integral variable $\sigma$ is the Coulomb branch parameter 
and the factor in the denominator is the one-loop determinant of the hypermultiplets\footnote{
The integral \eqref{eq:localization} can be done exactly as
$Z$ $=$
$
\frac{\sqrt{2\pi}}{2^{N_f}}
\frac{\Gamma (N_f +i\eta )\Gamma (N_f -i\eta )}{\Gamma (N_f) \left( \sinh{m}\right)^{N_f -1/2} }
 P_{-\frac{1}{2}+i\eta}^{\frac{1}{2}-N_f} \left( \cosh{m}   \right) 
$
with the associated Legendre polynomial $P_\ell^m (x)$ \cite{Russo:2016ueu}
but this form does not seem particularly useful for our purpose.
}.

Now we are interested in the 't Hooft-like limit:
\begin{\eq}
N_f \rightarrow \infty ,\quad \lambda \equiv \frac{\eta}{N_{f} } ={\rm fixed}.
\end{\eq}
For this purpose, it is convenient to write the partition function as
\begin{align}
Z
= \frac{1}{2^{N_f}}\int_{-\infty}^{\infty}\dd{\sigma}\:  e^{-S (\sigma)},
\end{align}
where $S (\sigma )$ is the ``action'' defined by
\begin{align}
S (\sigma ) = N_f \Bigl[  -i\lambda\sigma + \ln(\cosh\sigma+\cosh m)  \Bigr].
\label{eq:action}
\end{align}
In the large $N_f$ limit, the integral is dominated by saddle points satisfying
\begin{\eq}
S'(\sigma) = N_f \left( -i\lambda + \frac{\sinh{\sigma}}{\cosh{\sigma}+\cosh{m}} \right) =0 .
\end{\eq}
This equation is solved by
\begin{\eq}
\sigma_n^\pm 
= \log{\left( \frac{-\lambda\cosh{m} \pm i\Delta (\lambda ,m) }{i+\lambda} \right)} +2\pi i n \quad
(n\in \mathbb{Z}),
\end{\eq}
where
\begin{\eq}
\Delta (\lambda ,m) =\sqrt{1-\lambda^2 \sinh^2{m} }.
\end{\eq}
The action at the $n$-th saddle point is 
\begin{align}
S (\sigma_n^\pm)
&= N_f \Biggl[
-\frac{i\lambda}{2}\ln\left(\frac{-i+\lambda}{\phantom{+}i+\lambda}\frac{-\lambda\cosh m\pm i\Delta}{-\lambda\cosh m\mp i\Delta}\right)
+\ln(\frac{\cosh m\pm\Delta}{1+\lambda^2})
+2\pi  n\lambda 
\Biggr].
\end{align}
Note that the saddles and the action values are complex in general.
Also, note that the action can be written as
\begin{align}
S (\sigma_n^\pm)
&=S (\sigma_0^\pm)+2\pi n N_f \lambda.
\label{eq:saddle_n}
\end{align}
This implies that
the most dominant saddle point for real $\lambda$
is either $\sigma=\sigma_0^{+}$ or $\sigma=\sigma_0^{-}$, if it contributes to the integral.

The authors in \cite{Russo:2016ueu} have observed that
dominant saddles change at $\lambda =\lambda_c$ with
\begin{\eq}
\lambda_c \equiv \frac{1}{\sinh{m} } .
\label{eq:criticalPoint}
\end{\eq}
For the subcritical region $\lambda<\lambda_c$,
	only a single saddle $\sigma_0^+$ dominates the integral.
For the supercritical region $\lambda \geq\lambda_c$,
	two saddles $\sigma_0^+$ and $\sigma_0^-$ contribute to the integral with equal weights.
\cite{Russo:2016ueu} numerically checked that
	these saddle approximations agree with the exact analytic expression of \eqref{eq:localization} at the large $N_f$ limit.
The second derivative of the ``free energy'' jumps at $\lambda =\lambda_c$ as
\begin{\eq}
\frac{d^2 F}{d\lambda^2}
= \begin{cases}
\frac{N_f}{1+\lambda^2} \left( 1 +\frac{\cosh{m} }{\sqrt{1-\lambda^2 \sinh^2{m} } } \right) & \lambda <\lambda_c \cr
\frac{N_f}{1+\lambda^2} & \lambda \geq \lambda_c 
\end{cases} .
\end{\eq}
This implies that the system exhibits 
a second-order phase transition at $\lambda =\lambda_c$.

\section{Lefschetz thimble structures}
\label{sec:thimble}
In general, saddles with smaller $\Re S$ give larger weights.
However, such saddles do not necessarily contribute to the path integral.
This is because
	its original integration contour may not be deformed to the Lefschetz thimbles (steepest descent paths) associated with such saddles.
Also note that
	the ``free energy'' of a partition function on a general manifold is not necessarily real
	since it is not interpreted as a thermodynamic one.
This means that
	contributing saddles cannot be determined
	only by requiring the free energy to be real.
For these reasons,
	we should study Lefschetz thimble structures to describe quantum phase transitions.

In this section,
we interpret the quantum phase transition in terms of Lefschetz thimbles of the integral \eqref{eq:localization}
	obtained by SUSY localization.
This provides a more precise justification for the arguments  in \cite{Russo:2016ueu}
	reviewed in the previous section.
For this purpose,
	we first extend the Coulomb branch parameter $\sigma \in {\mathbb R}$ to complex values $z \in {\mathbb C}$,
	since saddle points and the associated Lefschetz thimbles are complex-valued in general. 
As we have seen in the last section,
	we have infinitely many saddle points $\sigma_n^\pm$ satisfying the saddle point equation.
Contributing saddles are determined by looking at the Lefschetz thimbles (or the steepest descents)
	obtained by deforming the original contour 
	without changing the value of the integral.
This can depend on the original integral contour,
	the parameters $(\lambda ,m)$ and properties of the (dual) Lefschetz thimbles as explained below.

The Lefschetz thimbles $\mathcal{J}_n^\pm$
associated with the saddle points $\sigma_n^\pm$ 
are defined as solutions of the differential equation called the flow equation,
\begin{align}
\left. \frac{d z}{ds} \right|_{\mathcal{J}_n^\pm} & 
=  \, \overline{\frac{\partial S[z]}{\partial z}}
\end{align}
with the initial conditions 
\begin{align}
  \lim_{s \rightarrow -\infty} z(s) = \sigma_n^\pm ,
\label{eq:ini_cond_thim}
\end{align}
where $s$ is the flow parameter along the Lefschetz thimbles.
%
%
%
Using the flow equation, we can easily prove the following properties
\begin{align}
 \left. \frac{d}{ds} \, {\rm Re} S[z(s)] \right|_{\mathcal{J}_n^\pm} \, \ge \, 0 
  \quad{\rm and}\quad
\left.  \frac{d}{ds} \, {\rm Im} S[z(s)] \right|_{\mathcal{J}_n^\pm} \, = \, 0 ,
\end{align}
which indicate that
integrals along Lefschetz thimbles are rapidly convergent and non-oscillating.
We can express the original contour ${\cal C}_{\mathbb R}$
as a linear combination of the Lefschetz thimbles 
\begin{align}
{\cal C}_{\mathbb R} \,
=\, \sum_\pm \sum_{ n=-\infty}^\infty k_n^\pm {\cal J}_n^\pm .
\end{align}
If $k_n^\pm$ is nonzero, 
it implies that $\sigma_n^\pm$ contributes to the integral
while we have no contributions from the saddle points with $k_n^\pm =0$.
It is known that each expansion coefficient $k_n^\pm$ is an integer
since $k_n^\pm$ is identified with the intersection number 
between the original contour ${\cal C}_{\mathbb R}$
and the dual thimble (or the steepest ascent contour) $\mathcal{K}_n^\pm$ 
associated with $\sigma_n^\pm$,
which is defined by
\begin{align}
\left. \frac{d z}{ds} \right|_{\mathcal{K}_n^\pm}  
= \overline{\frac{\partial S[z]}{\partial z}} \quad\quad
{\rm with}\   \lim_{s \rightarrow +\infty} z(s) = \sigma_n^\pm . 
\end{align}
In general, $k_n^\pm$ depends on $(g,m)$ but its dependence is not continuous
since $k_n^\pm$ is an integer.
Typically $k_n^\pm$ is a constant or a step function, 
and the latter case leads us to a Stokes phenomenon.

\subsection{Real positive $N_f$}
\begin{figure}[t]
\centering
\includegraphics[width=75mm]{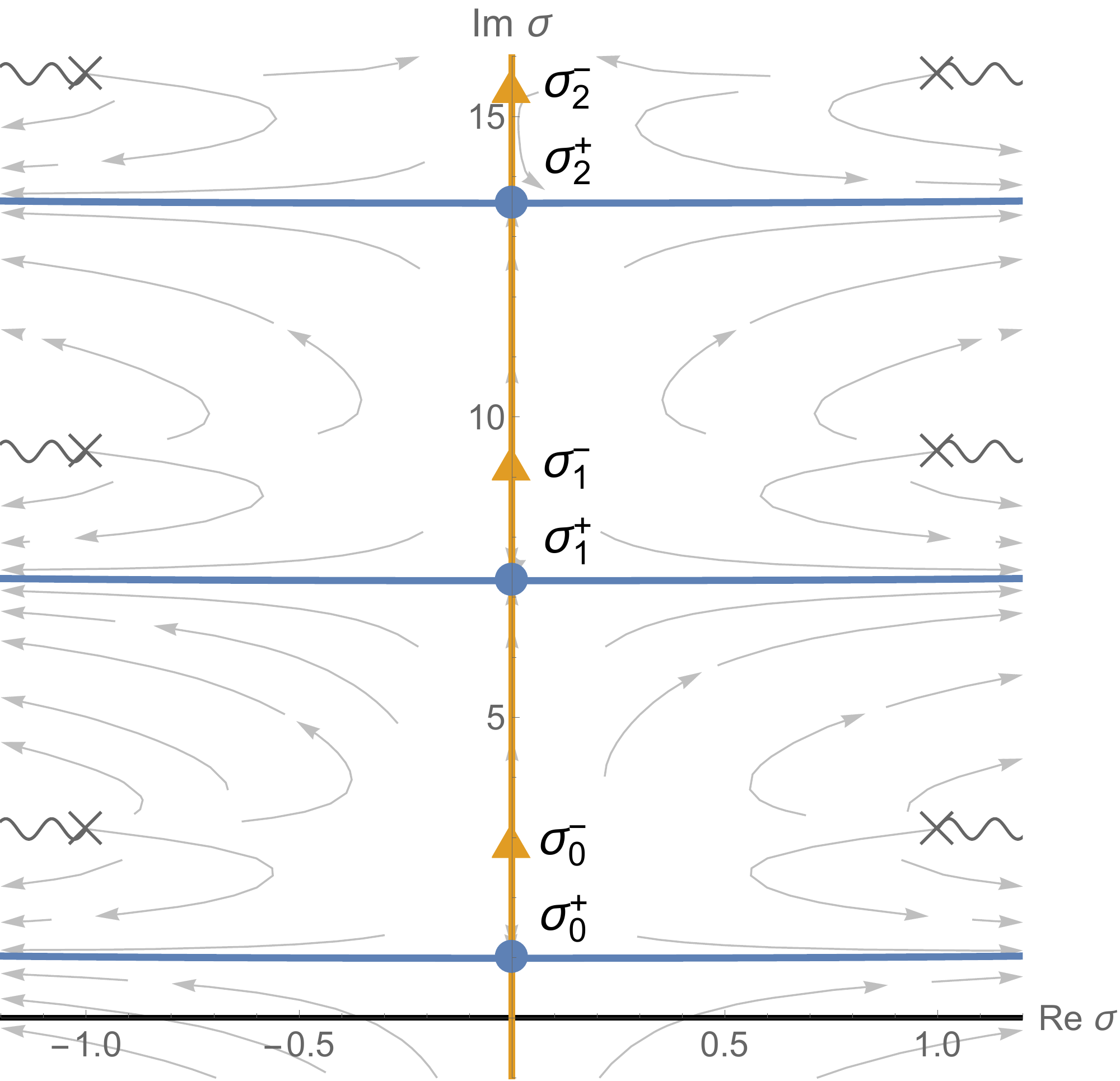}
\hspace{8mm}
\includegraphics[width=75mm]{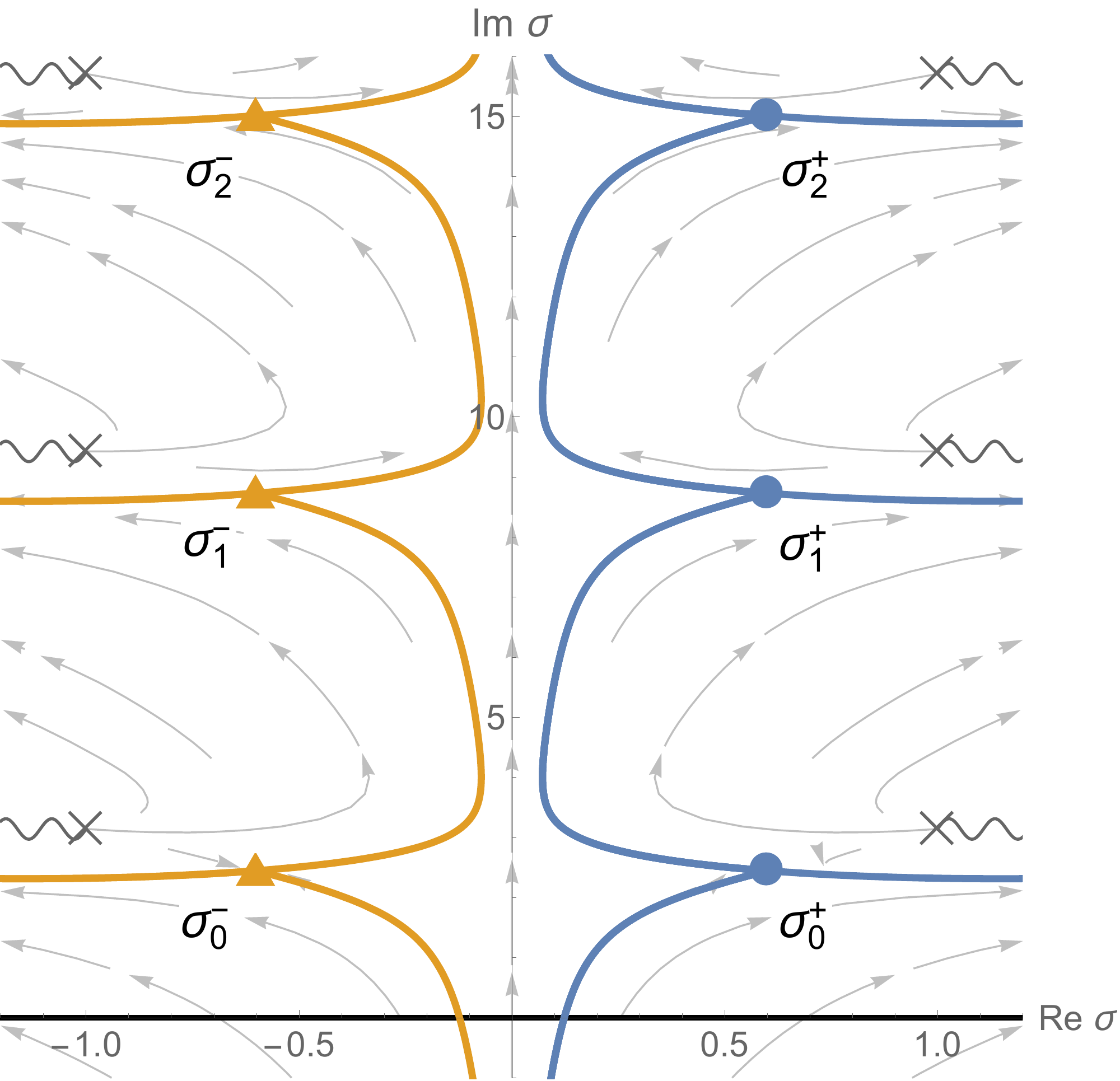}
\caption{
Illustrations of the Lefschetz thimble structures of the integral \eqref{eq:localization} for ${\rm arg}(N_f)=0$
(In these figures, $m=1$, for which $\lambda_c \simeq 0.85$).
The blue circle and orange triangle symbols
	indicate the saddles $\sigma_n^{+}$ and $\sigma_n^{-}$, respectively.
The Lefschetz thimbles associated with them are drawn as the lines with the same colors.
The gray wavy lines and cross symbols
	indicate the branch cuts and singularities of the action \eqref{eq:action},
	which essentially come from the logarithm.
[Left] At $\lambda=0.4<\lambda\crit$ as a representative of the subcritical region.
[Right] At $\lambda=1.2>\lambda\crit$ as a representative of the supercritical region.
}
\label{fig:thimble1}
\end{figure}

First, let us briefly see the Lefschetz thimble structures for real positive $N_f$ 
	i.e.~${\rm arg}(N_f )=0$. 
We have numerically solved the flow equations 
	and drawn the results at some representative values of $(m,\lambda )$ in Fig.~\ref{fig:thimble1}.
%
%
%
We immediately see that
the Lefschetz thimbles pass multiple saddle points
	both in the subcritical ($\lambda<\lambda_c$) and supercritical ($\lambda\geq\lambda_c$) regions.
Although we have explicitly shown the results only at the two values of $(m,\lambda )$,
we have checked that this feature remains to hold unless the parameters cross the phase boundary.
The thimble structures imply that
the decomposition in terms of the thimbles is not well-defined at ${\rm arg}(N_f)=0$
and the Stokes coefficient has a discrete change. 
In other words, the present case ${\rm arg}(N_f )=0$ is on Stokes lines.

The appearance of the Stokes lines here is natural
because we have infinitely many saddle points with the same imaginary part of the action at
${\rm arg}(N_f )=0$
although this is not sufficient but necessary to have the Stokes lines.
%
%
%
This can be explicitly seen as follows.
In the subcritical region $\lambda <\lambda_c$,
one can easily show that all the saddle points are purely imaginary and
their actions are real:
\begin{\eq}
{\rm Im}( S_n^\pm ) =0 \quad {\rm for}\ \lambda <\lambda_c .
\end{\eq}
In the supercritical region $\lambda >\lambda_c$,
the imaginary parts of the actions at the saddles are nonzero 
but they satisfy
\begin{\eq}
{\rm Im } (S_n^\pm  ) = -{\rm Im} (S_n^\mp ) \ \  {\rm and}\ \  
{\rm Im } (S_n^\pm  ) = {\rm Im} (S_0^\pm ) 
\quad {\rm for}\ ^\forall n\ {\rm and}\ \lambda >\lambda_c .
\end{\eq}
The above structure essentially comes from the fact that
the action at the saddles depends on $n$
only via the term ``$2\pi n \lambda N_f$'' as seen from \eqref{eq:saddle_n}.
This motivates us to take complex $N_f$ to go beyond the Stokes lines
and understand the thimble structures more precisely.

\subsection{Complex $N_f$}
\label{sec:complexNf}
\begin{figure}[t]
\centering
\includegraphics[width=75mm]{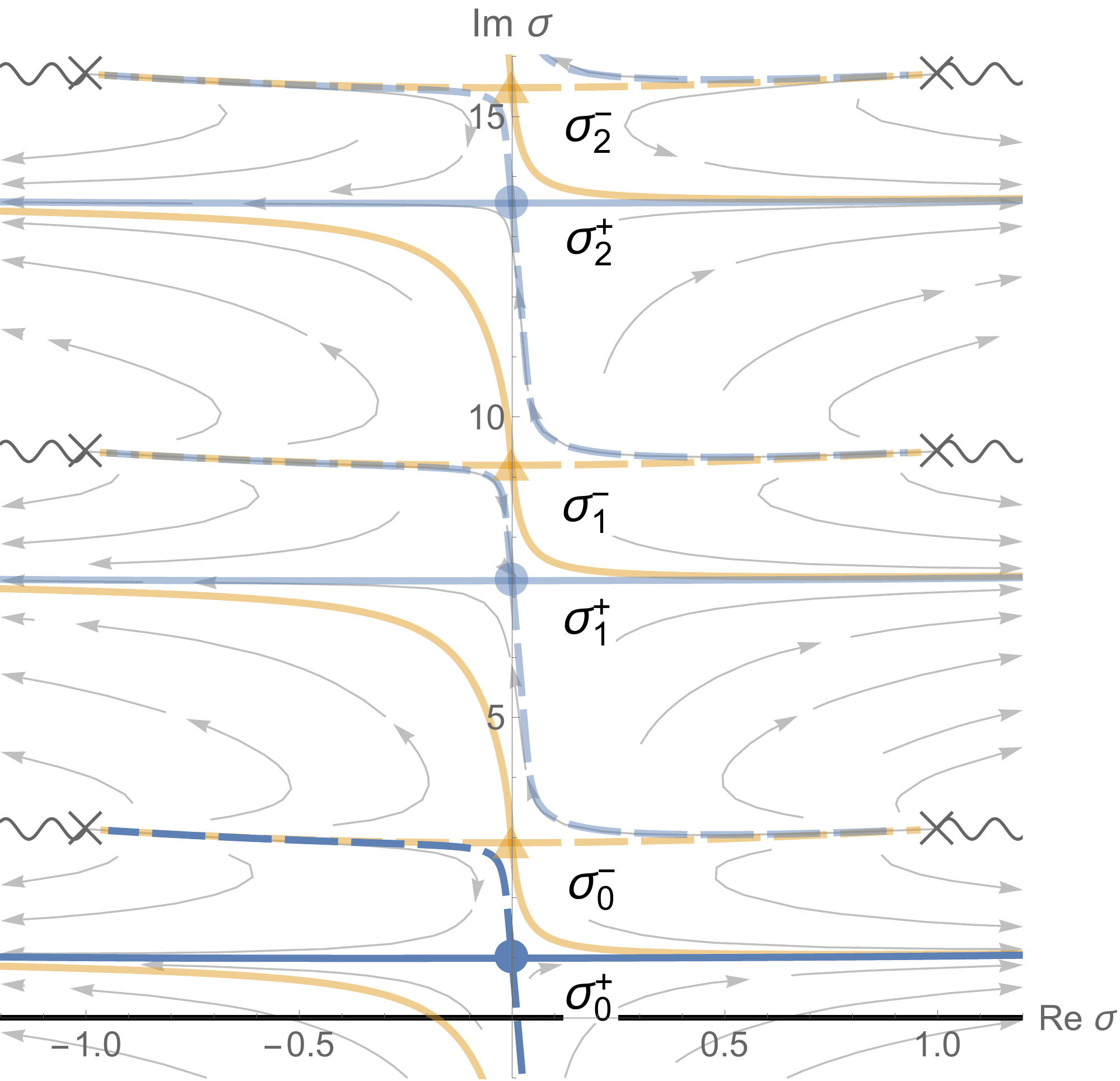}
\hspace{8mm}
\includegraphics[width=75mm]{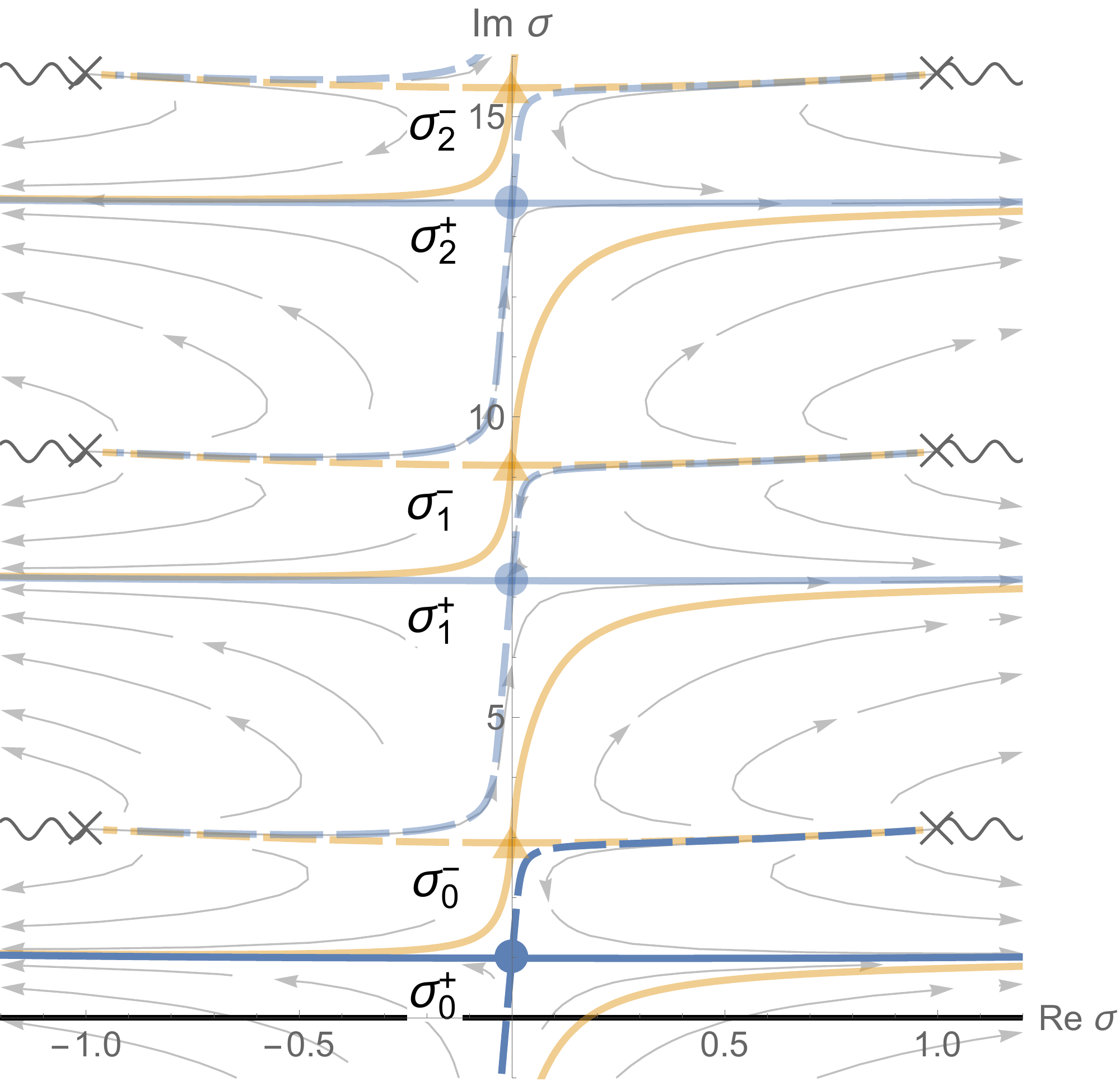}
\caption{
Illustrations of the Lefschetz thimble structure
	for the subcritical region $\lambda=0.4<\lambda\crit$ with $m=1$.
Small phases $\theta=-0.025$ (left) and $\theta=+0.025$ (right) are given
	to illustrate behaviors around $\theta =0$.
The solid lines show the Lefschetz thimbles
while the dashed lines are the dual thimbles
	associated with saddle points.
The opaque saddles and thimbles contribute to the integral,
while the translucent ones do not.
In this case, the contributing saddle $\sigma_0^+$ does not change for any small $\theta$.
}
\label{fig:thimble2}
\end{figure}

Let us take $N_f$ to be complex while keeping $\lambda$ real\footnote{
To keep $\lambda $ real, we should also take ${\rm arg}(\eta )=\theta $
since $\lambda =\eta /N_f$.
}:
\begin{align}
\theta \equiv {\rm arg}(N_f ),
\end{align}
and study the Lefschetz thimble structures.
In the main text we study the thimble structures only around $\theta =0$.
See App.~\ref{sec:Lefschetz_thimble_larger_arg} for the non-small $\theta$ case.

Let us first focus on the subcritical region $\lambda<\lambda\crit$ presented in Fig.~\ref{fig:thimble2}.
Regardless of the sign of $\theta$,
	the dual thimble $\mathcal{K}_0^+$ intersects once with the original integration contour ${\cal C}_{\mathbb R}$.
This means that
	the original integral contour ${\cal C}_{\mathbb R}$ can be deformed to the thimble $\mathcal{J}_0^+$.
Indeed, we can apply Cauchy's integral formula since the integrand decreases at infinity of the upper-half plane.
Thus, we find a unique thimble decomposition 
\begin{align}
    {\cal C}_{\mathbb R} = k_0^+\mathcal{J}_0^+ \quad \mbox{with} \quad
    k_0^+ = 1.
\end{align}
In other words, there is no Stokes phenomenon in the subcritical region.

\begin{figure}[t]
\centering
\includegraphics[width=65mm]{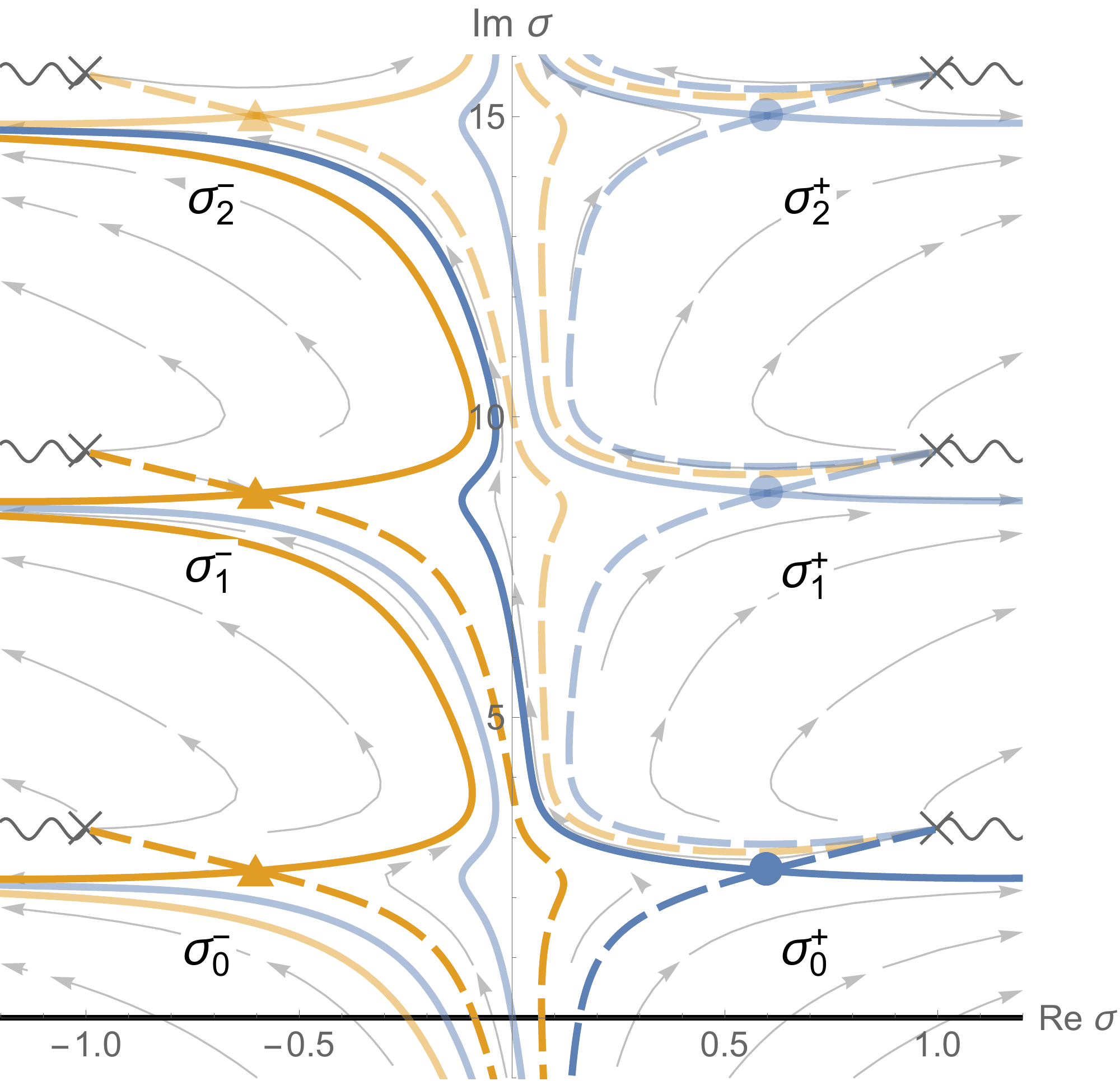}
\hspace{8mm}
\includegraphics[width=65mm]{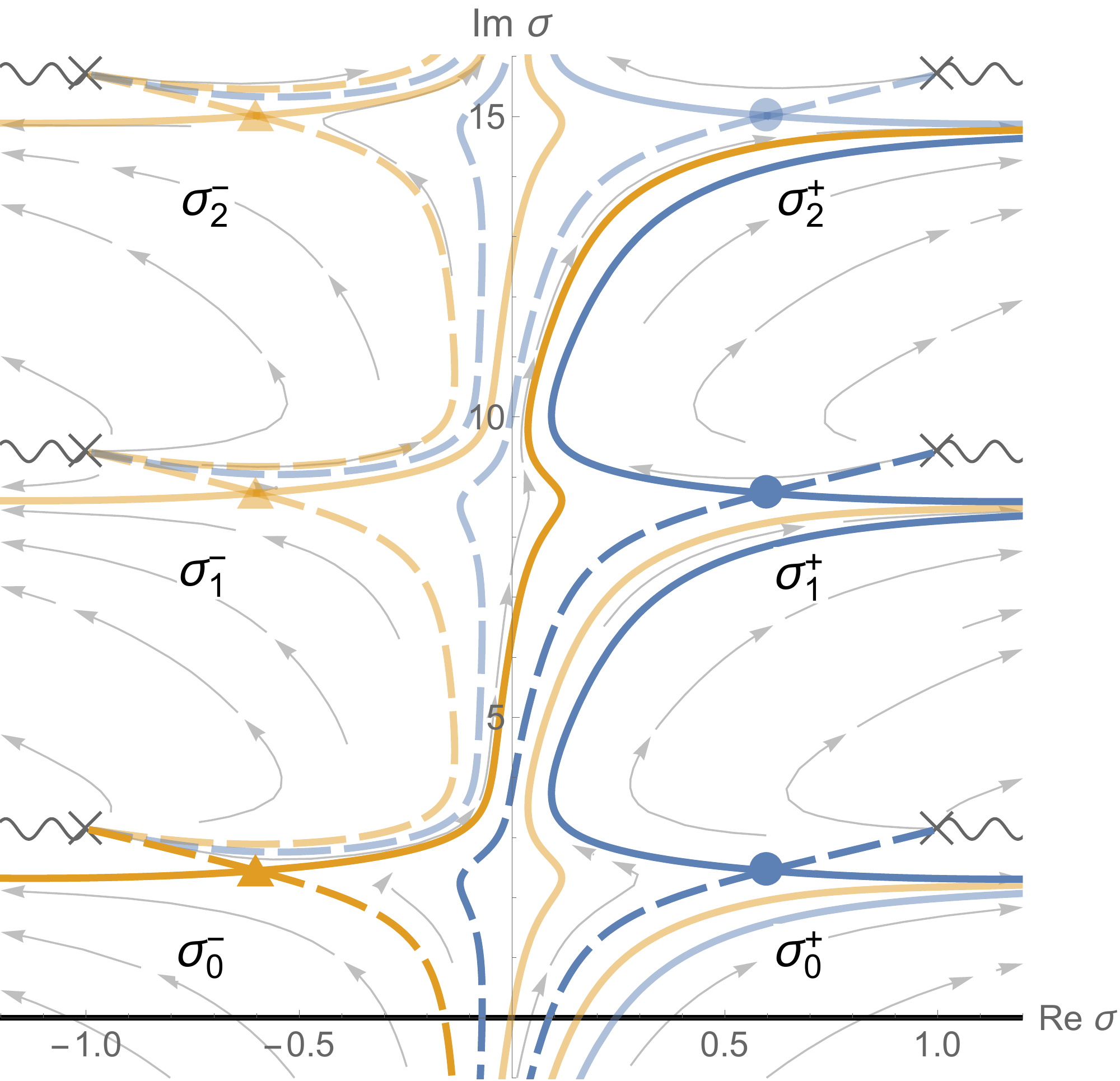} \\
\vspace{5mm}
\includegraphics[width=65mm]{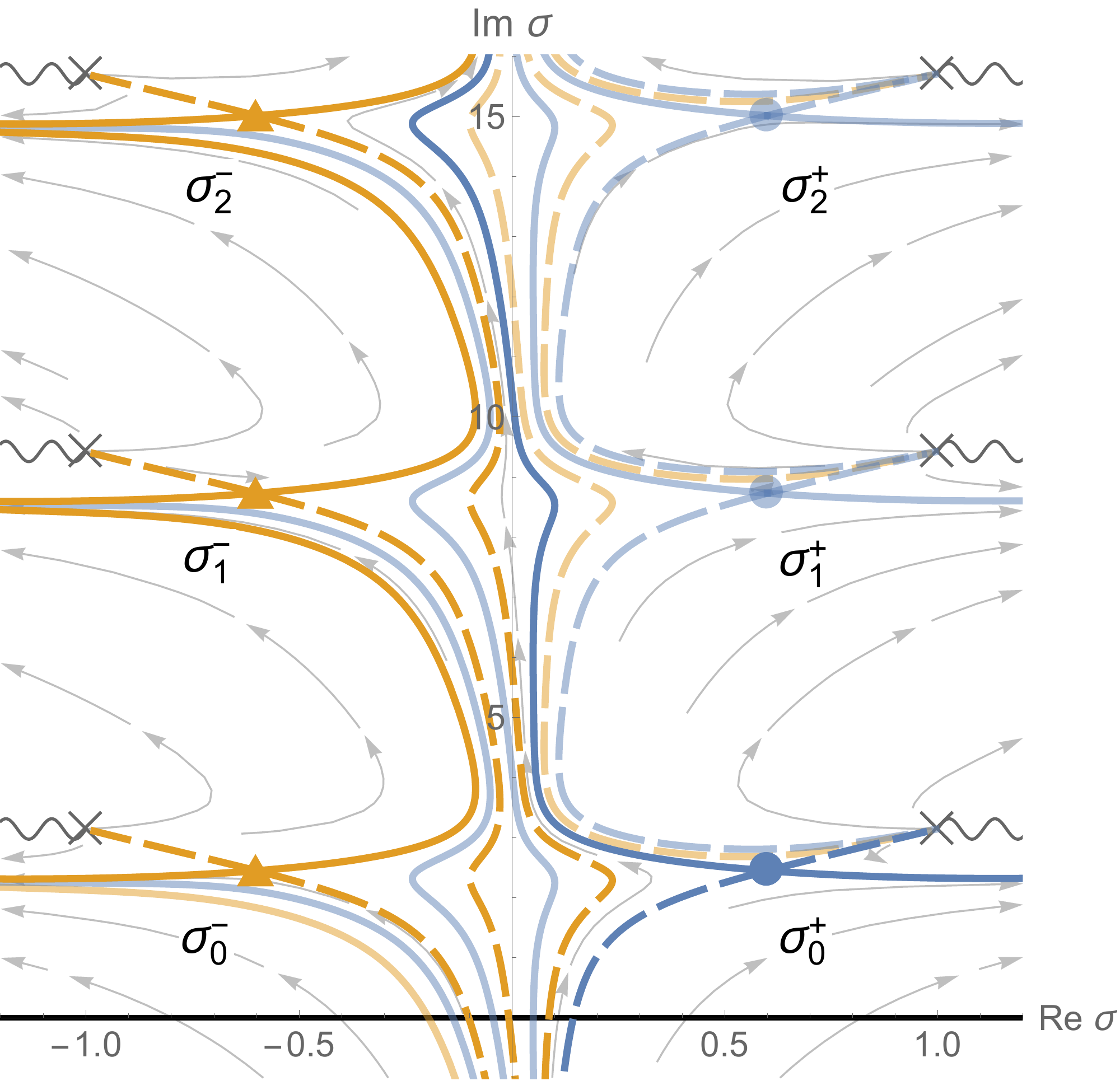}
\hspace{8mm}
\includegraphics[width=65mm]{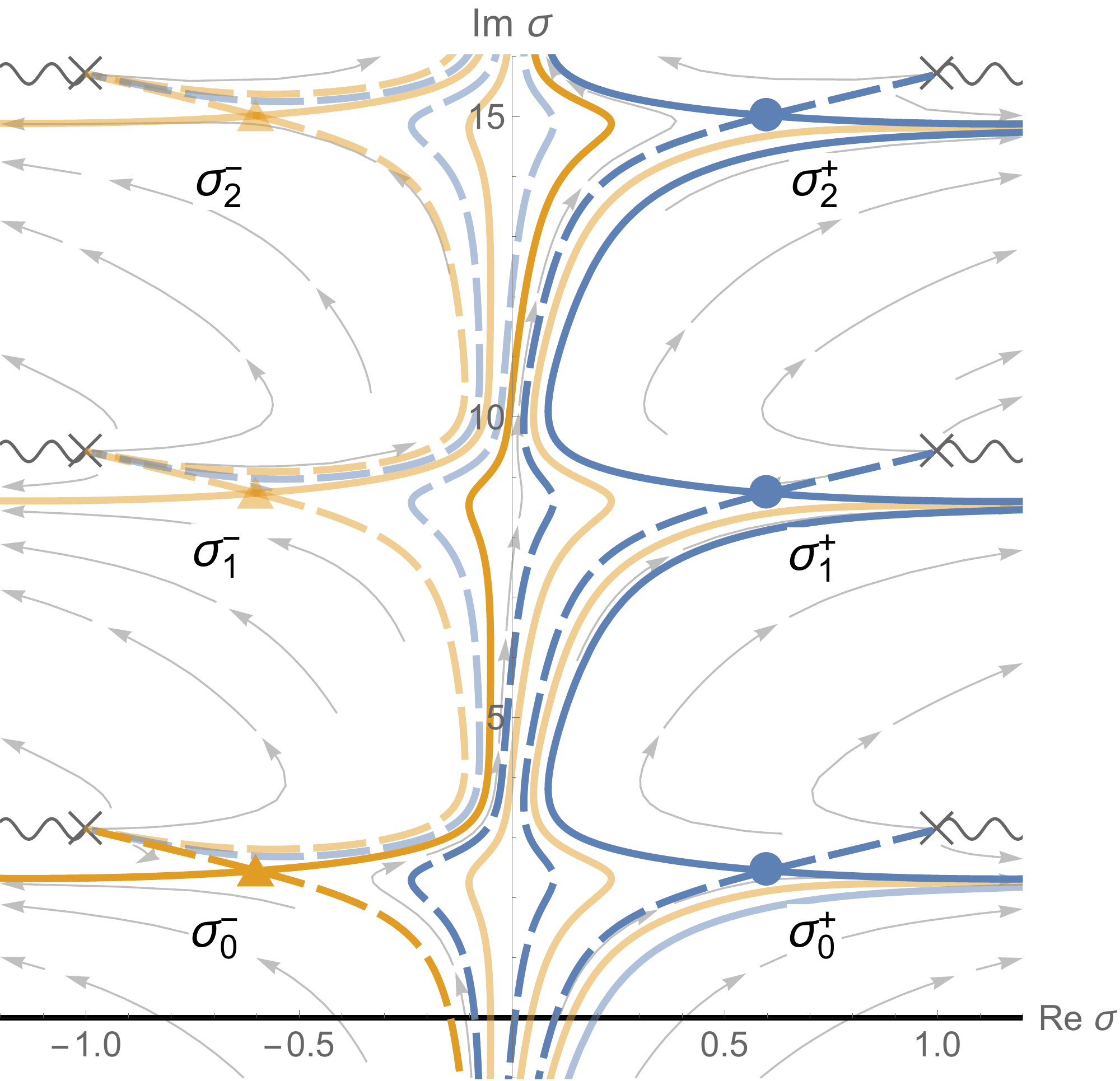}
\caption{
Illustrations of the thimble structure for the supercritical region $\lambda = 1.2 > \lambda\crit$ with $m=1$.
Small phases
	$\theta=-0.025$ (left-top),
	$\theta=+0.025$ (right-top),
	$\theta=-0.015$ (left-bottom), and
	$\theta=+0.015$ (right-bottom)
are given to illustrate behaviors around $\theta =0$.
Contrary to the subcritical region, the contributing saddles change discontinuously.
As the phase approaches $\theta\rightarrow \pm 0$,
	saddle points $\sigma_n^{\pm}$ with larger $n$ contribute to the path integral.
}
\label{fig;thimble3}
\end{figure}

%
The structures in the supercritical region $\lambda > \lambda\crit$ are shown in Fig.~\ref{fig;thimble3}.
We observe qualitatively different behaviors compared with the subcritical region\footnote{
It is common that the saddle points $\sigma_n^\pm$ with $n<0$ never contribute.
}.
Firstly,
multiple saddle points contribute to the integral.
In particular, two saddle points $\sigma_0^\pm$ always contribute to it
	as their dual thimbles always intersect with the original integral contour ${\cal C}_{\mathbb R}$.
This justifies the arguments by \cite{Russo:2016ueu} reviewed in Sec.~\ref{sec:transition}.
Note that this fact is a priori nontrivial since the saddles are complex.
Secondly,
the intersection numbers $k^{\pm}_{n\geq 1}$ jump discontinuously as the phase $\theta$ is changed.
The thimble structures depend not only on the sign of $\theta$ but also on the absolute value $|\theta |$.
Specifically, a common feature for each sign $\pm$ of $\sigma_n^\pm$ is that
	the saddle points $\sigma_{n\geq 1}^\pm$ do not contribute for ${\rm sign}(\theta ) =\mp $.
The dependence on the absolute value $|\theta |$ is more intricate.
At $\theta = +0.025$ shown in Fig.~\ref{fig;thimble3} [right top],
	we see that $\sigma_1^+$ contributes
	while $\sigma_n^+$ with $n\geq 2$ do not contribute.
Similarly, at $\theta = + 0.015$ shown in Fig.~\ref{fig;thimble3} [right bottom],
	we see that $\sigma_1^+$ and $\sigma_2^+$ contribute
	but $\sigma_n^+$ with $n\geq 3$ do not contribute.
As we further decrease $|\theta |$,
	we have found the following structure (although we do not explicitly show the plots).
For finite $\theta >0$ ($\theta <0$),
	we have contributions from the saddle points $\sigma_n^+$ with $1\leq n \leq M_+ (\theta )$
	($\sigma_n^-$ with $1\leq n \leq M_- (\theta  )$)
	where $M_\pm (\theta )$ is an integer such that
\begin{\eq}
M_\pm (\theta ) \rightarrow \infty \quad {\rm as}\ \ \theta\rightarrow  \pm 0.
\end{\eq}
In summary, the above analysis suggests that the thimble structure in the supercritical region is
\begin{align}
    {\cal C}_{\mathbb R}
    = k_0^+\mathcal{J}_0^+
    + k_0^-\mathcal{J}_0^-
    + \sum_{\pm,n\geq1}k_n^{\pm}\mathcal{J}_n^{\pm},
\end{align}
where
\begin{align}
    k_0^+ = k_0^- = 1, \quad
    k_{n\geq 1 }^+ = \left\{\begin{array}{ll}
        0 & (\theta= -0 ) \\
        1 & (\theta= +0 )
    \end{array}\right., \quad
    k_n^- = \left\{\begin{array}{ll}
        1 & (\theta= -0 ) \\
        0 & (\theta= +0 )
    \end{array}\right..
\end{align}
This indicates an infinite number of Stokes phenomena at $\theta =0$ in the supercritical region.

\subsection{Phase transition and thimble structures}
\label{sec:transition_thimble}
We clarify a relation between the phase transition and the above Lefschetz thimble analysis.
Whether the saddles contribute to the integral was a priori nontrivial
	since all of them are complex for $\lambda>0$.
Our Lefschetz thimble analysis showed that
	only a single saddle $\sigma_0^+$ contributes to the integral in the subcritical region $\lambda<\lambda\crit$,
	while multiple saddles $\sigma_n^{\pm}$ with $n\geq0$ contribute in the supercritical region $\lambda\geq\lambda\crit$.
Among them, only two saddles $\sigma_0^{\pm}$ survive the large-flavor limit.
Thus, the dominant saddles jump at the critical point $\lambda=\lambda\crit$
	from $\sigma_0^+$ to $\sigma_0^{\pm}$,
	which causes the phase transition.
These provide a more precise interpretation of the phase transition in terms of the Lefschetz thimble analysis.
All of these behaviors come from (anti-)Stokes phenomena and
therefore these motivate us to study, in more detail, a relation between the phase transition and (anti-)Stokes phenomena.
This will be summarized in Sec.~\ref{sec:phase_trans_stokes_pheno}.

Also, note that the periodicity of the action \eqref{eq:action} along the imaginary axis causes an infinite number of Stokes phenomena at $\arg(N_f)=0$.
Such property is typical in sphere partition functions of $\mathcal{N}\geq 4$ supersymmetric gauge theories with FI terms and without diagonal Chern-Simons terms.
These infinite number of Stokes phenomena are inevitably related to the phase transition
as we will see more details in Sec.~\ref{sec:phase_trans_stokes_pheno}.

Finally, we shortly provide some preparations for the next section.
We have collected the thimble structures for larger values of $\arg(N_f)$
as summarized in App.~\ref{sec:Lefschetz_thimble_larger_arg}.
For non-small $\arg(N_f)$,
we have encountered a subtlety essentially coming from the logarithmic branch cuts in the action \eqref{eq:action}:
when a thimble crosses the branch cuts once, the action changes its value by $\pm 2\pi i N_f $
where the sign depends on a direction of the crossing.
Note that this modifies the condition for having Stokes phenomena
as the imaginary part of the action is changed.
App.~\ref{sec:Lefschetz_thimble_larger_arg} demonstrates that 
the Stokes phenomena due to this effect indeed happen in our problem.
For instance, in the subcritical region $(\lambda ,m) =(0.4, 1)$,
we have numerically found that 
the Lefschetz thimble associated with the saddle $\sigma_0^+$ crosses one of the branch cuts once and
then passes the neighboring saddle $\sigma_1^+$ around $\arg(N_f)\simeq -1.190$.
The appearance of this Stokes phenomenon cannot be understood 
without taking the effect of the branch cuts into account as follows.
The existence of the branch cut implies that
the condition for having Stokes phenomena between the saddles $\sigma_n^+$ and $\sigma_0^+$ is
modified as
\begin{align}
\Im \left[ S(\sigma_{n}^+ )-S(\sigma_0^+) +2\pi i N_f \mathbb{Z} \right]  = 0  ,
\end{align}
which is solved by
\begin{align}
\tan{\theta} = -\frac{1}{n\lambda} \mathbb{Z} .
\end{align}
This condition specifically for the above case corresponds to 
$\theta =\left. {\rm Arctan}(-1/\lambda ) \right|_{\lambda =0.4} \simeq -1.190$,
which agrees with the result in App.~\ref{sec:Lefschetz_thimble_larger_arg}.
Similarly in the supercritical region $(\lambda ,m) =(1.2 ,1)$,
we have found the Stokes phenomenon between $\sigma_1^-$ and $\sigma_0^+$
coming from the effect of the branch cuts for $\theta\simeq = -0.039$.

From the viewpoint of the resurgence theory,
	the information of Stokes phenomena is encoded in the $1/N_f$ expansion of the partition function \eqref{eq:localization}.
Stokes phenomena observed in the figures in App.~\ref{sec:Lefschetz_thimble_larger_arg}
	are associated with the Borel singularities of the $1/N_f$ expansion.
Relations between Stokes phenomena and Borel singularities in one-dimensional integrals are shown in \cite{Berry:1991,Boyd:1993,Boyd:1994}
although some of the assumptions there are violated in the SQED
due to the logarithmic branch cuts in the action.
We will discuss corresponding Borel singularities in the next section.

\section{Borel singularities and resurgence structure}
\label{sec:Borel}
%
In this section,
we consider the $1/N_f$ expansion of the partition function \eqref{eq:localization} and
study its resurgence structure from the viewpoint of 
the Borel resummation method.
We numerically compute the $1/N_f$ expansion up to 50th order
and then study the structures of the Borel singularities.
We confirm that the locations of the Borel singularities are consistent with the Lefschetz thimble structure.
The resurgence structure of trans-series with respect to $\eta$ or $\lambda$,
	instead of $1/N_f$,
	is discussed in App.~\ref{sec:trans_finite_eta} and App.~\ref{sec:trans_finite_lam}.

\subsection{Numerical study of Borel singularities}
\label{sec:Borel_pade}
Let us focus on the $1/N_f$ expansion around the saddle point $\sigma=\sigma_0^+$.
It can be computed in the standard way 
and the expansion takes the form (see App.~\ref{sec:app_expansion} for details)
\begin{align}
	\int_{\mathcal{J}_0^+}\dd{\sigma}\:
	e^{-N_f\tilde{S}(\sigma)}
&=	e^{-N_f\tilde{S}(\sigma_0^+)}
 \sqrt{\frac{2\pi}{N_f \tilde{S}\pp(\sigma_0^+)}}
	\sum_{\ell =0}^{\infty} \frac{a_\ell}{N_f^{\ell  }},
\end{align}
where 
\begin{\eq}
S(\sigma)=N_f \tilde{S}(\sigma) .
\end{\eq}
The coefficients are given by
\begin{align}
	a_\ell
	= \sum_{n=0}^{2\ell}
	\frac{(-1)^n\Gamma\left(\frac{1}{2}+\ell +n\right)\tilde{c}_{2\ell -n}(n)}
	{\Gamma(1/2)\Gamma(n+1)}
\label{eq:a_l}
\end{align}
where
\begin{align}
	\tilde{c}_k(n)
	= \left.\left(
	\sum_{k\p=0}^{\infty}c_{k\p}\epsilon^{k\p}
	\right)^n\right|_{\epsilon^k}, \quad
	c_k
	= \frac{2^{\frac{k+3}{2}}} 	{(k+3)! }
	 \frac{\tilde{S}^{(k+3)}(\sigma_0^+)}	{\bigl( \tilde{S}\pp(\sigma_0^+) \bigr)^{\frac{k+3}{2}}}.
\end{align}

This implies that the coefficient $a_\ell$ grows factorially and
the formal $1/N_f$ expansion is not convergent. 
Therefore we apply the Borel resummation technique to control the divergence.
Let us write the perturbation series as
\begin{align}
F\left(\frac{1}{N_f}\right) = \sum_{\ell =0}^{\infty}\frac{a_l}{N_f^\ell},
\end{align}
and define its Borel transformation by
\begin{align}
\mathcal{B}F(t) = \sum_{l=0}^{\infty}\frac{a_\ell}{\Gamma (\ell+1)}t^\ell
= \sum_{\ell =0}^{\infty}b_\ell t^\ell.
\label{eq:Borel_trans}
\end{align}
Then, the Borel resummation of the function $F(1/N_f)$ is given by
\begin{align}
F_C\left(\frac{1}{N_f}\right)
= N_f\int_C\dd{t}\:
e^{-N_f t} \widetilde{\mathcal{B}F}(t),
\label{eq:Borel_sum}
\end{align}
where $\widetilde{\mathcal{B}F}(t)$ is a simple analytic continuation of the series
\eqref{eq:Borel_trans} and
the integration contour $C$ is chosen so that $\arg(N_f t)=0$.
The Borel resummation discontinuously changes
if the integration contour crosses a singularity of $\mathcal{B}F(t)$.
From the viewpoint of resurgence,
the Borel singularities must correspond to Stokes phenomena
summarized in App.\;\ref{sec:Lefschetz_thimble_larger_arg}.

\subsubsection{Pad\'{e} approximation and its improvement}
As seen from \eqref{eq:Borel_trans},
the Borel transformation is defined in terms of an infinite number of the perturbative coefficients.
In practice, we often encounter the situation where
we know only the finite number of the coefficients and
have to estimate (the analytic continuation of) the Borel transformation from the limited perturbative data in some way.
One of the standard ways to do this is the so-called Borel-Pad\'{e} approximation,
where we replace the Borel transformation $\widetilde{\mathcal{B}F}(t)$ in \eqref{eq:Borel_sum}
by its Pad\'{e} approximation.
The Pad\'{e} approximation with degrees $(m,n)$ is defined by a rational function
\begin{align}
\mathcal{P}_{m,n}(t) = \frac{P_m (t)}{Q_n (t)} ,
\label{eq:Pade_def}
\end{align}
where $P_m (t)$ and $Q_n (t)$ are polynomials of degrees $m$ and $n$, respectively.
The explicit forms of the polynomials are determined such that
the small-$t$ expansion of $\mathcal{P}_{m,n}(t)$ agrees with the Borel transformation up to a desired order $L$: 
\begin{align}
\sum_{\ell =0}^{L}b_\ell t^\ell = \mathcal{P}_{m,n}(t) + \mathcal{O}(t^{L+1}) 
\quad {\rm with}\ m+n=L .
\end{align}
While there are various possible choices of $(m,n)$ given $L$,
it is empirically known that
the Pad\'{e} approximation often has better accuracy when $m$ and $n$ are close.
Therefore we here take $m=n=L/2$ with even $L$ and
do not pursue $(m,n)$-dependence.
This particular case is called the diagonal Pad\'{e} approximation.
In practice, we will present results for $L=50$ as a representative.

In general, the Pad\'{e} approximation is good at approximating meromorphic functions since it has only pole-type singularities.
For cases with branch cuts,
the Pad\'{e} approximation typically becomes worse 
as it is impossible to express branch cuts in terms of a rational function in the exact sense.
It is known that
when the Pad\'{e} approximation works for the cases with branch cuts,
there appear sets of dense poles around locations of the branch cuts\footnote{
In other words, many poles in the Pad\'{e} are consumed to resemble the jump around the branch cuts and
we typically need larger $(m,n)$.
}.
It is also known that
the Pad\'{e} approximation generically gives better descriptions for singularities closer to the origin $t=0$.
In particular, the location of the closest singularity is expected to be predicted when $(m,n)$ is larger
because this information is closely related to the radius of the convergence of the small-$t$ expansion.
In other words, it is typically hard to detect Borel singularities far away from the origin
when $(m,n)$ is not-so-large.
%

There are various ways to improve the Pad\'{e} approximation \eqref{eq:Pade_def}.
Here we use 
one of the improvements called the Pad\'{e}-Uniformized approximation \cite{Costin:2020pcj},
which can be used when we know information on the location of a branch cut 
in the Borel transformation $\widetilde{\mathcal{B}F}(t)$.
This is constructed as follows\footnote{
Another way to treat branch cuts is to use functions with branch cuts for approximation (e.g. fractional power of some simple functions). 
See \cite{Sen:2013oza,Beem:2013hha,Alday:2013bha,Honda:2014bza,Honda:2015ewa,Chowdhury:2016hny} for such approaches.
}.
Suppose that the function $\widetilde{\mathcal{B}F}(t)$ has a branch cut ending at $t=s$.
We send the Borel $t$-plane to a $u$-plane by the uniformization map
\begin{align}
t\mapsto 
u(t) = -\ln \left( 1-\frac{t}{s} \right) ,
\label{eq:uniform}
\end{align}
which can be inverted as
\begin{align}
u \mapsto t(u) = s(1-e^{-u}) .
\end{align}
Note that the singularity at $t=s$ in the $t$-plane is mapped to infinity in the $u$-plane.
Then we construct the standard Pad\'{e} approximation in the $u$-plane,
meaning that we construct a rational approximation $\mathcal{P}_{m,n}(u)$ such that
\begin{align}
\mathcal{P}_{m,n}(u) 
=\sum_{\ell =0}^{L}b_\ell^\prime u^\ell  + \mathcal{O}(u^{L+1}) 
\quad {\rm with}\ m+n=L ,
\end{align}
where the coefficient $b_\ell'$ is defined to satisfy
\begin{align}
\sum_{\ell =0}^{L}b_\ell^\prime u^\ell  
= \sum_{\ell =0}^{L} b_\ell \left( t(u) \right)^\ell + \mathcal{O}(t^{L+1})  .
\end{align}
Finally, we come back to the $t$-plane and
approximate the Borel transformation as
\begin{align}
\widetilde{\mathcal{B}F}(t) \simeq  \mathcal{P}_{m,n}(u(t) ) . 
\end{align}
The uniformization map sends the branch cut singularity to infinity, where the standard Pad\'{e} approximation does not see.
Instead, it sends a region away from the branch cut singularity to a region around the origin.
Thus, we can avoid pole resources of the Pad\'{e} approximation being wasted on the branch cut.
Also, thanks to the logarithm of the uniformization map,
the order of Borel singularities affect only the scale of $u$.
For example, $1/(1-t)^{1/n}$ is mapped just to $e^{u/n}$
and logarithmic singularities are mapped to regular points.
Another good example is $\ln(1-t)$, which is simply mapped to $-u$.
For the above reasons,
the Pad\'{e}-Uniformized approximation is safer than the standard Pad\'{e} approximation
in application to our problem.

\subsubsection{Subcritical region}
\begin{figure}[t]
\centering
\includegraphics[width=74mm]{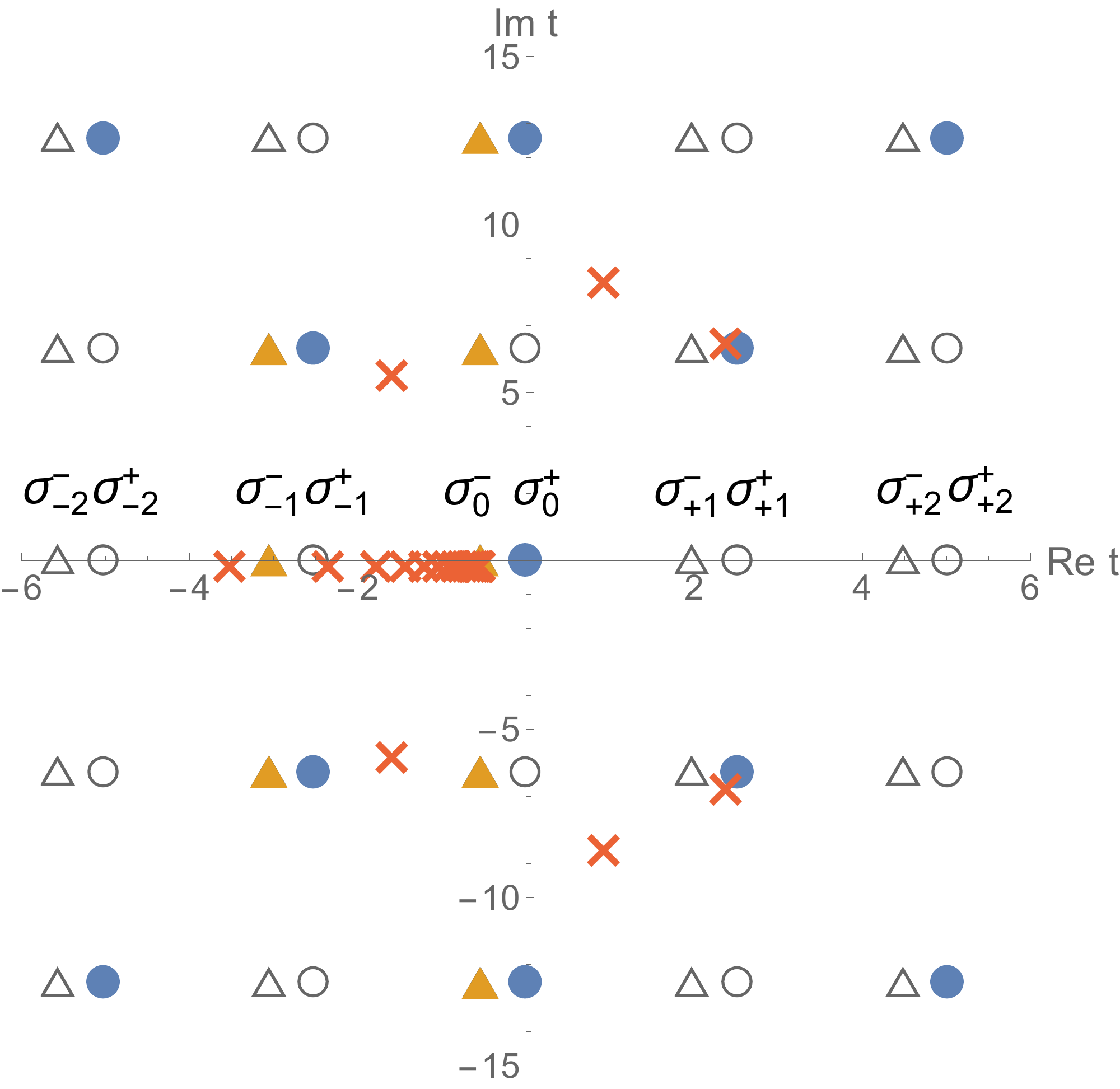} \hspace{8mm}
\includegraphics[width=74mm]{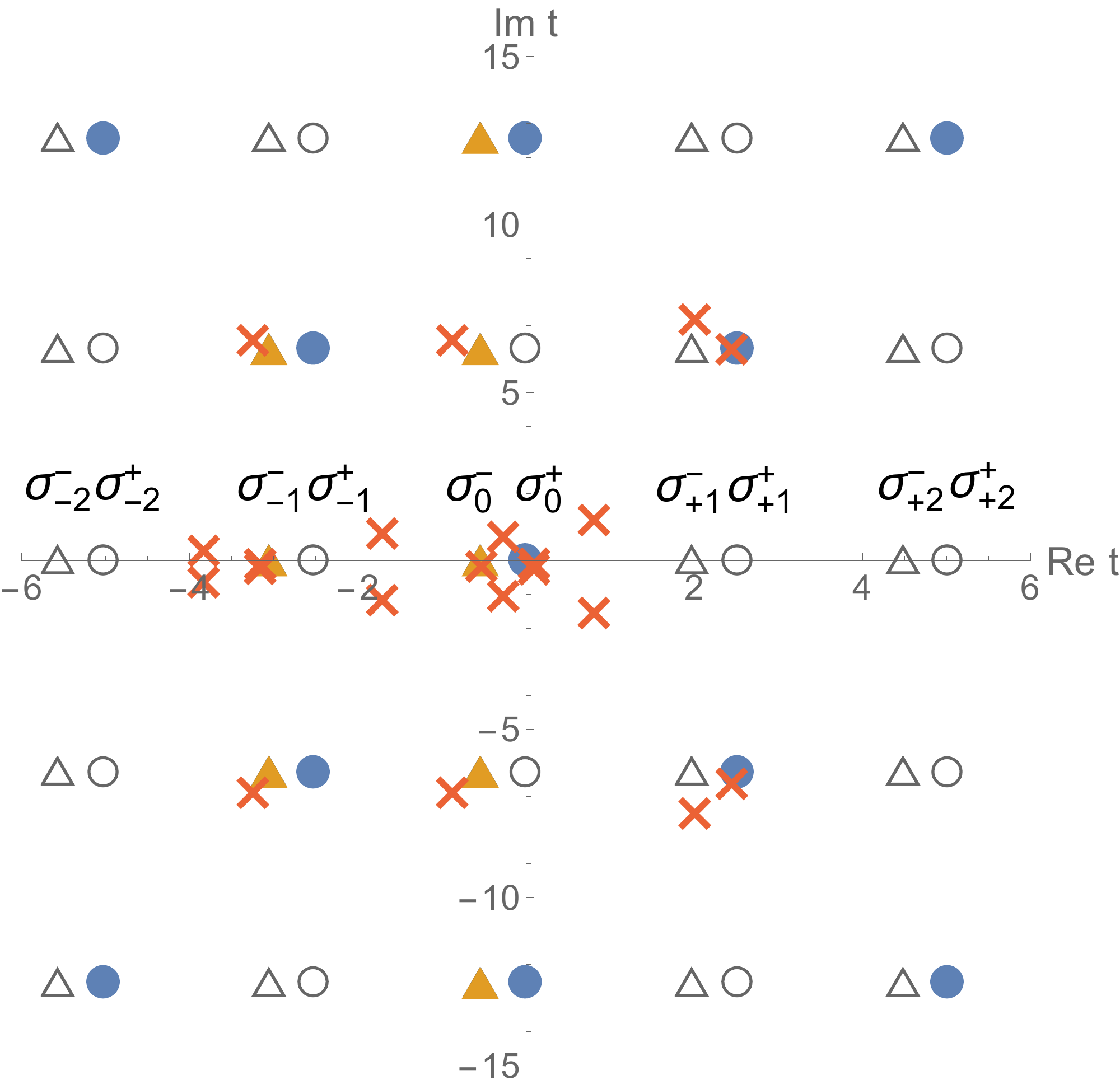}
\caption{
Illustrations of the Borel plane structure
for the subcritical region $\lambda=0.4<\lambda\crit$
obtained by the standard Pad\'{e} approximation (left panel)
and the Pad\'{e}-Uniformized approximation (right panel).
The red cross symbols indicate the location of poles found by the approximations.
The circle/triangle symbols on the real axis indicate 
the values of the saddle point actions relative to the action of $\sigma_0^+$
which are specifically $N_ft=S(\sigma_{n}^{\pm})-S(\sigma_0^+)$.
The same symbols beyond the real axis are their counterparts on different Riemann sheets obtained by shifting the ones on the real axis by $2\pi i\mathbb{Z}$. The colored symbols among them denote the saddles at which Stokes phenomena are expected to occur
from the Lefschetz thimble analysis.
}
\label{fig:pade_m1_lam0p4}
\end{figure}

%
Let us focus on the subcritical region $\lambda<\lambda\crit$.
In the left panel of Fig.~\ref{fig:pade_m1_lam0p4},
we present the locations of poles of the standard Pad\'{e} approximation 
for the Borel transformation $\widetilde{\mathcal{B}F}(t)$,
which is expected to approximate the Borel singularities.
The red crosses indicate the poles of the Pad\'{e} approximant 
while the other symbols denote the values of the saddle point actions
subtracted by the one of $\sigma_0^+$ 
and their counterparts on different Riemann sheets\footnote{
Recall the arguments at the end of Sec.~\ref{sec:transition_thimble}.
The thimble structures for non-small $\arg(N_f)$ studied in App.~\ref{sec:Lefschetz_thimble_larger_arg} imply that
we have to take care of situations that thimbles cross the branch cuts and then the action is shifted by $2\pi i N_f \mathbb{Z}$.
}.
The colored symbols among them denote the saddles at which Stokes phenomena are expected to occur
from the Lefschetz thimble analysis in the last section.
In other words, a symbol associated with a saddle $\sigma_n^{\pm}$ is colored
if there exists $\arg(N_f)$ such that\footnote{
For example, as discussed in the end of Sec.~\ref{sec:transition_thimble} and
demonstrated in App.~\ref{sec:Lefschetz_thimble_larger_arg},
this condition for $(\lambda ,m)=(0.4 , 1)$  is satisfied when $\arg(N_f)=-1.190$.
Therefore the circles on a ray $\arg (t)=-\arg(N_f)=1.190$ are colored.
}
\begin{align}
    \Im[S(\sigma_{n}^{\pm})-S(\sigma_0^+) +2\pi i N_f \mathbb{Z}  ] = 0. 
\end{align}
From the viewpoint of resurgence,
we expect that the Borel singularities are located at these color symbols.

In the left panel of Fig.~\ref{fig:pade_m1_lam0p4},
we see that a bunch of poles appear around the point corresponding to the saddle $\sigma_0^-$ and are stretched along the negative real axis.
According to general expectation on the Pad\'{e} approximation,
this signals that the Borel transformation has a branch cut type singularity 
ending on the point corresponding to $\sigma_0^-$ along the negative real axis.
This is consistent with the expectation that we have Stokes phenomena with $\sigma_0^-$.
We also see a good agreement between
the location of the poles and the action values
at $N_ft=S(\sigma_1^+)\pm2\pi i -S(\sigma_0^+)$
as expected from the thimble analysis.
However, it seems that
we do not have a similar agreement for the other saddles
in particular 
when we go away from the origin.
One reason is that the Pad\'{e} approximation becomes worse outside the convergence radius.
Another reason is that there is a branch cut on the Borel $t$-plane.
The Pad\'{e} approximant has limited pole resources to resemble the genuine Borel plane structure.
In our case, there are only $L/2=25$ poles.
If there is a branch cut, a lot of poles are consumed to resemble it.
Indeed, a lot of poles are accumulated on the negative real axis.
Thus, it seems more appropriate to use the Pad\'{e}-Uniformized approximation
using the input that we have a branch cut ending on the point $N_ft=S(\sigma_0^-)\pm2\pi i -S(\sigma_0^+)$.

In the right panel of Fig.~\ref{fig:pade_m1_lam0p4},
we show the result of the Pad\'{e}-Uniformized approximation 
where we have eliminated the expected branch cut by the uniformization map \eqref{eq:uniform}.
We first see that
there is no longer a bunch of dense poles,
which appeared in the standard Pad\'{e} approximant.
This confirms that the uniformization map has successfully removed the branch cut.
Because of this, we expect that
the Pad\'{e}-Uniformized approximation has a better description of other singularities.
Indeed the result shows better agreements
between the locations of poles and the expected Borel singularities around the origin.
In particular, note
that there is no Borel singularity on the positive real axis.
This is consistent with the Lefschetz thimble structure around $\arg(N_f)=0$
as shown in Fig.~\ref{fig:thimble1}.
However, we see some poles around the real axis which do not coincide with action values.
It seems that they are artifacts by the Pad\'{e} approximation.
For details, see App.~\ref{app:Borel_comments}.

\subsubsection{Supercritical region}
\begin{figure}[t]
\centering
\includegraphics[width=70mm]{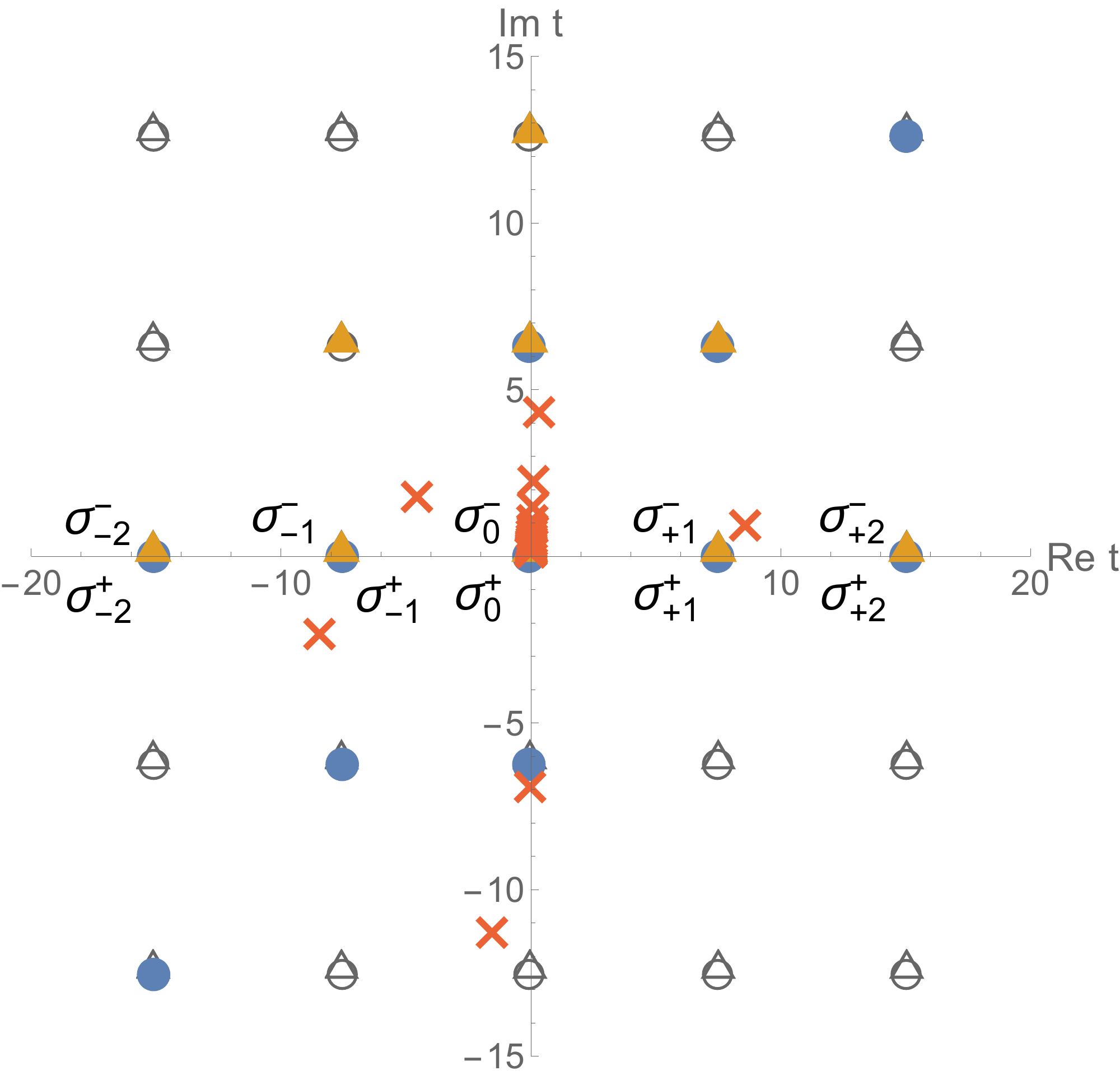} \hspace{8mm}
\includegraphics[width=70mm]{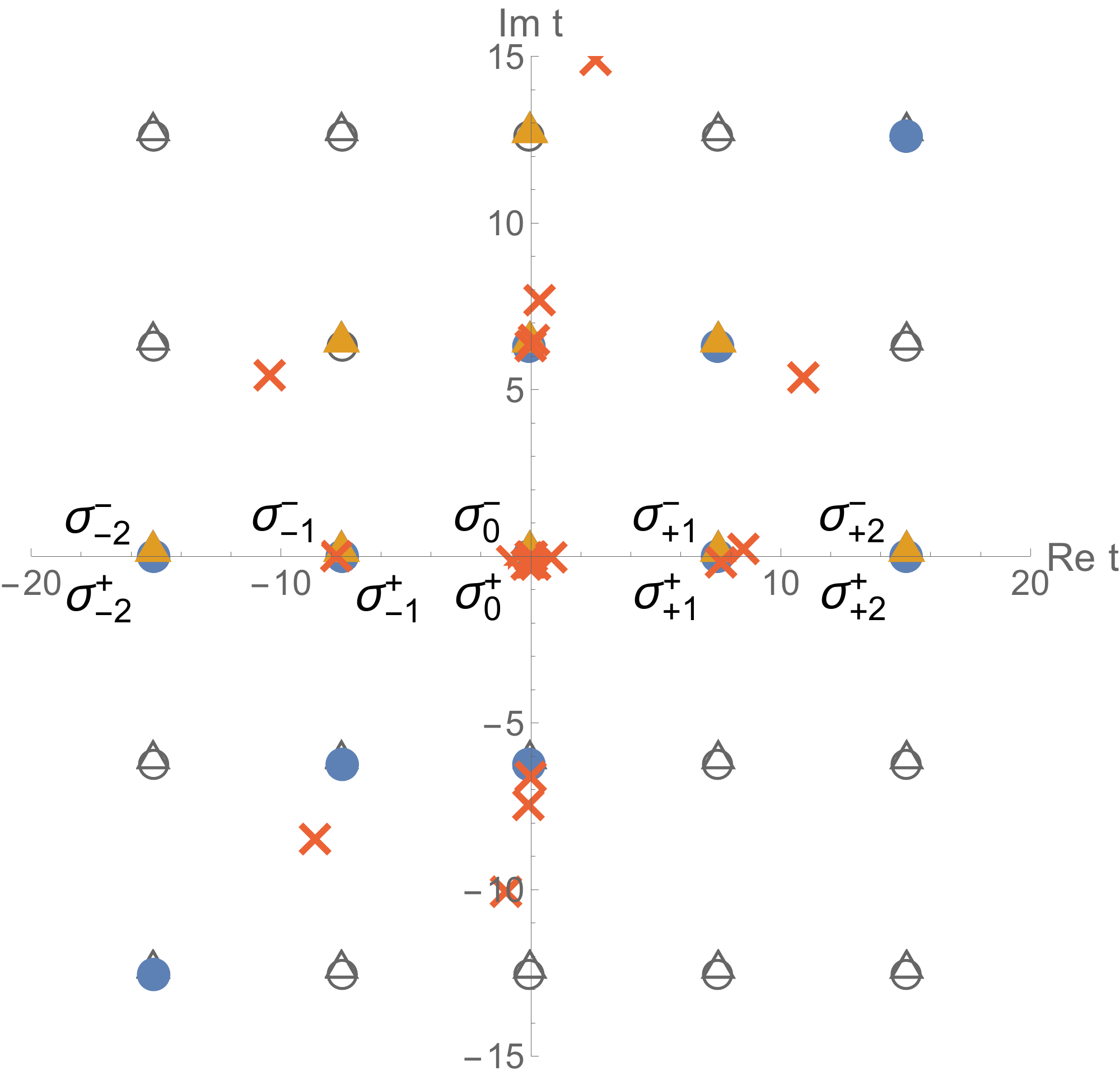}\\
\vspace{5mm}
\includegraphics[width=70mm]{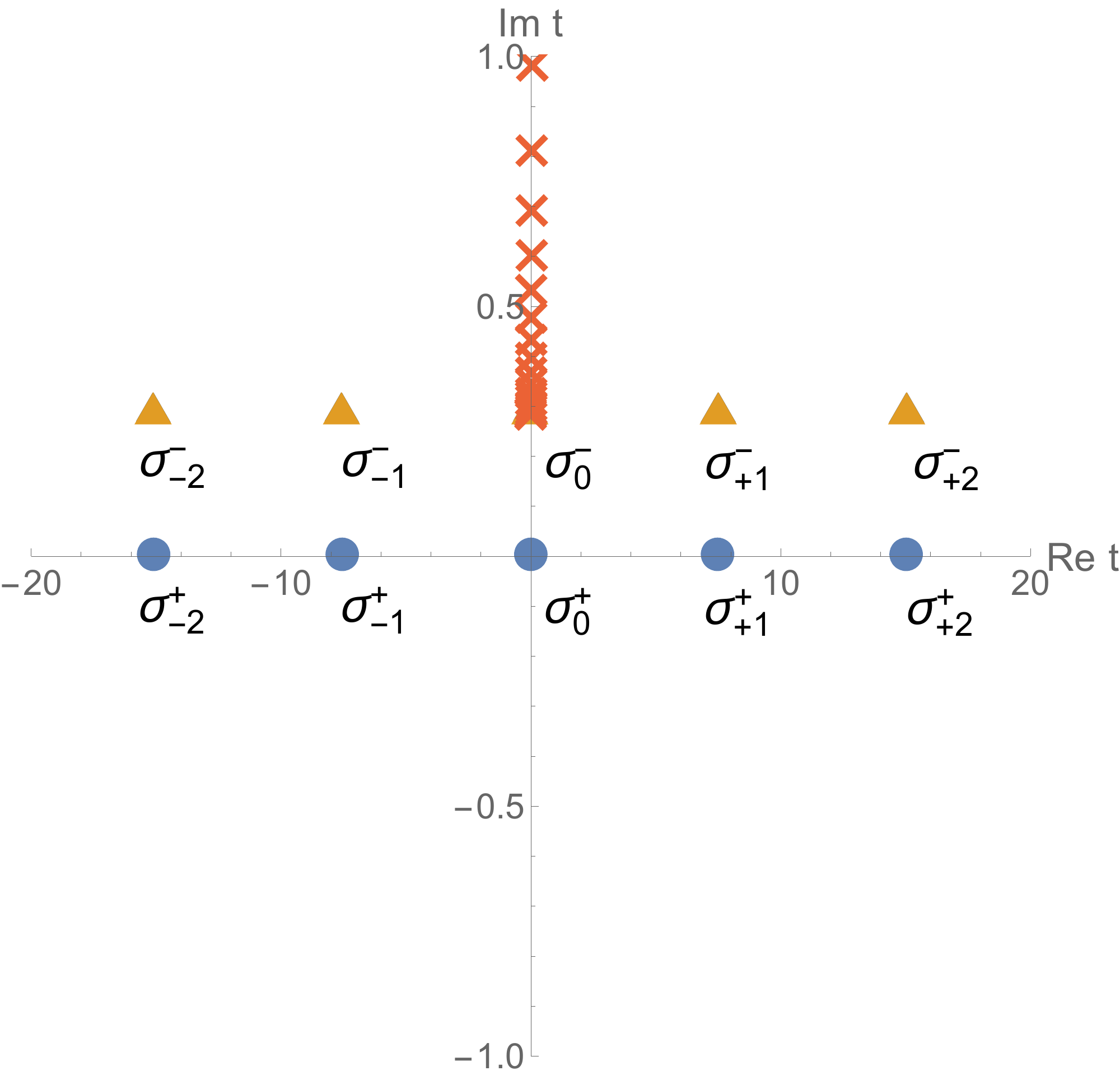} \hspace{8mm}
\includegraphics[width=70mm]{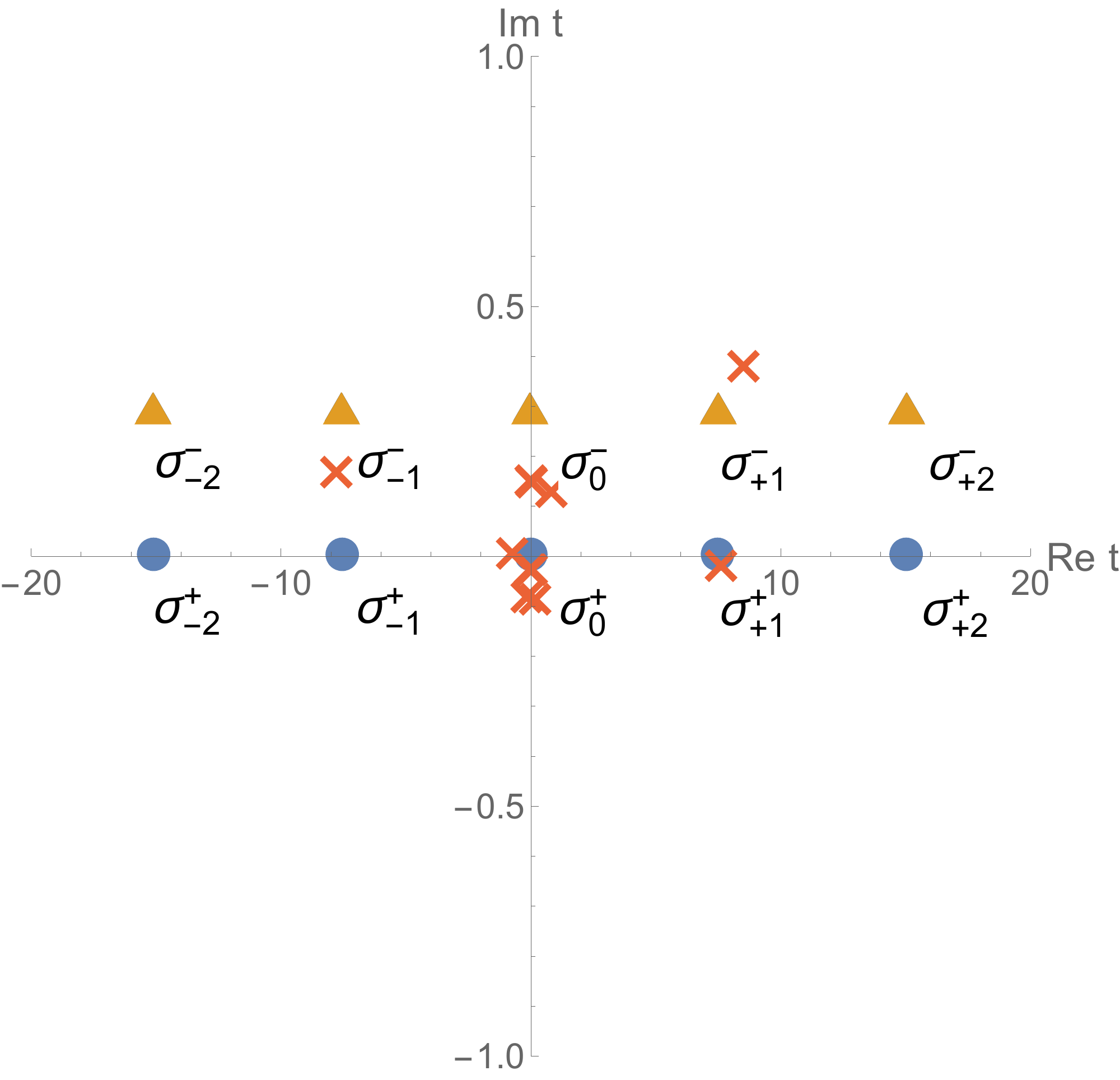}
\caption{
Illustrations of the Borel plane structure for the supercritical region $\lambda=1.2>\lambda\crit$
obtained by the Pad\'{e} approximation (left panels)
and by the Pad\'{e}-Uniformized approximation (right panels).
The lower panels are the zoomed versions of the upper ones.
}
\label{fig:pade_m1_lam1p2}
\end{figure}

%
The results for the supercritical region $\lambda\geq\lambda\crit$
are shown in 
Fig.~\ref{fig:pade_m1_lam1p2}.
As in the subcritical case,
the left panels are the result of the standard Pad\'{e} approximation
while the right panels denote the one of the Pad\'{e}-Uniformized approximation.
The upper and lower panels are essentially the same 
but we plot them in different scales for convenience.
Note that the actions of the saddles $\sigma_n^+$ and $\sigma_n^-$ are different by purely imaginary values
even on the same Riemann surface.

Let us first focus on the result of the standard Pad\'{e} approximation shown in the left panels of 
Fig.~\ref{fig:pade_m1_lam1p2}.
We easily see that 
there are a bunch of poles around the point corresponding to the saddle $\sigma_0^-$ again
but now they are stretched along the upper imaginary axis in contrast to the subcritical case.
This again implies that the Borel transformation has a branch cut ending on the point corresponding to $\sigma_0^-$ along the upper imaginary axis.
While this agrees with the expectation from the resurgence,
we do not see good agreements beyond that saddle.
Therefore we again improve the Pad\'{e} approximation assuming the information on the branch cut
as in the subcritical region. 

The right panels of Fig.~\ref{fig:pade_m1_lam1p2} show the result of the Pad\'{e}-Uniformized approximation.
The expected branch cut has been eliminated by the uniformization map \eqref{eq:uniform} and
one can check that a bunch of dense poles are indeed absent in this case.
We now see better agreements: there are poles around expected locations of the Borel singularities. 
This is consistent with the Lefschetz thimble structure around $\arg(N_f)=0$ 
as shown Fig.~\ref{fig:thimble2}.
However, we still have missing singularities away from the origin.
For details, see App.~\ref{app:Borel_comments}.

\subsection{Analytical study of Borel singularities for large $\lambda$}
In Sec.~\ref{sec:Borel_pade},
we have numerically found the Borel singularities at $2\pi i \mathbb{Z}$ in the supercritical region
as demonstrated in the right panel of Fig.~\ref{fig:pade_m1_lam1p2}.
We interpret that 
this class of singularities corresponds to the saddle $\sigma_0^+$ on different Riemann sheets.
Here we provide an analytical justification for that:
We analytically prove that
the Borel transformation of $1/N_f$ expansion around the saddle point $\sigma_n^\pm$ has
singularities at $2\pi i  \mathbb{Z}$ in the large $\lambda$ limit.

Let us consider the large $\lambda$ limit $\lambda \gg \lambda\crit$ and $\lambda \gg 1$.
The saddle point $\sigma_n^\pm$ in this limit is expanded as
\begin{\eq}
	\sigma_\pm^n 
=  \mp m +(2n+1)\pi i +\frac{i}{\lambda} +\mathcal{O}\left(  \frac{1}{\lambda^2} \right) .
\end{\eq}
The action values at these saddle points are
\begin{\eq}
S(\sigma_\pm^n )
	=N_f \Bigl[ 
			\pm i\lambda m +(2n+1)\pi\lambda -\log{\lambda} +1+\log{(i\sinh{m})} 
 	\Bigr] 
 		+\mathcal{O}\left( \frac{1}{\lambda} \right) .
\label{eq:action_asymptotic}
\end{\eq}
%
%
%
We are interested in the perturbative coefficients in the leading order of the large-$\lambda$ limit.
Let us expand the action around the saddle points:
\begin{\eq}
S(\sigma)  
	= S(\sigma_\pm^n)  
		+\sum_{n=2}^\infty \frac{1}{n!} S^{(n)}(\sigma_\pm^n ) \delta \sigma^n ,
\end{\eq}
where we regard $\delta\sigma =\mathcal{O}( N_f^{-1/2} )$ and
the first few derivatives of the action are
\begin{\eqa}
S (\sigma )
&=& N_f \Bigl[ -i \lambda \sigma +\log{\left( \cosh{\sigma} +\cosh{m} \right) } 
\Bigr] ,\NN\\
S^{(1)} (\sigma )
&=& N_f \Biggl[ -i \lambda  +\frac{\sinh{\sigma}}{\left( \cosh{\sigma} +\cosh{m} \right) } \Biggr]  ,\NN\\
S^{(2)} (\sigma )
&=& N_f \Biggl[ \frac{\cosh{\sigma}}{\left( \cosh{\sigma} +\cosh{m} \right) } 
 -\frac{\sinh^2{\sigma}}{\left( \cosh{\sigma} +\cosh{m} \right)^2 } \Biggr] .
\end{\eqa}
Noting
\begin{\eq}
\left. \frac{1}{\cosh{\sigma} +\cosh{m}} \right|_{\sigma =\sigma_\pm^n} 
\simeq \mp i \frac{\lambda}{\sinh{m}} ,
\end{\eq}
we can approximate $S^{(n)}(\sigma_\pm^n )$ as
\begin{\eq}
S^{(n)}(\sigma_\pm^n )
\simeq \left.  - N_f (-1)^n (n-1)!  
\left( \frac{\sinh{\sigma}}{\cosh{\sigma} +\cosh{m}}  \right)^n 
\right|_{\sigma =\sigma_\pm^n}
= -N_f (n-1)!  (i\lambda )^n .
\end{\eq} 
Therefore the action becomes
\begin{\eqa}
S[\sigma ]  
&\simeq & S[\sigma_\pm^n ]  
-N_f \sum_{n=2}^\infty \frac{1}{n}  (i\lambda \delta \sigma )^n  \NN\\
&=& S[\sigma_\pm^n ] +N_f \left( i\lambda \delta \sigma + \log{(1 -i\lambda \delta \sigma  )} \right).
\end{\eqa}
Then the perturbative series in the large $\lambda$ limit is generated by
\begin{\eq}
F(N_f ;\lambda )
	= \int_{-\infty}^\infty \dd{\delta\sigma}\:
		e^{-N_f \left( i\lambda \delta\sigma + \log{(1 -i\lambda \delta\sigma  )} \right)} .
\end{\eq}
One can rewrite this integral in the form of a Laplace transformation as in the Borel resummation formula 
if we make a change of variable as
\begin{\eq}
t= i\lambda \delta\sigma + \log{(1 -i\lambda \delta\sigma  )} .
\end{\eq}
Noting that this equation is rewritten as
\begin{\eq}
-e^{t-1}  =(  i\lambda \delta\sigma -1) e^{i\lambda \delta\sigma -1}  ,
\end{\eq}
we can write the solution as
\begin{\eq}
\label{eq:large-lam_map}
\delta\sigma =\frac{1}{i\lambda} \left( 1 +W(-e^{t-1} ) \right) ,
\end{\eq}
where $W(z)$ is the Lambert $W$ function defined as a solution of the following equation
\begin{\eq}
z = u e^u  \quad \Leftrightarrow\quad  u=W(z) .
\end{\eq}
Then, using 
\begin{\eq}
\frac{dW(z)}{dz} = \frac{W(z)}{z (1+W(z))} ,
\end{\eq}
we find
\begin{\eq}
F(N_f ;\lambda )
= \frac{1}{i\lambda } \int dt\ e^{-N_f t}  \frac{W(-e^{t-1}) }{1+W(-e^{t-1}) }
\end{\eq}
It seems natural to identify
\begin{\eq}
\widetilde{\mathcal{B}F}(t)
=\frac{1}{i\lambda } \frac{W(-e^{t-1}) }{1+W(-e^{t-1}) }.
\end{\eq}
Some important features of the Lambert $W$ function are 
\setlist{nolistsep}
\begin{itemize}[topsep=2pt, itemsep=2pt]
	\item $W(z)$ has a branch cut along $(-\infty, -e^{-1} )$.
	\item $W(-e^{-1}) =-1$.
	\item Small $z$ expansion of $W(z)$ has a radius of convergence $e^{-1}$.
\end{itemize}
Thus the  Borel transformation in the large $\lambda$ limit has
the branch cut singularities at
\begin{\eq}
t =2\pi i \mathbb{Z}.
\end{\eq}
One might wonder why we now do not have singularities beyond the imaginary axis 
which appeared in the numerical study represented in Fig.~\ref{fig:pade_m1_lam1p2}.
This is because of the large $\lambda$ limit:
the singularities beyond the the imaginary axis go to infinity as $\lambda\rightarrow\infty$.
It is most transparent in the formula \eqref{eq:action_asymptotic} for the asymptotic behaviors for the action.

We can see that the above Borel singularities come from the branch cuts in the $\sigma$-plane as follows.
The variable $t$ of the Borel plane is related to the $\sigma$-plane by the map \eqref{eq:large-lam_map}.
Therefore
the origin $t=0$ is associated with a saddle $\delta\sigma = 0$
while
the infinity is associated with the branch cut singularity $\delta\sigma = 1/i\lambda$.
An interval $[0,2\pi i]$ is associated with a closed loop
	which starts from the saddle $\delta\sigma=0$
	and runs around the branch cut singularity $\delta\sigma = 1/i\lambda$
	back to the saddle.
Since there is a logarithmic branch cut on the $\delta\sigma$-plane,
we reach the next Riemann sheet once we move along the closed loop.
Thus the Borel singularities at $t = 2\pi i \mathbb{Z}$ 
are associated with the saddles on the different Riemann sheets.
Such a relation should hold even when $\lambda$ is not large
as long as we are in the supercritical region.

The above structures technically come from the fact that 
the action \eqref{eq:action} has the periodic structure and the logarithm branch cuts.
Physically this type of factor is originated from one-loop contributions of hypermultiplets 
in the localization formula of $S^3$ partition functions \cite{Kapustin:2009kz,Hama:2010av,Jafferis:2010un}.
This indicates that the above structures hold not only for the SQED but also for more general supersymmetric gauge theories.

\section{Lessons from 3d $\mathcal{N}=4\:$ SQED}
\label{sec:phase_trans_stokes_pheno}
In this section, 
given the lessons from the SQED obtained in the previous sections, 
we provide a more generic discussion on relations between the resurgence and phase transitions.
In particular, we discuss
	how the order of phase transitions are described
	from the viewpoint of (anti-)Stokes phenomena.

\subsection{Phase transitions as collisions of saddles}
\begin{figure}
	\centering
	\includegraphics[width=70mm]{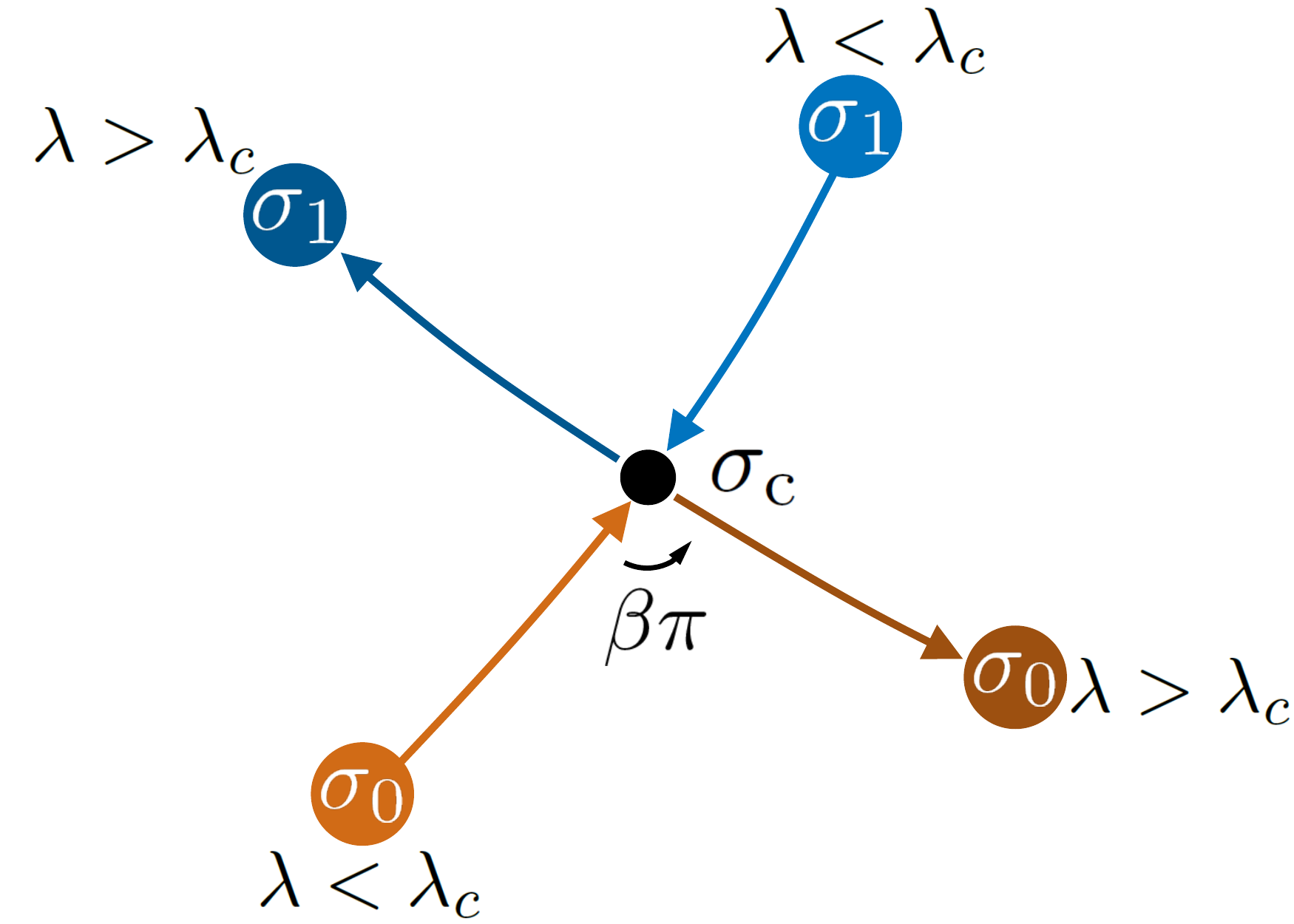}
	\caption{
An illustration of collision and scattering of saddles for $n=2$.
		As the parameter $\lambda$ is varied through the critical point $\lambda=\lambda_c$,
			two saddles $\sigma_0, \sigma_1$ collide at $\sigma=\sigma_c$
			and scattered with an angle $\beta\pi$.
	}
	\label{fig:saddle_collision}
\end{figure}

Let us consider a generic theory whose partition function is described by a one-dimensional integral of the form
\begin{align}
e^{-NF(\lambda)}
= \int\dd{\sigma}\:
e^{-N\tilde{S}(\lambda;\sigma)}
\label{eq:gen_model}
\end{align}
where $\tilde{S}(\lambda;\sigma)$ is the ``action'' and
$(N,\lambda )$ are some parameters specifying the theory.
Suppose that the theory undergoes a phase transition at $\lambda=\lambda\crit$
in the limit $N\rightarrow\infty$,
accompanying a collision and a scattering of $n$ saddles at $\sigma=\sigma\crit$.
We do not consider phase transitions simply by anti-Stokes phenomena
which have been often discussed in the context of the Lefschetz thimble analysis.
Here we show that the order of phase transition is determined by
the scattering angle of saddles.
More specifically, we prove the following statement:
if the $n$-saddles collide and scatter with a scattering angle $\beta\pi$
as we vary the parameter $\lambda$ through the critical point $\lambda =\lambda_c$
(as illustrated in Fig.~\ref{fig:saddle_collision}),
then we have the phase transition of the order $\lceil (n+1)\beta \rceil$,
where $\lceil x \rceil$ is the smallest integer larger than or equal to $x$.

Before moving onto the proof, 
let us recall some basics on phase transition.
We have an $p$-th order phase transition at $\lambda =\lambda_c$
when the $p$-th derivative of the ``free energy'' $F(\lambda )$ becomes singular at $\lambda =\lambda_c$
given its non-singular lower derivatives, that is
\begin{align}
\abs{F^{(p)}(\lambda\crit\pm0)} = \infty \quad
\text{or}\quad
F^{(p)}(\lambda\crit-0)
\neq F^{(p)}(\lambda\crit+0) ,
\end{align}
with
\begin{align}
	\abs{F^{(k<p)}(\lambda\crit\pm0)} < \infty \quad
	\text{and}\quad
	F^{(k<p)}(\lambda\crit-0)
	=F^{(k<p)}(\lambda\crit+0)  .
\end{align}
The order $p$ is related to a behavior of the free energy around the critical point as follows.
Suppose that the free energy is expanded around the critical point as
\begin{align}
F(\lambda =\lambda\crit+\delta\lambda)
\simeq  \left\{
\begin{array}{ll}
C + A(\delta\lambda)^{\gamma} & {\rm for}\ \delta\lambda<0 \\
C + B(\delta\lambda)^{\gamma} & {\rm for}\  \delta\lambda>0
\end{array}\right. ,
\end{align}
where $A,B,C$ are complex constants and $\gamma>0$.
Note that the free energy
is not necessarily real since it is not necessarily interpreted as the thermodynamic one 
for QFT on a generic manifold\footnote{
It is interpreted as the thermodynamic one when QFT is put on a manifold including $S^1$.
}.
If $\gamma \notin \mathbb{Z}$,
	the phase transition is of the order $p=\lceil \gamma \rceil$.
Similarly, if $\gamma \in \mathbb{Z}$ and $A\neq B$,
then the order of the phase transition is $p=\lceil \gamma \rceil$
while there is no phase transition for $A = B$.
In what follows, we consider only the case with $A\neq B$ and the exponent $\gamma$ independent of the sign of $\delta\lambda$.
This is a consequence of a saddle collision as we will see soon.

Now we provide the 
proof.
As mentioned above, we are interested in the situation that
the $n$-saddles collide and scatter at $\sigma=\sigma\crit$ in varying the parameter $\lambda$ thorough the critical point $\lambda\crit$.
This means that the saddle point equation at $\lambda =\lambda\crit$ 
has the root with the degeneracy $n$.
Therefore the action around $\sigma=\sigma\crit$ is expanded as\footnote{
Note that the configuration $\sigma=\sigma\crit$ is not a saddle point for $\lambda \neq \lambda\crit$ generically.
}
\begin{align}
\tilde{S}(\lambda;\sigma =\sigma\crit )
= a_0(\lambda)
+a_1(\lambda)\delta\sigma
+\cdots
+a_{n+1}(\lambda)\delta\sigma^{n+1}
+\cdots.
\end{align}
The coefficient $a_i(\lambda)$ is constrained by the condition that the saddles collide at $\lambda =\lambda\crit$ as
\begin{align}
a_i(\lambda) = c_i\delta\lambda^{\alpha_i} + \cdots.
\end{align}
Without loss of generality,
one can shift the action by an appropriate constant to make $a_0(\lambda)$ independent of $\lambda$:
\begin{align}
a_0(\lambda) = c_0.
\end{align}
Furthermore, using the condition that the saddle point equation has the root with the degeneracy $n$,
we find
\begin{align}
&a_i(\lambda\crit) = 0 \quad
(i=1,\dots,n), \\
&a_{n+1}(\lambda\crit) \neq 0.
\end{align}
Combining the above conditions for the coefficient leads us to
\begin{align}
&c_i=0, \quad
\text{or} \quad
c_i\neq0,\: \alpha_i>0
\quad
(i=1,\dots,n), \\
&c_{n+1}\neq0,\: \alpha_{n+1}=0.
\end{align}
Solving the saddle point equation
\begin{align}
0 = a_1(\lambda) + \cdots + (n+1)a_{n+1}(\lambda)\delta\sigma^{n} + \cdots
\end{align}
around the collision point $\delta\sigma=0$, 
the saddle point around the critical point is simply written as
\begin{align}
\delta\sigma_m
\simeq
s_m\delta\lambda^{\beta}\quad
(m=0,\dots,n-1),
\end{align}
where $s_m$ is a some constant and 
\begin{align}
    \beta = \min\left(\frac{\alpha_1}{n},\frac{\alpha_2}{n-1},\dots,\frac{\alpha_n}{1}\right).
\end{align}
Around the critical point $\delta\lambda=0$, each saddle acquires a phase $(-1)^{\beta}$.
This implies that the $n$ saddles collide and scatter with an angle $\beta\pi$.
Then the action at the saddle $\sigma_m$ takes the value
\begin{align}
\tilde{S}_m
\simeq
c_0
+T_m (\delta\lambda)^{(n+1)\beta} ,
\label{eq:action_delta_lambda}
\end{align}
with a constant $T_m$.

At the phase transition point,
there is a jump of contributions saddle points in various ways.
For example,
in the case where the contributing saddles jump as $\sigma_0 \rightarrow \sigma_1$,
the free energy changes as
\begin{align}
F
\simeq\left\{
\begin{array}{ll}
c_0 + T_0 (\delta\lambda)^{(n+1)\beta} & {\rm for}\ \delta\lambda<0 \\
c_0 + T_1 (\delta\lambda)^{(n+1)\beta} & {\rm for}\ \delta\lambda>0
\end{array}\right. .
\end{align} 
In the case where contributing saddles jump as $\sigma_0\rightarrow \sigma_0,\dots,\sigma_{n-1}$,
the free energy changes as
\begin{align}
F
\simeq\left\{
\begin{array}{ll}
c_0 + T_0 (\delta\lambda)^{(n+1)\beta} & {\rm for}\ \delta\lambda<0 \\
c_0 + (T_0+\cdots+T_{n-1}) (\delta\lambda)^{(n+1)\beta} & {\rm for}\ \delta\lambda>0
\end{array}\right. .
\end{align}
In any case,
the phase transition is of the order $\lceil (n+1)\beta \rceil$ 
and this completes the proof.
Our argument also shows a connection between the order of the phase transition and the anti-Stokes line.
The formula \eqref{eq:action_delta_lambda} for the action shows that
the anti-Stokes line is given by $\Re\left[(\delta\lambda)^{(n+1)\beta}\right]=0$.
Thus one can also read off the order of the phase transition by looking at the anti-Stokes line.

\subsection{Thimbles and Borel singularities around critical points}
\label{sec:Airy}
In this subsection, we demonstrate the discussion in the last subsection
using the integral representation of the Airy function
whose ``action'' is given by
\begin{align}
	\tilde{S}(\lambda;\sigma)
	= \frac{i\sigma^3}{3}-i\lambda\sigma .
\end{align}
We refer to this example as the Airy-type model.
As we will see soon, 
this example corresponds to $n=2, \:\alpha=1/2$,
and has common features with the SQED \eqref{eq:localization} 
in the context of the argument in this section.

\subsubsection{Lefschetz thimbles}

The Airy-type model corresponds to
\begin{align}
c_0 = 0, \quad c_1 = -i, \quad \alpha_1 = 1, \quad
c_2 = 0, \quad c_3 = \frac{i}{3}, \quad \alpha_3 = 0.
\end{align}
The saddle points in this example are simply given by
\begin{align}
\sigma_\pm		= \pm \lambda^{1/2} ,
\end{align}
which indicates that the two saddles collide at $\sigma =0$ for $\lambda =0$.
Therefore, in the notation of the last subsection, we have
\begin{align}
	\sigma\crit = 0, \quad \lambda\crit = 0 ,\quad
\delta\sigma_\pm		= \pm \delta\lambda^{1/2} ,
\end{align}
and
\begin{align}
n=2 ,\quad \alpha =\frac{1}{2} .
\end{align}
The action at the saddle $\sigma_\pm$ is given by
\begin{align}
	\tilde{S}_\pm
&= \mp \frac{2i}{3}(\delta\lambda)^{3/2} .
\end{align}
This leads us to the standard Airy-type Stokes graph
\begin{align}
\text{Stokes line:}\ \Im[i (\delta\lambda)^{3/2}] = 0, \qquad
\text{Anti-Stokes line:}\ \Re[i (\delta\lambda)^{3/2}] = 0.
\end{align}
We can find the (dual) thimbles by solving
\begin{align}
\mathcal{J}_\pm :\;
\dv{\sigma}{s}
= \left. \overline{\dv{S(\sigma)}{\sigma}} \right|_{\sigma\simeq\sigma_\pm}, \quad
\mathcal{K}_\pm :\;
\dv{\sigma}{s}
= - \left. \overline{\dv{S(\sigma)}{\sigma}} \right|_{\sigma\simeq\sigma_\pm} .
\end{align}

\begin{figure}[t]
	\centering
	\includegraphics[width=65mm]{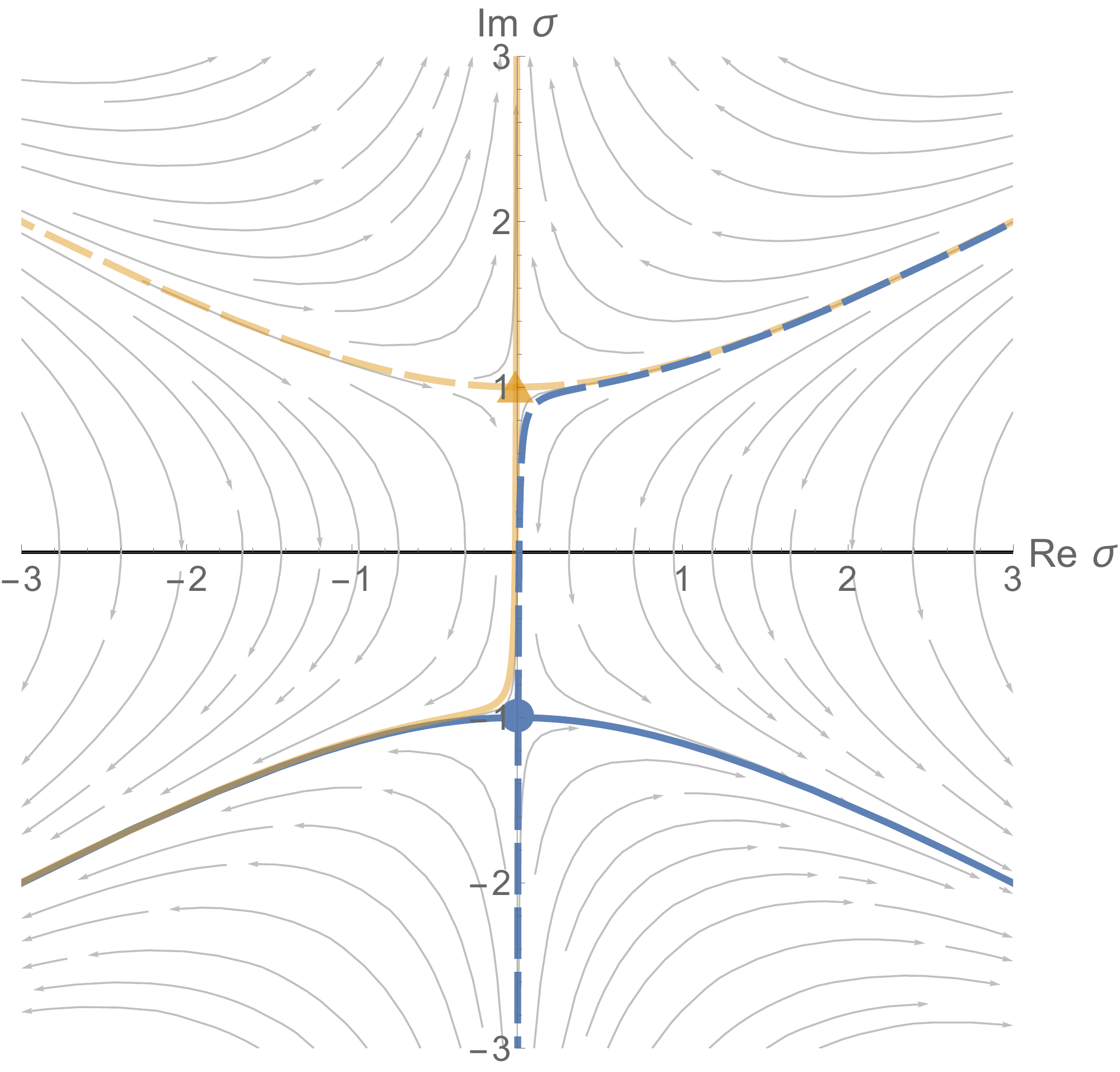} \hspace{8mm}
	\includegraphics[width=65mm]{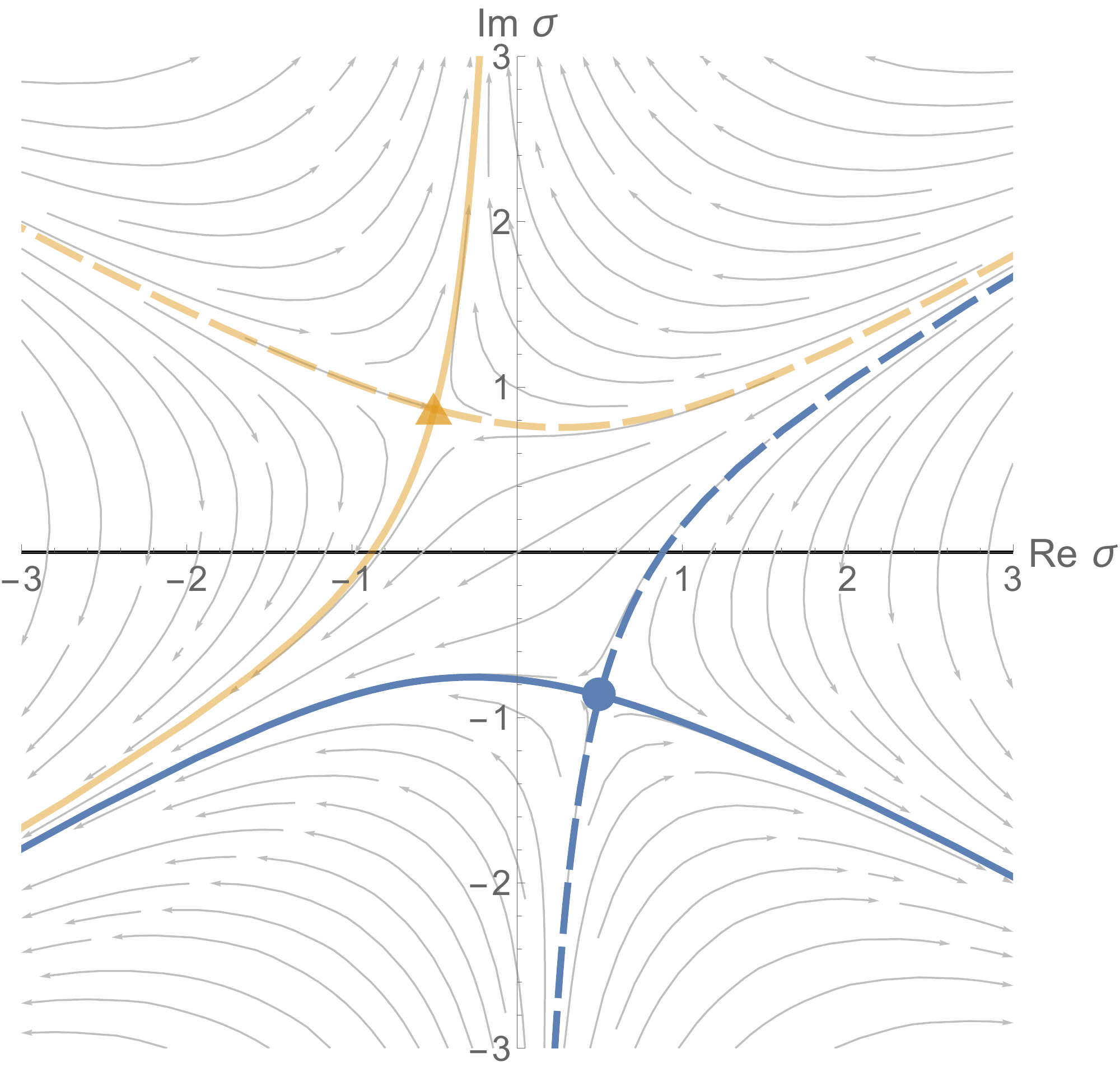} \\
	\vspace{5mm}
	\includegraphics[width=65mm]{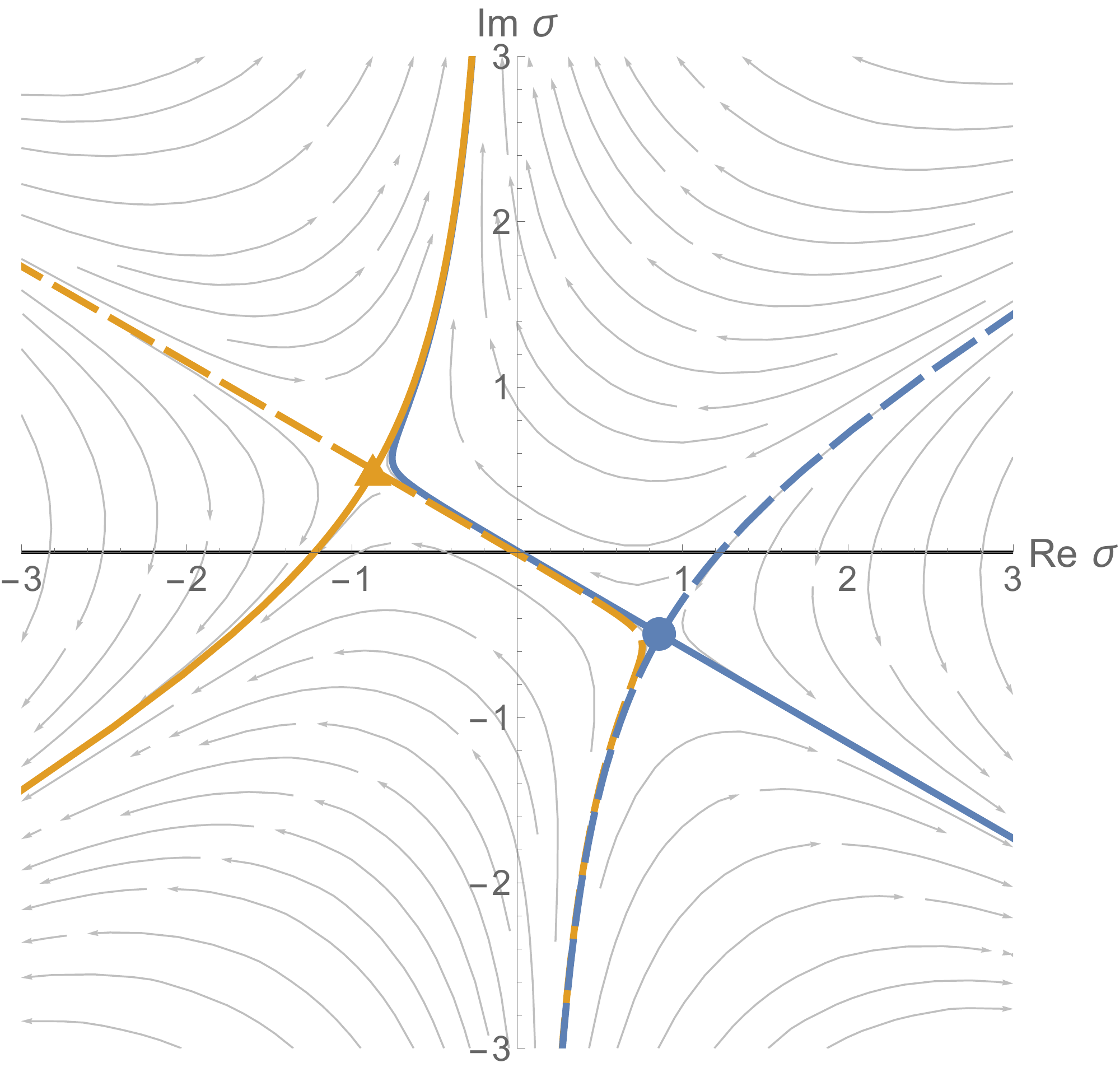} \hspace{8mm}
	\includegraphics[width=65mm]{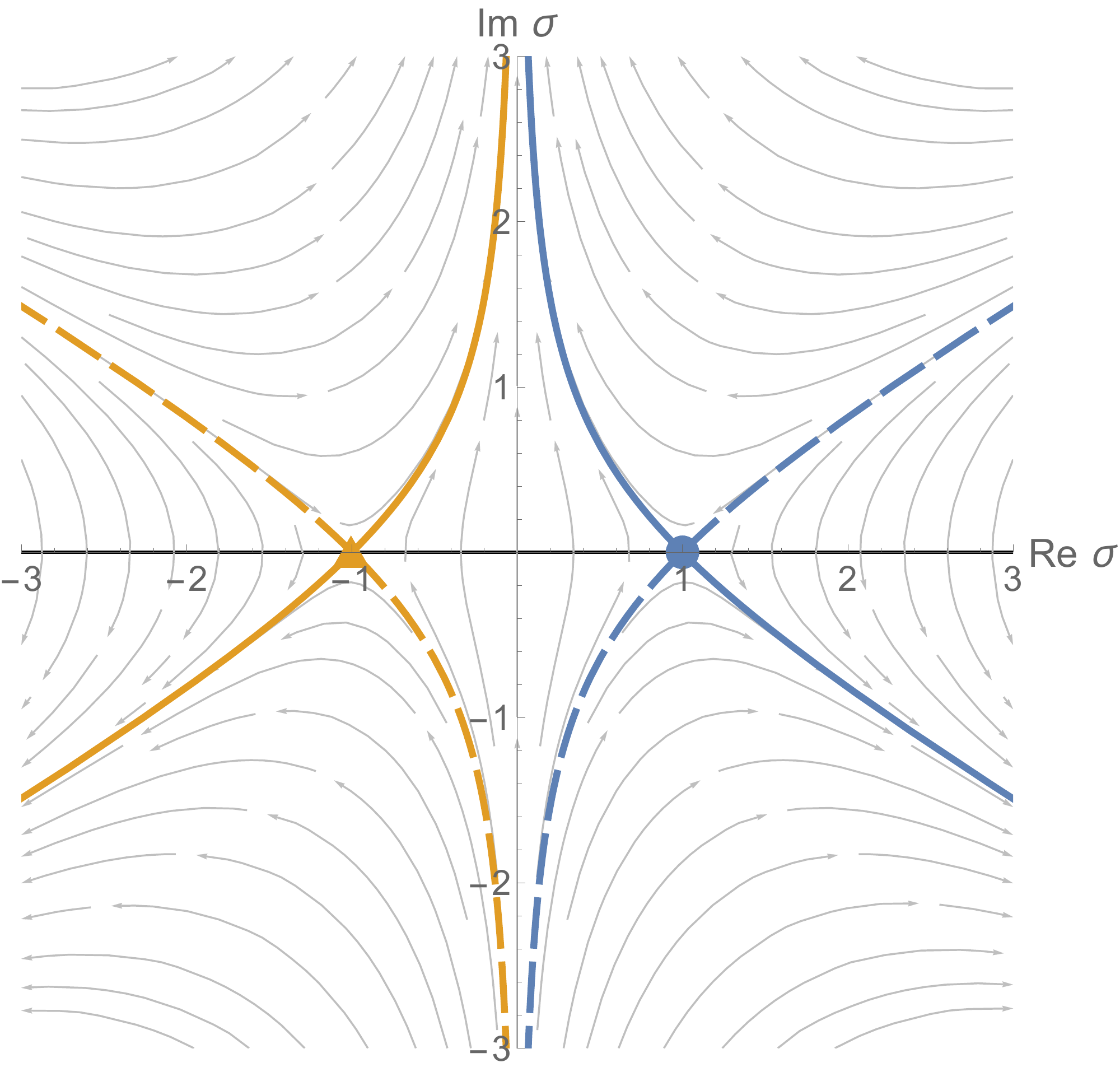}
	\caption{
		Illustrations of the thimble structures of the Airy-type model for $\abs{\delta\lambda}=1$.
		The blue circle and orange triangle symbols indicate the saddle $\sigma_0$ and $\sigma_1$, respectively.
		The Lefschetz thimbles associated with them are drawn as the lines with the same colors.
		The dual thimbles are drawn as the dashed lines.
		The opaque saddles and thimbles contribute to the integral,
			while the translucent ones do not.
		The phase of $\delta\lambda$ is
			$\arg (\delta\lambda )= -\pi + 0.01$ [top left],
			$-2\pi/3$ [top right],
			$-\pi/3 + 0.01$ [bottom left],
			and $0$ [bottom right].
	}
	\label{fig:thimble_airy}
\end{figure}

In Fig.\;\ref{fig:thimble_airy}, we show 
how the thimbles change as increasing the phase $\arg (\delta\lambda )$.
As $\arg (\delta\lambda )$ is increased from a negative value, 
we encounter a Stokes phenomenon at $\arg (\delta\lambda )= -\pi$,
an anti-Stokes phenomenon at $\arg (\delta\lambda )= -2\pi/3$,
a Stokes phenomenon at $\arg (\delta\lambda )= -\pi/3$ and
an anti-Stokes phenomenon at $\arg (\delta\lambda )= 0$.
In particular, we observe the jump of the contributing saddles at $\arg (\delta\lambda ) = -\pi/3$ (as well as $\arg (\delta\lambda ) = +\pi/3$):
we have a contribution only from $\sigma=\sigma_+$ for $\arg (\delta\lambda ) < -\pi/3 $
while we have contributions from the two saddles $\sigma=\sigma_+$ and $\sigma=\sigma_-$ at $\arg (\delta\lambda ) = -\pi/3 +0$.
This is a manifestation of the Stokes phenomenon.
The free energy also jumps as
\begin{align}
	F
		\simeq \left\{
			\begin{array}{lll}
				\tilde{S}_+
					&= \frac{2}{3}(-\delta\lambda)^{3/2}
						& {\rm for}\ \delta\lambda < 0 \\
				\tilde{S}_+ + \tilde{S}_-
					&= 0
						& {\rm for}\ \delta\lambda > 0
			\end{array}
		\right. ,
\end{align}
which implies the second-order phase transition.

Next, let us increase $\delta\lambda$ from $-1$ to $1$,
keeping $\Im\delta\lambda = 0$.
As $\delta\lambda$ goes from $-1$ to $0$,
the two saddles (in the left top panel of Fig.~\ref{fig:thimble_airy}) approach the origin along the imaginary axis.
At $\delta\lambda=0$, they collide and change their directions.
As $\delta\lambda$ goes from $0$ to $+1$,
the two saddles (in the bottom right panel of Fig.~\ref{fig:thimble_airy}) depart the origin along the real axis.
In other words, the two saddles collide with an angle $\pi/2$ at the phase transition.
Also, we remark that, during the phase transition,
we cross the anti-Stokes line $\arg (\delta\lambda ) = -2\pi/3$
and the Stokes line $\arg (\delta\lambda )= -\pi/3$.
Thus, the second-order phase transition is understood as a phenomenon
in which an anti-Stokes and a Stokes phenomenon occur simultaneously.
To summarize, the second-order phase transition in the Airy-type model is interpreted as follows
\begin{enumerate}[label=\roman*.,topsep=2pt, itemsep=2pt]
	\item Contributing saddles jump as $\sigma_+ \rightarrow \sigma_+, \sigma_-$.
	\item The two saddles collide and scatter with a scattering angle $\pi/2$.
	\item A Stokes phenomenon and an anti-Stokes phenomenon occur simultaneously.
\end{enumerate}

\subsubsection{Borel singularities}
The ``partition function'' of the Airy-type model is defined as
\begin{align}
	Z(\lambda)
		= \int \dd{\sigma}\:
				e^{-N\tilde{S}(\lambda;\sigma)} .
\end{align}
Let us consider the $1/N$ expansion around the ``trivial'' saddle $\sigma_+ = \lambda^{1/2}$.
Using the formula in App.~\ref{sec:app_expansion},
the perturbative series is formally given by
\begin{align}
	\int_{\mathcal{J}_+} \dd{\sigma}\:
	e^{-N\tilde{S}(\lambda;\sigma)}
	= \sqrt{ \frac{\pi}{i\lambda^{1/2}N} }
			e^{ \frac{2i\lambda^{3/2}}{3}N}
			F \left( \frac{1}{N} \right),
\end{align}
where
\begin{align}
	F \left( \frac{1}{N} \right)
		= \sum_{l=0}^{\infty} \frac{a_l}{N^l}, \quad
				a_l = \frac{ \Gamma(3l+1/2) }{ 3^{2l}(-i\lambda^{3/2})^l \Gamma(1/2)\Gamma(2l+1) }.
\end{align}
Note that the coefficient grows factorially.
The analytic continuation of its Borel transformation is
\begin{align}
	\widetilde{\mathcal{B}F}(t)
		= {}_2F_1 \left(\frac{1}{6},\frac{5}{6},1; \frac{3t}{4(-i\lambda^{3/2})}\right).
\end{align}
This function has a Borel singularity (branch cut singularity) at
\begin{align}
	t = \frac{4(-i\lambda^{3/2})}{3}.
\end{align}
This Borel singularity corresponds to the ``non-trivial saddle'' $\sigma_- = -\lambda^{1/2}$,
	and it collides with the origin corresponding to the ``trivial saddle'' at the critical point $\lambda=0$.
The scattering angle is $-3\pi/2 \sim \pi/2$.
After the collision ($\lambda>0$),
	the Borel singularity is on the imaginary axis.
This means that there occurs an anti-Stokes phenomenon: $\Re \tilde{S}_+ = \Re \tilde{S}_- = 0$.
Thus, the collision of saddles is appropriately encoded in the perturbative series
as expected by the resurgence theory.

\subsection{Second-order phase transition in the SQED revisited}

In this section, we revisit the second-order phase transition in the SQED
based on the previous subsections
to clarify more the relationship between the phase transition and resurgence.

\subsubsection{Lefschetz thimble analysis}
From the Lefschetz thimble analysis in Sec.~\ref{sec:thimble},
we have seen that
the SQED around the second-order phase transition point has the following properties
\begin{enumerate}[label=\roman*.,topsep=2pt, itemsep=2pt]
	\item Contributing saddle points jump as $\sigma_0^+ \rightarrow \sigma_0^+, \sigma_0^-$.
	\item The two saddles collide and scatter with a scattering angle $\pi/2$.
	\item An infinite number of Stokes phenomena associated with saddles $\sigma_{n>0}^\pm$ occur.
\end{enumerate}
%
%
The first two points are common with the Airy-type model in the last subsection.
This is because the ``action'' of the SQED \eqref{eq:action} has a similar expansion 
to one of the Airy-type models around the critical point.
Thus, the second-order phase transition in the SQED is interpreted in a similar way as the Airy-type model.
%
%
The third point is particular for the SQED.
The difference essentially comes from the fact that
the SQED has infinite number of saddles periodically distributed along the imaginary axis.
Once thimbles run along the imaginary axis after a phase transition,
	they inevitably path through the periodic saddles.
Such behavior of thimbles causes an infinite number of Stokes phenomena.
Technically the appearance of the periodic saddles is due to the ${\rm cosh}$ factors
originated from the one-loop determinant of the hypermultiplets 
in the SUSY localization of the $S^3$ partition function.
Therefore we expect that 
the above features appear also in other SUSY gauge theories on $S^3$.

\subsubsection{Borel resummation}
In the language of the Borel resummation,
the second-order phase transition has the following features
\begin{enumerate}[label=\Roman*.,topsep=2pt, itemsep=2pt]
	\item In the supercritical region, the two Borel singularities line up along the imaginary axis on the Borel plane.
	\item The two Borel singularities collide and scatter with a scattering angle $\pi/2$ as we cross the critical point.
	\item The $1/N_f$-expansion becomes Borel non-summable along the positive real axis in the supercritical region.
\end{enumerate}
(\Romannum{1}), (\Romannum{2}) and (\Romannum{3}) here correspond to
	(\romannum{1}), (\romannum{2}) and (\romannum{3}) of the Lefschetz thimble analysis, respectively,
as expected from the resurgence theory.
The first point means that 
the saddle points associated with the Borel singularities along the imaginary axis
have the same real part of the actions and 
therefore contribute to the integral with the equal weights in the supercritical region.
The second point is a counterpart of the collision of the two saddles from the viewpoint of the Borel resummation.
The relation between Borel singularities and saddle points implies that
the collision of the two saddles in the $\sigma$-plane leads to one of the two Borel singularities in the $t$-plane.
Thus we can also decode the order of the phase transition purely from how the Borel singularities collide.
The third point means that the thimbles cross the multiple saddle points for $\arg (N_f ) = 0$
as shown in Fig.~\ref{fig:thimble1}.
In the case of the SQED, the Borel non-summability detects the infinite number of periodic saddles 
which come from the contribution from the hypermultiplets.

Finally, let us see the Stokes graph.
The second-order phase transition is interpreted in terms of the Stokes graph as follows.
\begin{itemize}
	\item The anti-Stokes line is given by $\Re[(\delta \lambda)^{3/2}]=0$.
	\item A Stokes phenomenon and an anti-Stokes phenomenon associated with saddles $\sigma_{0}^{\pm}$ occur simultaneously.
\end{itemize}
These points are analogous to the Airy-type model discussed in Sec.~\ref{sec:Airy}.
The only difference is that
the infinite number of Stokes phenomena associated with $\sigma_{n>0}^\pm$ occur simultaneously.

\section{Conclusions and discussion}
\label{sec:discussions}
We have studied the resurgence structure of a quantum field theory with a phase transition to uncover relations between resurgence and phase transitions.
In particular we have focused on the three dimensional $\mathcal{N}=4$ SQED, 
which undergoes the second-order quantum phase transition in the large-flavor limit \cite{Russo:2016ueu}.
We have approached the problem from the viewpoints of the Lefschetz thimbles and Borel resummation.
In the Lefschetz thimble approach,
we have specifically studied the thimble structures of the integral representation of the partition function obtained by the supersymmetric localization \cite{Kapustin:2009kz,Hama:2010av,Jafferis:2010un}.
We have first justified the assumption in \cite{Russo:2016ueu} that all the dominant complex saddles contribute to the integral by the Lefschetz thimble analysis.
Then we have found that 
there are a collision of the two saddles and a jump of the contributing saddle points 
as we cross the critical value of the parameter $\lambda =\eta /N_f$.
While this is the Stokes phenomenon,
we have seen that an anti-Stokes phenomenon also occurs at the same time.
Thus we interpret the second-order phase transition as the simultaneous Stokes and anti-Stokes phenomena.
Our result also shows that
the phase transition accompanies an infinite number of Stokes phenomena associated with the other saddles.
This behavior technically comes from the fact that
the action \eqref{eq:action} has the periodic structure 
which is physically originated from one-loop contributions of hypermultiplets 
in the localization formula of $S^3$ partition functions.
This indicates that the above structures hold not only for the SQED but also for more general supersymmetric gauge theories.

In the Borel resummation approach,
we have seen that
the thimble structures are appropriately mapped to
the Borel plane structures of the large-flavor expansion
as expected from the resurgence theory.
We have found the Borel singularities,
two of which correspond to the two saddles.
The two Borel singularities line up vertically along the imaginary axis after the phase transition.
It is a sign that the two saddles contribute to the integral with equal weights.
At the phase transition, the two Borel singularities collide as the two saddles.
The scattering angle of the Borel singularities at the collision is related to the order of the phase transition.
We have also seen that
the large-flavor expansion becomes Borel non-summable along the positive real axis in the supercritical region,
due to an infinite number of the Borel singularities.
This reflects the infinite number of Stokes phenomena.


Given the lessons from the SQED, 
we have provided more generic discussion on relations between the resurgence and phase transitions.
We have considered the one-dimensional integral of the form \eqref{eq:gen_model}
and shown that
if the $n$-saddles collide and scatter with a scattering angle $\beta\pi$
as we vary the parameter $\lambda$ through the critical point $\lambda =\lambda_c$,
then we have the phase transition of the order $\lceil (n+1)\beta \rceil$.
Our argument has also shown that
we have anti-Stokes phenomena at the critical point 
where the anti-Stokes line is given by $\Re\left[(\delta\lambda)^{(n+1)\beta}\right]=0$.
This implies that
one can read off the order of the phase transition also by looking at the anti-Stokes line.
We have also argued that 
the above behaviors are naturally translated into the language of the Borel plane.
This means that
the order of phase transitions can be determined also 
by tracking how Borel singularities move as varying the parameter.
This implies that we can read off information on phase structures 
purely in terms of perturbative expansions.
The above results apply to more general theories as long as they reduce to the form \eqref{eq:gen_model}.

Finally, we have revisited the second-order quantum phase transition in the SQED from the above viewpoints.
In the case of the SQED,
the two saddles $\sigma_0^+$ and $\sigma_0^-$ collide and scatter with  $\pi /2$
as we cross the critical point $\lambda =\lambda_c$.
Therefore we have $(n,\beta )=(2,1/2)$ in the formula $\lceil (n+1)\beta \rceil$
for the order of the phase transition and
this agrees with the fact that the second-order phase transition occurs.
From the viewpoint of Stokes graphs, the second-order phase transition is essentially described as the standard Airy-type graph.
This clarifies that the second-order phase transition is understood as a phenomenon
where the Stokes and anti-Stokes phenomena occur at the same time.
This is a clear contrast to the common understanding that a first-order phase transition is associated with an anti-Stokes phenomenon.
All of the above results support that resurgence works for describing the second-order phase transition in the SQED.

We have obtained a good news which may be useful to develop studies of resurgence on the technical side.
Originally, the correspondence between saddles and Borel singularities in one dimensional integrals was shown in \cite{Berry:1991,Boyd:1993,Boyd:1994}.
It is not guaranteed that 
we can naively apply the correspondence to the SQED
because some of their assumptions are violated due to the logarithmic branch cuts in the action.
Nevertheless our results suggest that the correspondence still holds even in the SQED.
This seems to imply that
one can extend the correspondence beyond the class of integrals studied in \cite{Berry:1991,Boyd:1993,Boyd:1994}.
It would be interesting to pursue this direction.


We believe that our results give a good step to understand 
connections between phase transitions and resurgence.
Yet there are still various questions and tasks which should be addressed as next steps.
First, 
it is important to understand the physical meaning of the second order phase transition in the SQED.
For instance, we have not understood yet
whether there is a change of symmetries around the critical point,
whether the critical point describes some conformal field theory and so on.
Second,
we have not identified interpretations of the saddle points in the SQED
in the language of the original path integral.
It seems that
they are closely related to the complex supersymmetric solutions found in \cite{Honda:2017qdb}
as discussed in App.~\ref{app:CSS}.
We need further studies to clarify the relations more precisely.
Third,
it would be interesting to study 
relations between Lee-Yang zeros and the Stokes graph.
While the authors in \cite{Kanazawa:2014qma} found that
Lee-Yang zeros are on anti-Stokes curve in the zero dimensional Gross-Neveu-like model,
the SQED studied in this paper does not seem to have such a property.
This may suggest that the situation in the SQED is different from the zero dimensional Gross-Neveu-like model.
Fourth,
it would be illuminating to study resurgence structures with respect to other parameters in the SQED such as the FI parameter $\eta$.
There may be interesting relations to the resurgence structure of the large flavor expansion
as in the two-dimensional pure $U(N)$ Yang-Mills theory on lattice,
where there were found interesting connections among expansions by $1/N$, Yang-Mills coupling and 't Hooft coupling \cite{Buividovich:2015oju,Ahmed:2017lhl,Ahmed:2018gbt}.
Finally 
it is technically important	to improve the Pad\'{e}-Uniformized approximation 
for cases with multiple branch cuts in Borel planes.
It seems that
an improvement of the approximation is hindered by the non-trivial topology of Riemann sheets due to the branch cuts.
Such a problem often arises in the context of resurgence and therefore further studies are desired.

While this paper has focused on connections between resurgence and phase transitions,
more generally, 
it would be very interesting to explore relations 
between resurgence and phases themselves rather than their transitions.
It is known that
information on phases in quantum field theories are partially captured by 't Hooft anomalies, including phases beyond the Ginzburg-Landau or Nambu paradigm.
While 't Hooft anomalies are typically easy to calculate and
give quite robust information on phases,
it relies on existence of symmetries\footnote{
This is not necessarily true for ``anomalies in the space of coupling constants" which was recently proposed \cite{Cordova:2019jnf,Cordova:2019uob}. 
}.
In contrast,
analysis of resurgence does not require symmetries and gives detailed information 
while it is technically much more complicated.
Therefore they play complementary roles.
In this paper we have discussed that 
some features of phase transitions are captured by 
qualitative behaviors of the objects appearing in the analysis of resurgence.
It would be great if one can find similar connections for 't Hooft anomalies\footnote{
See \cite{Sueishi:2021xti} for a very recent work on quantum mechanics in a similar spirit.
}.
It might
open a door to a shining world of non-perturbative physics.

\subsection*{Acknowledgment}
The authors would like to thank
Okuto Morikawa, Naohisa Sueishi, Hiromasa Takaura and Yuya Tanizaki for valuable discussions.
%
Preliminary results of this work have been presented in 
the KEK workshop ``Thermal Quantum Field Theory and its Application'' (Aug. 2020),
the YITP workshop YITP-W-20-08 ``Progress in Particle Physics 2020'' (Sep. 2020), 
the JPS meetings (Sep. 2020 and Mar. 2021),
the YITP-RIKEN iTHEMS workshop YITP-T-20-03 ``Potential Toolkit to Attack Nonperturbative Aspects of QFT -Resurgence and related topics-'' (Sep. 2020),
the 15th Kavli Asian Winter School on Strings, Particles and Cosmology (Jan. 2021) and
the Osaka City University Workshop ``Randomness, Integrability and Representation Theory in Quantum Field Theory 2021" (Mar. 2021).
Discussions during the workshops were helpful to complete this work.
M.~H. is partially supported by MEXT Q-LEAP.
This work is supported in part by Grant-in-Aid for Scientific Research (KAKENHI) (B) Grant Number 18H01217 (T.~F., T.~M. and N.~S.). 
S.~K is supported by the Polish National Science Centre grant 2018/29/B/ST2/02457.

\appendix
\section{Details on large flavor expansion}
\label{sec:app_expansion}
In this appendix,
we compute the coefficients of the $1/N_f$ expansion of the partition function around a general saddle point $\sigma_\ast$.
First, to make $N_f$-dependence transparent, we introduce
\begin{\eq}
S(\sigma ) = N_f \tilde{S}(\sigma ).
\end{\eq}
Then we expand $\tilde{S}(\sigma )$ as
\begin{align}
\tilde{S}(\sigma)
= \tilde{S}( \sigma_\ast )+ \frac{\tilde{S} \pp( \sigma_\ast )}{2!}(\sigma-\sigma_\ast )^2
+\sum_{k=0}^{\infty}\frac{ \tilde{S}^{(k+3)}(\sigma_\ast )}{(k+3)!}(\sigma-\sigma_\ast )^{k+3},
\end{align}
and regard the last term
as a perturbation.
Next we rewrite the contribution from the saddle $\sigma_\ast$ to the integral \eqref{eq:localization} as
\begin{align}
\int_{\mathcal{J}_\ast  }\dd{\sigma}\: e^{-S (\sigma)}
&= e^{-N_f \tilde{S} (\sigma_\ast )}
\int_{\mathcal{J}_\ast }\dd{\sigma}\:
\sum_{n=0}^{\infty}
\frac{1}{n!}\left(
-\sum_{k=0}^{\infty}\frac{ N_f \tilde{S}^{(k+3)}(\sigma_\ast )}{(k+3)!}
(\sigma-\sigma_\ast )^{k+3}
\right)^n
e^{-\frac{N_f \tilde{S} \pp(\sigma_\ast )}{2!}(\sigma-\sigma_\ast )^2}
\notag\\
&=e^{-N_f \tilde{S} (\sigma_\ast )}
 \sqrt{\frac{2}{ N_f \tilde{S}\pp(\sigma_\ast )}}
\int_{-\infty}^{\infty}\dd{\xi}\:
\sum_{n=0}^{\infty}
\frac{(-1)^n}{n!}
\left(\sum_{k=0}^{\infty}c_k\frac{\xi^{k+3}}{N_f^{\frac{k+1}{2}}}\right)^n
e^{-\xi^2} ,
\end{align}
where $\mathcal{J}_\ast$ is the Lefschetz thimble associated with $\sigma_\ast$ and
\begin{align}
&c_k
= \frac{2^{\frac{k+3}{2}}} {(k+3)! }
 \frac{\tilde{S}^{(k+3)}(\sigma_\ast )} {\bigl( \tilde{S}\pp(\sigma_\ast ) \bigr)^{\frac{k+3}{2}}}.
\end{align}
To proceed, we also introduce
\begin{align}
\left(
\sum_{k\p=0}^{\infty}c_{k\p}\epsilon^{k\p}
\right)^n
\equiv  \sum_{k=0}^\infty \tilde{c}_k (n) \epsilon^k .
\end{align}
Then, exchanging the integration and summation,
the formal $1/N_f$ expansion is computed as
\begin{align}
& \ \ \  \   
e^{-N_f\tilde{S}(\sigma_\ast )} \sqrt{\frac{2}{N_f S\pp(\sigma_\ast )}}
\sum_{n,k=0}^{\infty}
\frac{(-1)^n \tilde{c}_k(n)}{n!}\frac{1}{N_f^{\frac{n+k}{2}}}
\int_{-\infty}^{\infty}\dd{\xi}\:
\xi^{3n+k} e^{-\xi^2} \notag\\
&=e^{-N_f\tilde{S}(\sigma_\ast )}  \sqrt{\frac{2\pi}{N_f\tilde{S}\pp(\sigma_\ast )}}
\sum_{n,k=0,\:3n+k=\text{even}}^{\infty}
\frac{(-1)^n \Gamma\left(\frac{1}{2}+\frac{3n+k}{2}\right)\tilde{c}_k(n)}
{\Gamma(1/2)\Gamma(n+1)}
\frac{1}{N_f^{\frac{n+k}{2}}} \notag\\
&= e^{-N_f\tilde{S}(\sigma_\ast )} \sqrt{\frac{2\pi}{N_f\tilde{S}\pp(\sigma_\ast )}}
\sum_{\ell =0}^{\infty} \frac{a_\ell}{N_f^\ell} ,
\end{align}
where
\begin{align}
a_\ell
= \sum_{n=0}^{2\ell}
\frac{(-1)^n\Gamma\left(\frac{1}{2}+\ell +n\right)\tilde{c}_{2\ell -n}(n)}
{\Gamma(1/2)\Gamma(n+1)}.
\end{align}

\section{Lefschetz thimble structures for larger $\arg(N_f)$}
\label{sec:Lefschetz_thimble_larger_arg}
In this appendix,
we study the thimble structures for larger values of $\theta = \arg(N_f)$ than the one in Sec.~\ref{sec:complexNf}
to understand the Stokes phenomena more precisely.
It appears that the larger $\arg(N_f)$ region is not directly related to the phase transition itself
since originally the parameters were real.
However, this is essential to understand the Borel plane structures
as discussed in Sec.~\ref{sec:Borel}.

As mentioned in the last of Sec.~\ref{sec:transition_thimble}, for non-small $\arg(N_f)$,
we have to take the effects of the branch cuts into account.
Namely, when thimbles cross the branch cuts, the action is shifted by $2\pi i N_f \mathbb{Z} $,
and this effect modifies the condition for having Stokes phenomena.
We will see soon that the Stokes phenomena due to this effect indeed occurs in this problem.

\begin{figure}
\centering
\includegraphics[width=75mm]{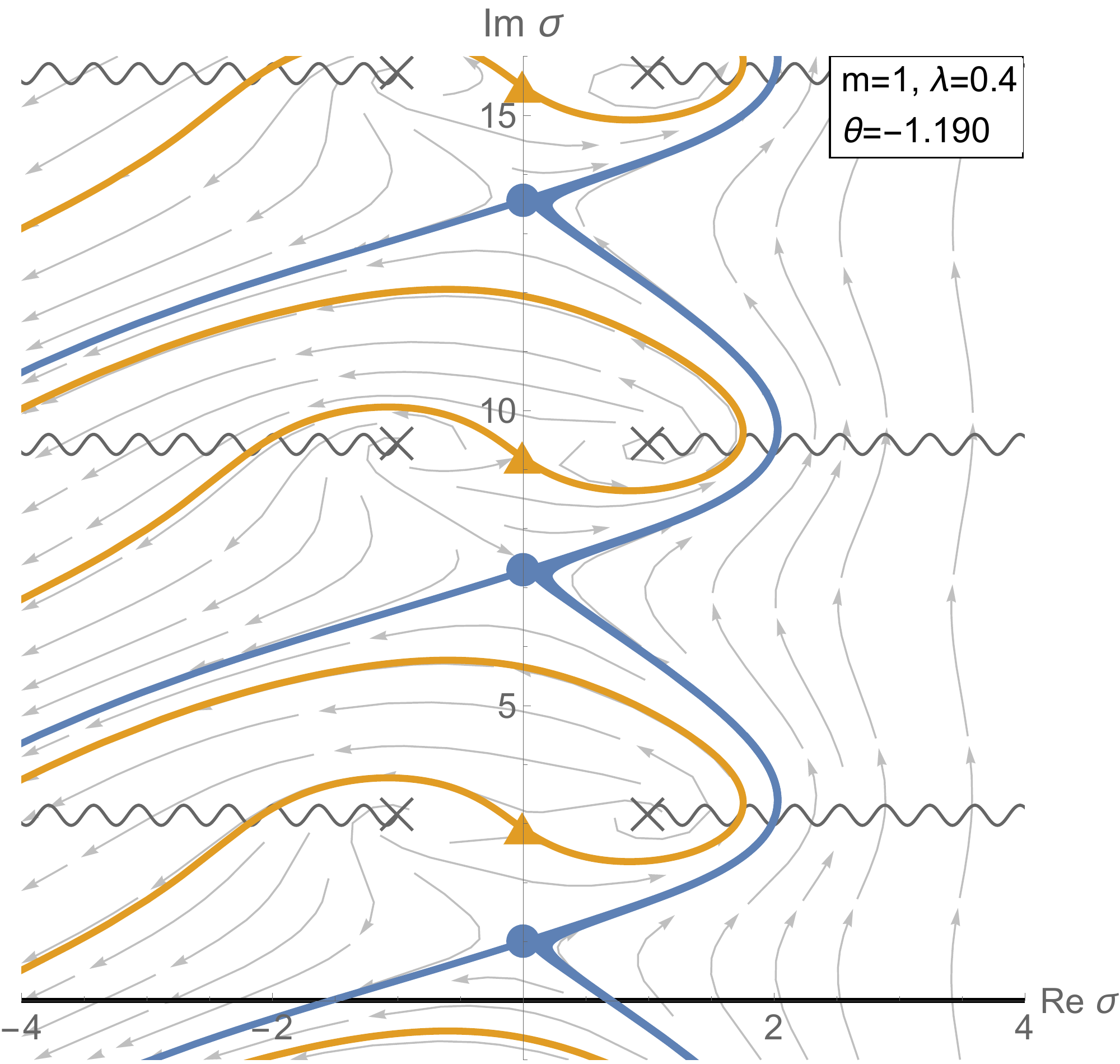} \hspace{8mm}
\includegraphics[width=75mm]{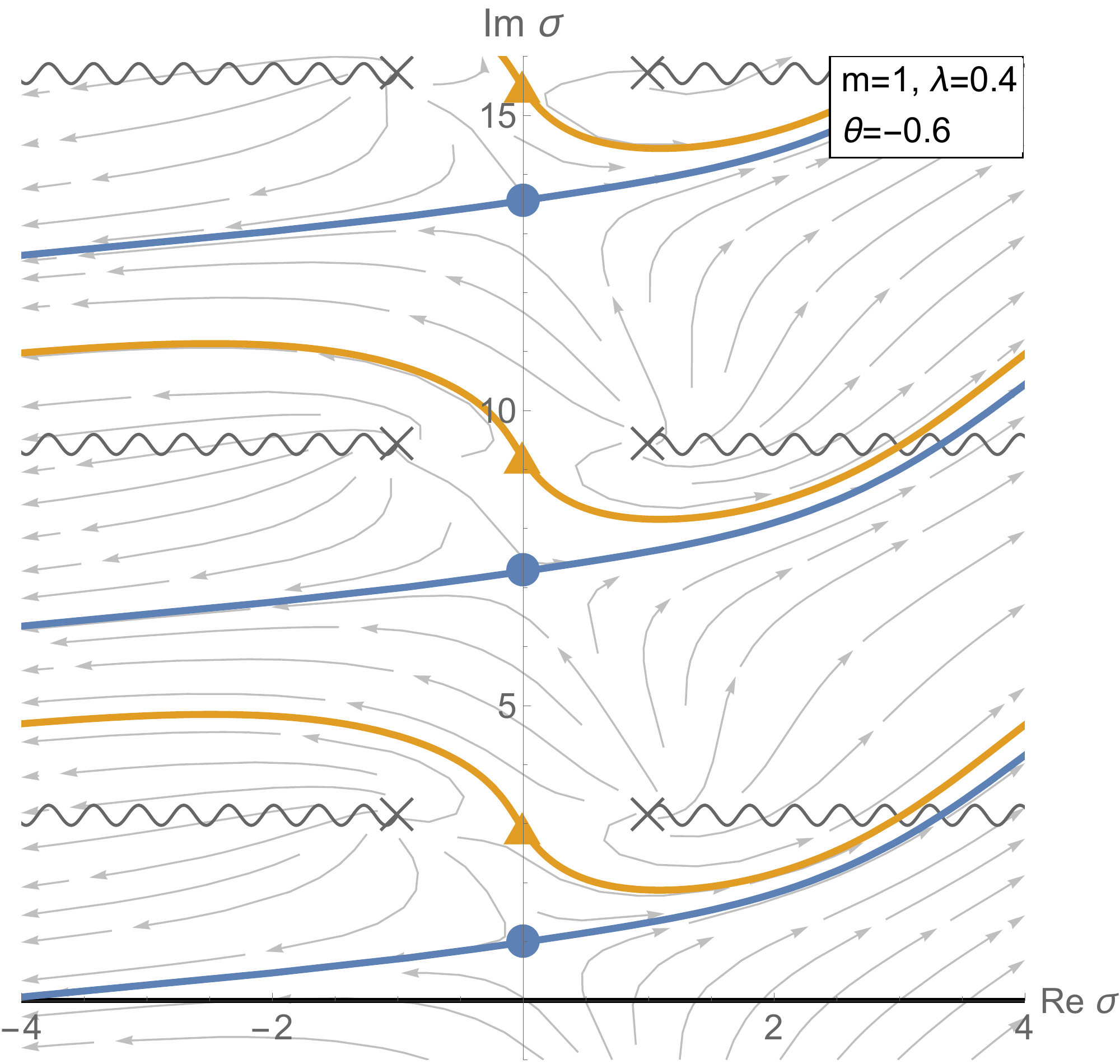} \\
\vspace{5mm}
\includegraphics[width=75mm]{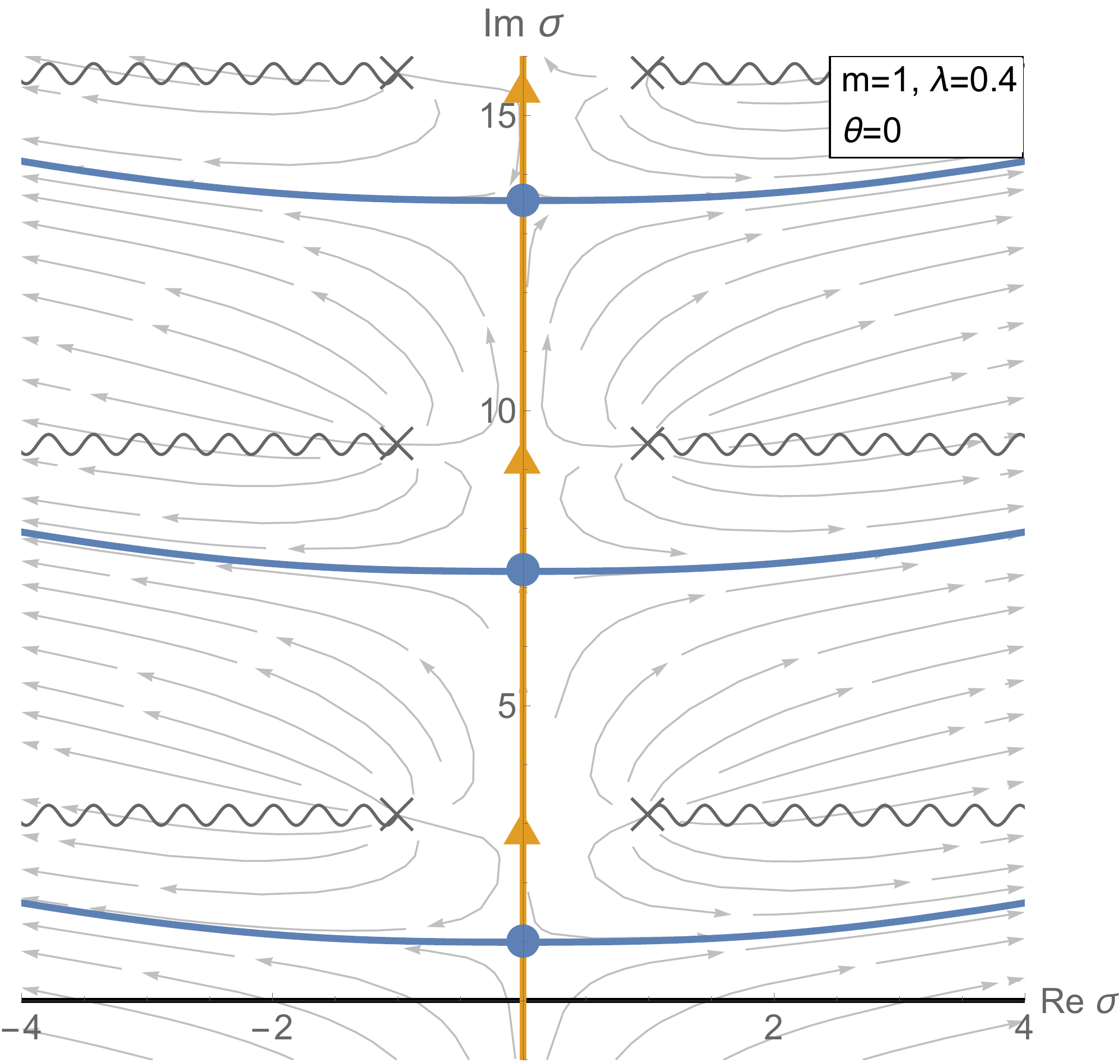}
\caption{
Illustrations of the Lefschetz thimble structures for the subcritical region $\lambda<\lambda\crit$
(in these figures, $m=1$, for which $\lambda=0.4$).
Larger phases are given $-\pi\leq\theta\leq0$
	so that we can observe Stokes phenomena (left ones)
	and thimble structures between them (right ones).
Reflecting figures along the imaginary axis
	corresponds to flipping the sign of $\theta$.
}
\label{fig:thimble_flow_m1_lam0p4_complete}
\end{figure}

In Fig.~\ref{fig:thimble_flow_m1_lam0p4_complete},
we summarize the Lefschetz thimble structures for $(\lambda ,m)=(0.4,1)$ with various $-\pi<\arg(N_f)<0$
as a representative of the subcritical region $\lambda<\lambda\crit$.
Reflecting these figures along the vertical axis corresponds to
	flipping the sign of $\arg(N_f)$.
Thus, these figures practically cover the full region $-\pi\leq\arg(N_f)<+\pi$.
In the figures on the left side, Stokes phenomena occur.
For example, at $\arg(N_f)=-1.190$,
	the Lefschetz thimble associated with a saddle $\sigma_0^+$ (blue line)
	passes also through other saddles $\sigma_{n>0}^+$.
This is nothing but a Stokes phenomenon.
One can easily check that
the codition for having Stokes phenomena: 
\begin{align}
\Im[S(\sigma_{n>0}^+)-S(\sigma_0^+) +2\pi i N_f \mathbb{Z}] = 0,
\end{align}
is satisfied by $\arg(N_f)=-1.190$.

\begin{figure}
\centering
\includegraphics[width=75mm]{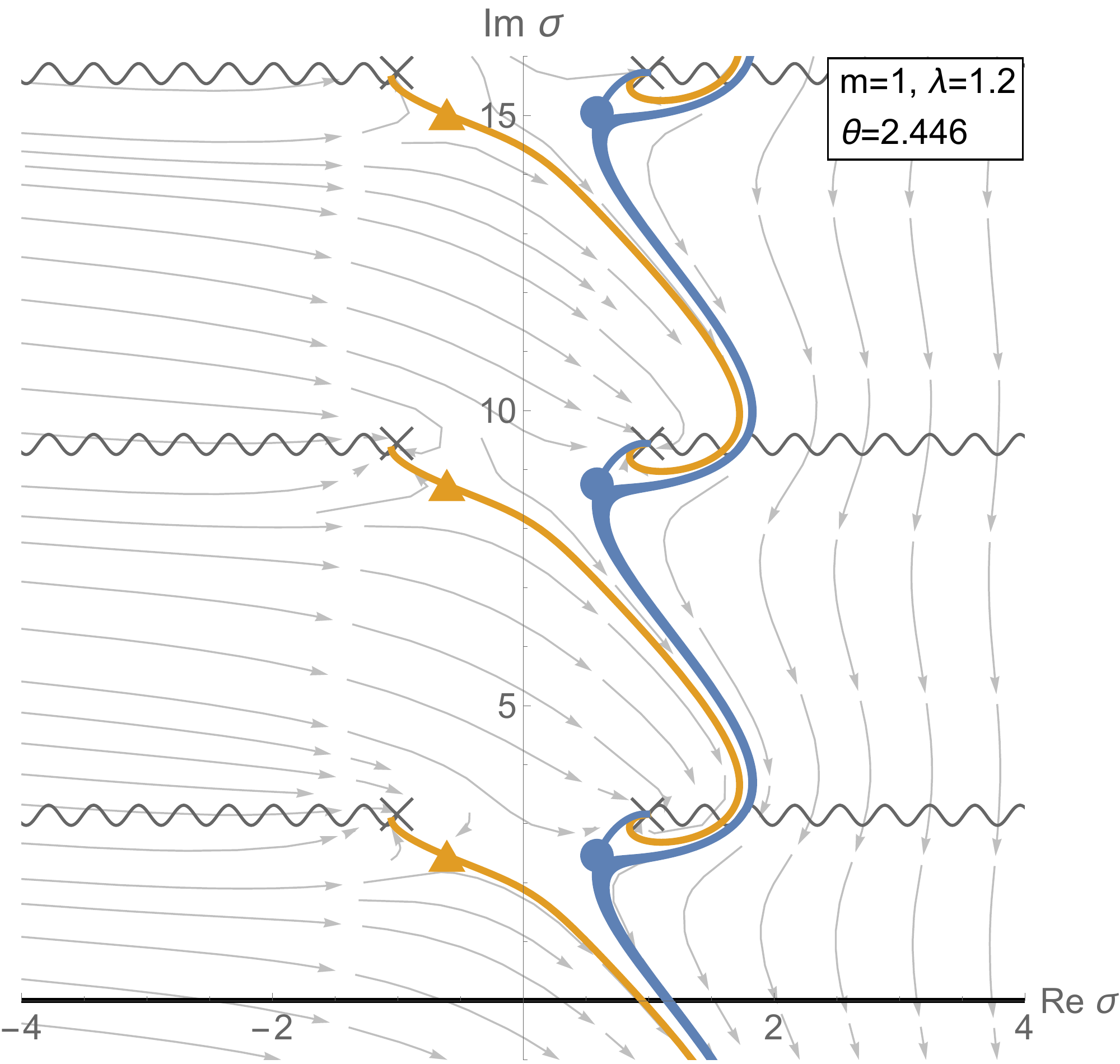} \hspace{8mm}
\includegraphics[width=75mm]{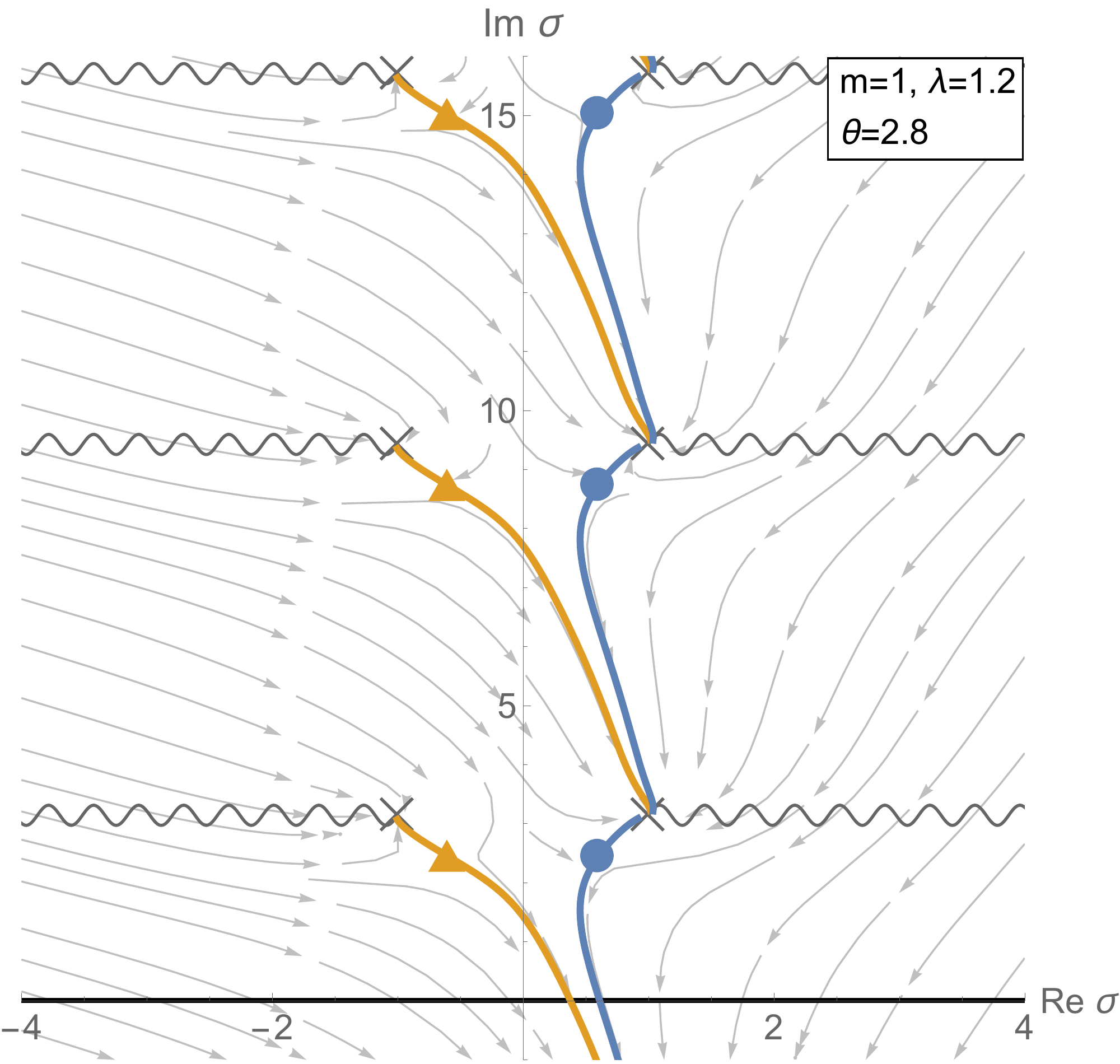}
\caption{
Illustrations of the Lefschetz thimble structures for the supercritical region $\lambda>\lambda\crit$
(in these figures, $m=1$, for which $\lambda=0.4$).
Larger phases are given $-\pi \leq \theta \leq 0$
	so that we can observe Stokes phenomena (left ones)
	and thimble structures between them (right ones).
}
\label{fig:thimble_flow_m1_lam1p2_complete}
\end{figure}

In Fig.~\ref{fig:thimble_flow_m1_lam1p2_complete},
we summarize the thimble structures for the supercritical region $\lambda>\lambda\crit$
with various $-\pi<\arg(N_f)<+\pi$ (specifically $(\lambda ,m)=(1.2,1)$ in the figures).
In the figures on the left side,
	Stokes phenomena occur.
For example, at $\arg(N_f)=-0.039$,
	the Lefschetz thimble associated with a saddle $\sigma_0^+$ (blue line)
	passes also through another saddle $\sigma_{1}^-$.
This means that
\begin{align}
\Im[S(\sigma_{1}^-)-S(\sigma_0^+) +2\pi i N_f \mathbb{Z}] = 0
\end{align}
holds at $\arg(N_f) = -0.039$.
This is also a Stokes phenomenon.
These structures are consistent with the Borel plane structures discussed in Sec.~\ref{sec:Borel}.

\begin{figure}
\centering
\includegraphics[width=75mm]{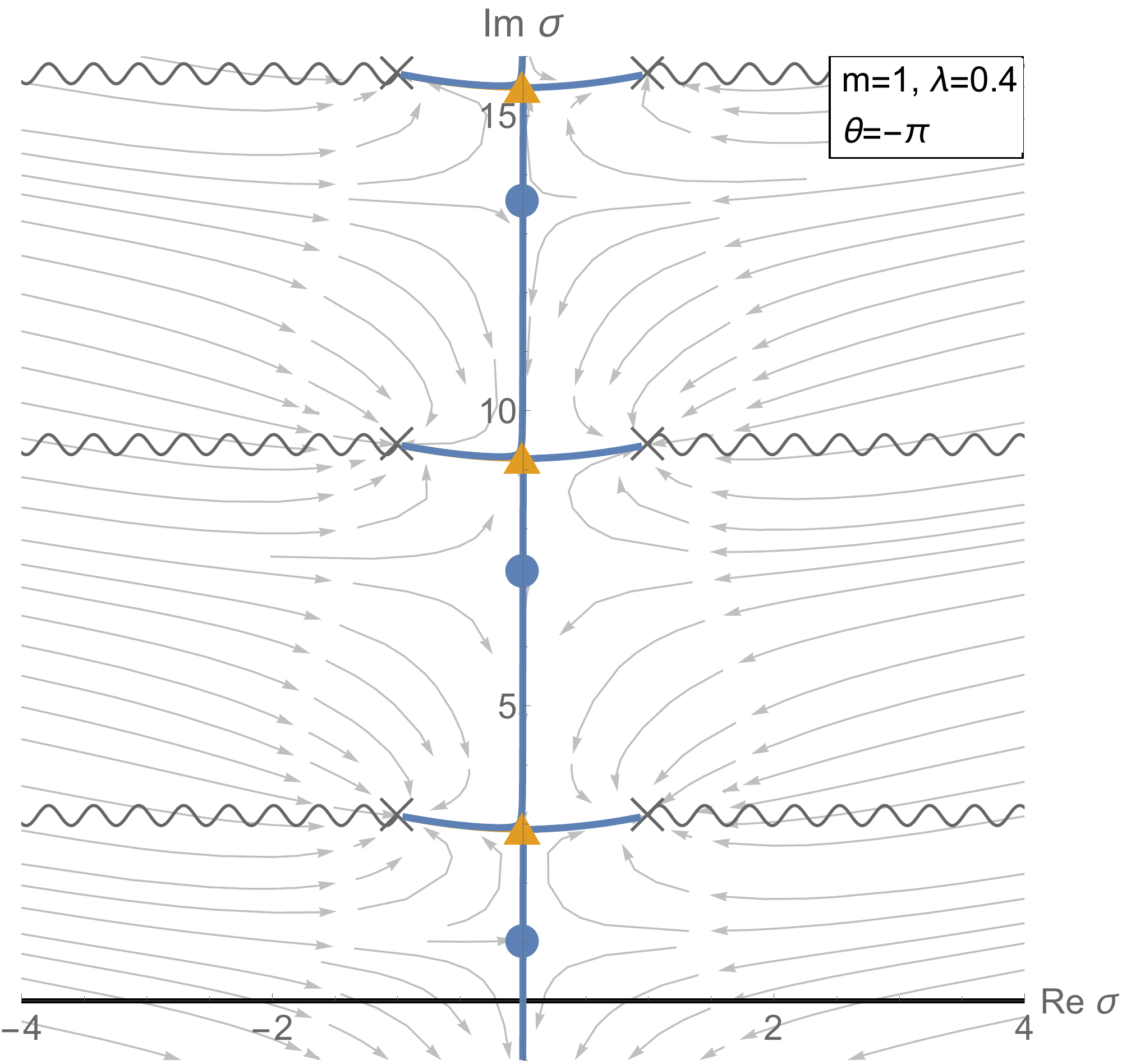} \hspace{8mm}
\includegraphics[width=75mm]{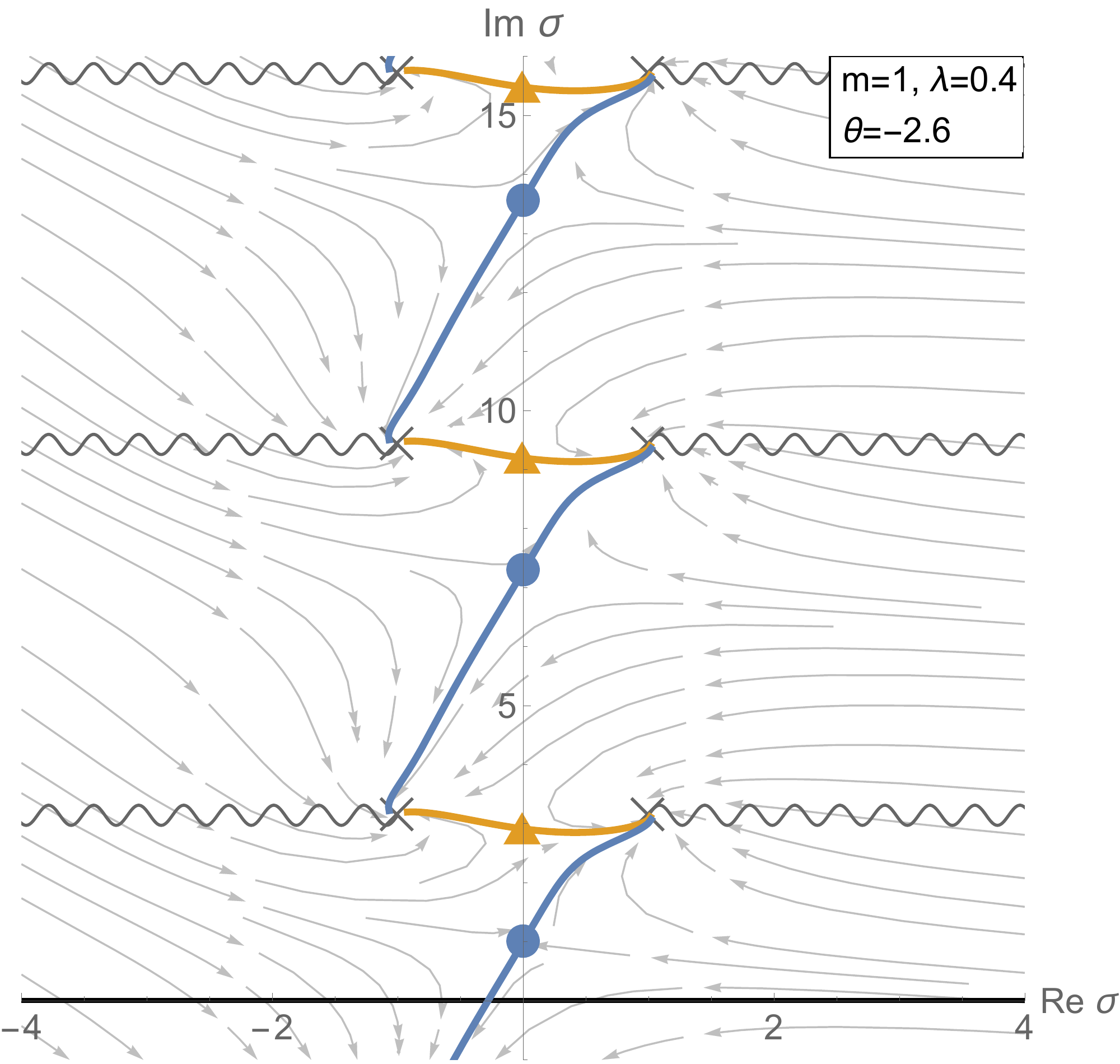} \\
\vspace{5mm}
\includegraphics[width=75mm]{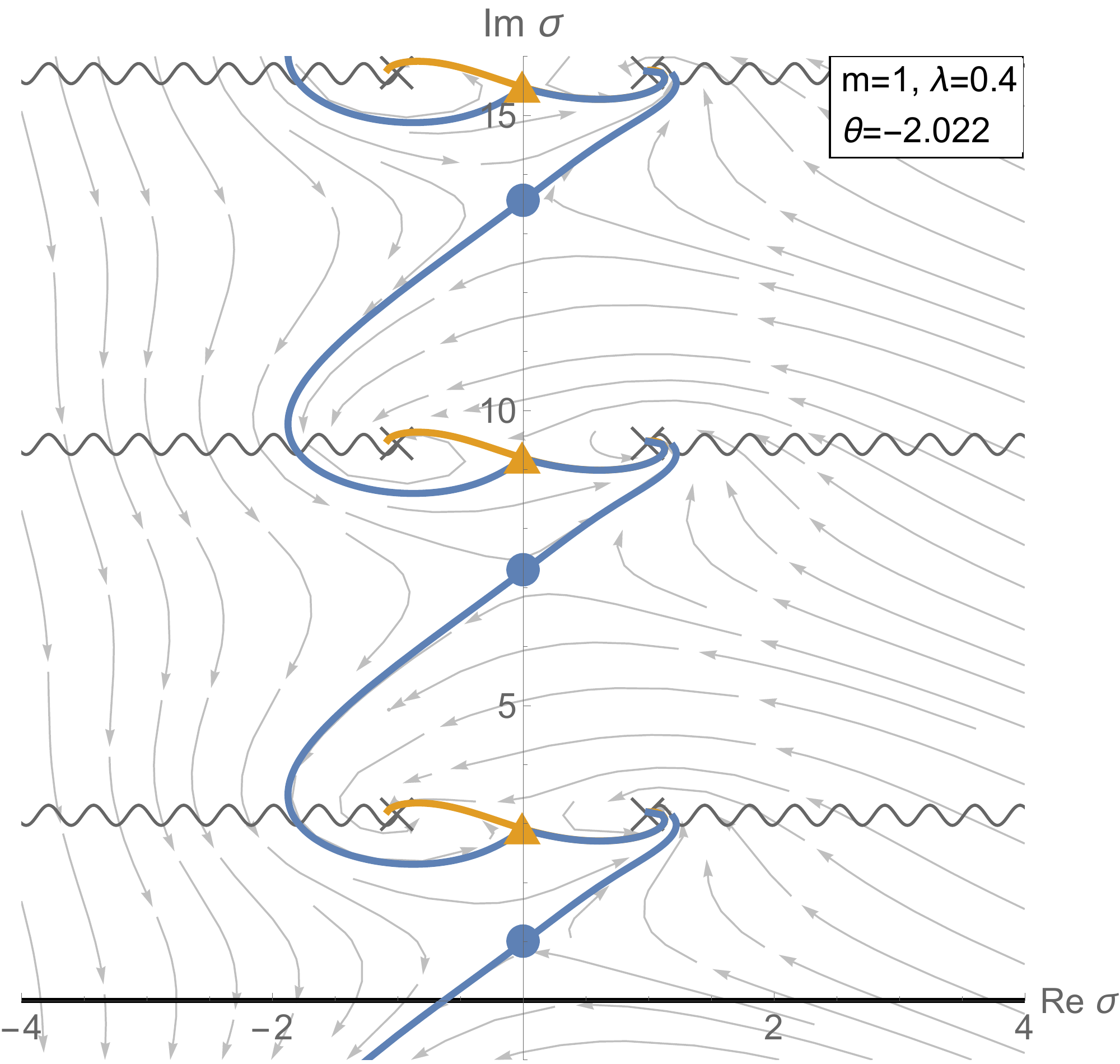} \hspace{8mm}
\includegraphics[width=75mm]{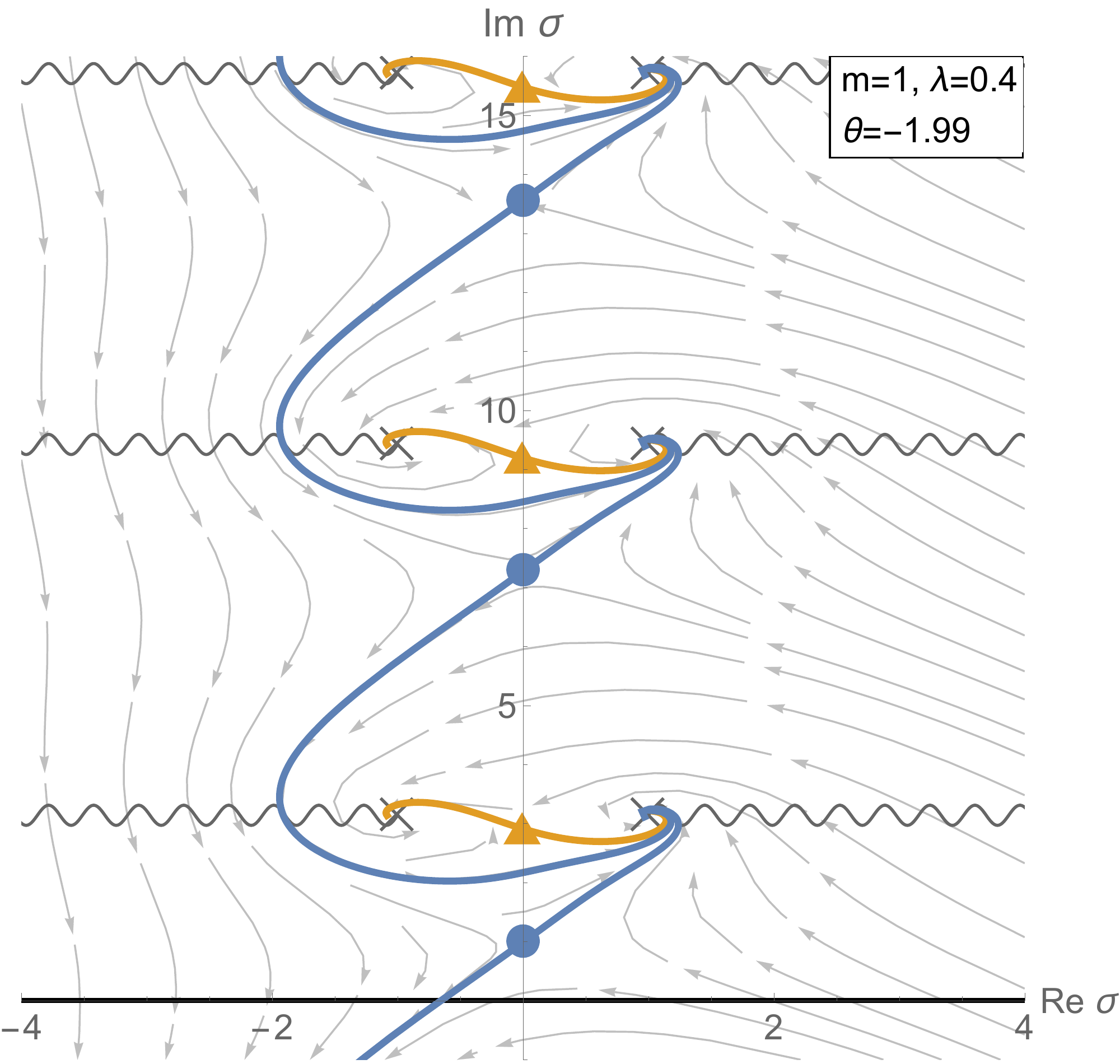} \\
\vspace{5mm}
\includegraphics[width=75mm]{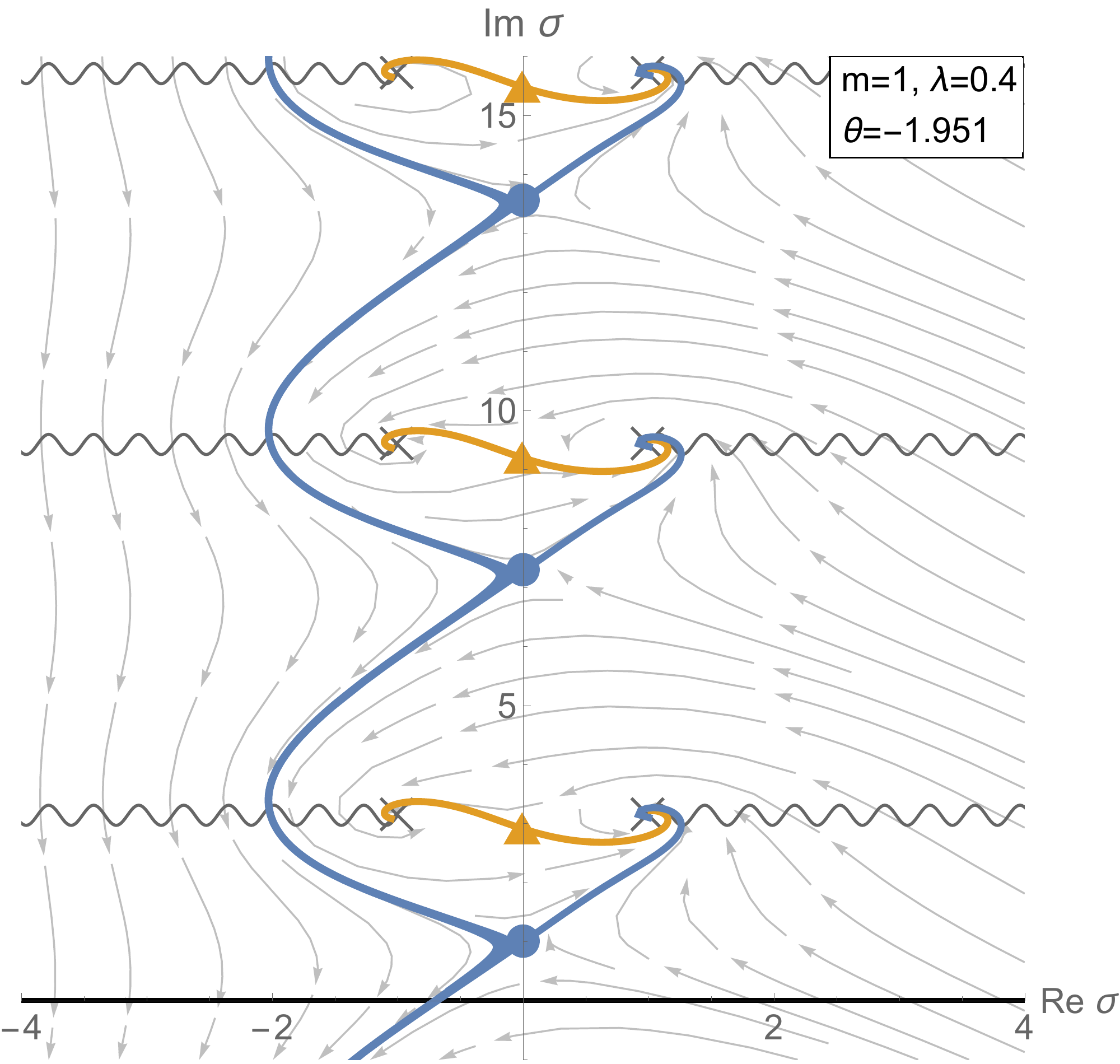} \hspace{8mm}
\includegraphics[width=75mm]{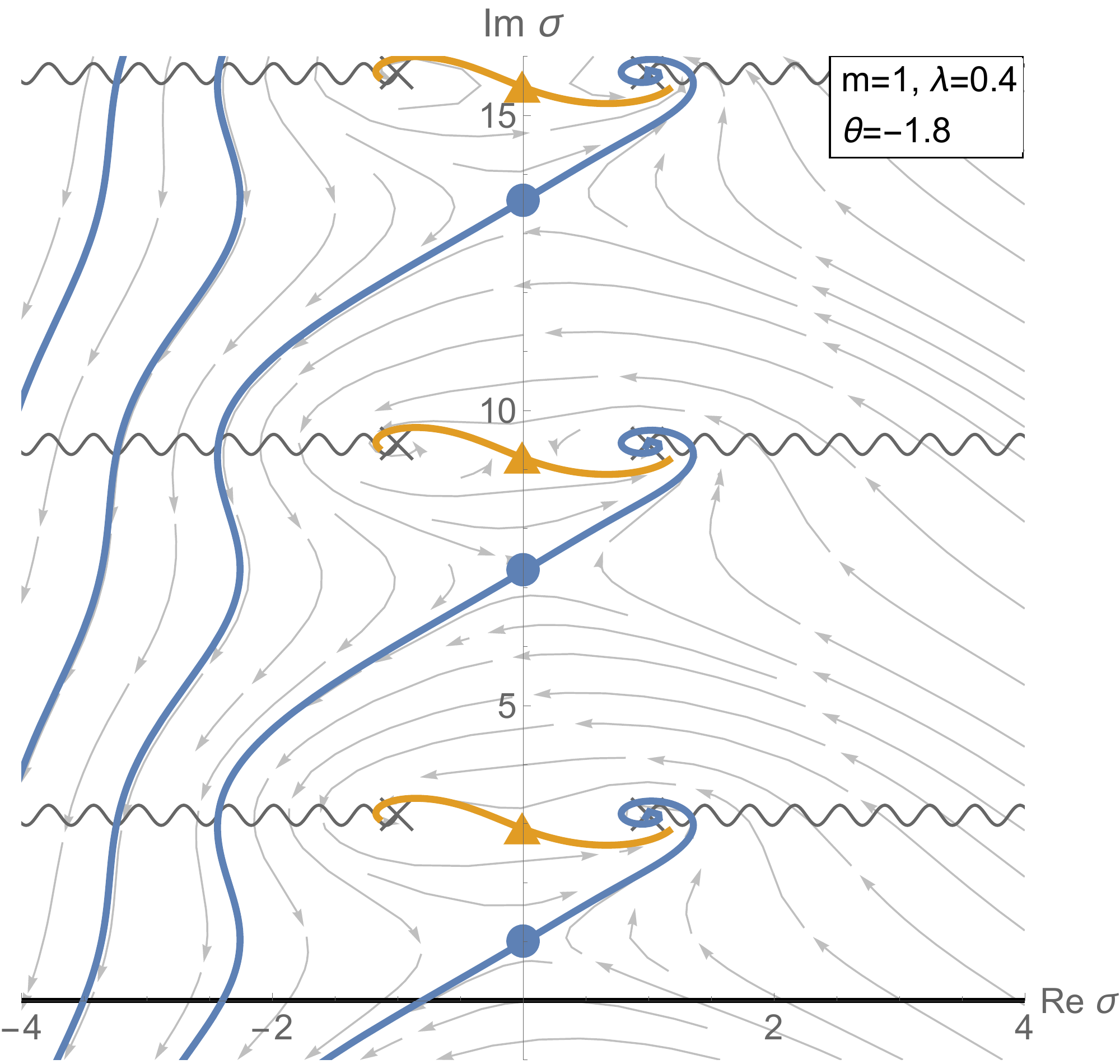}
\end{figure}
\begin{figure}
\centering
\includegraphics[width=75mm]{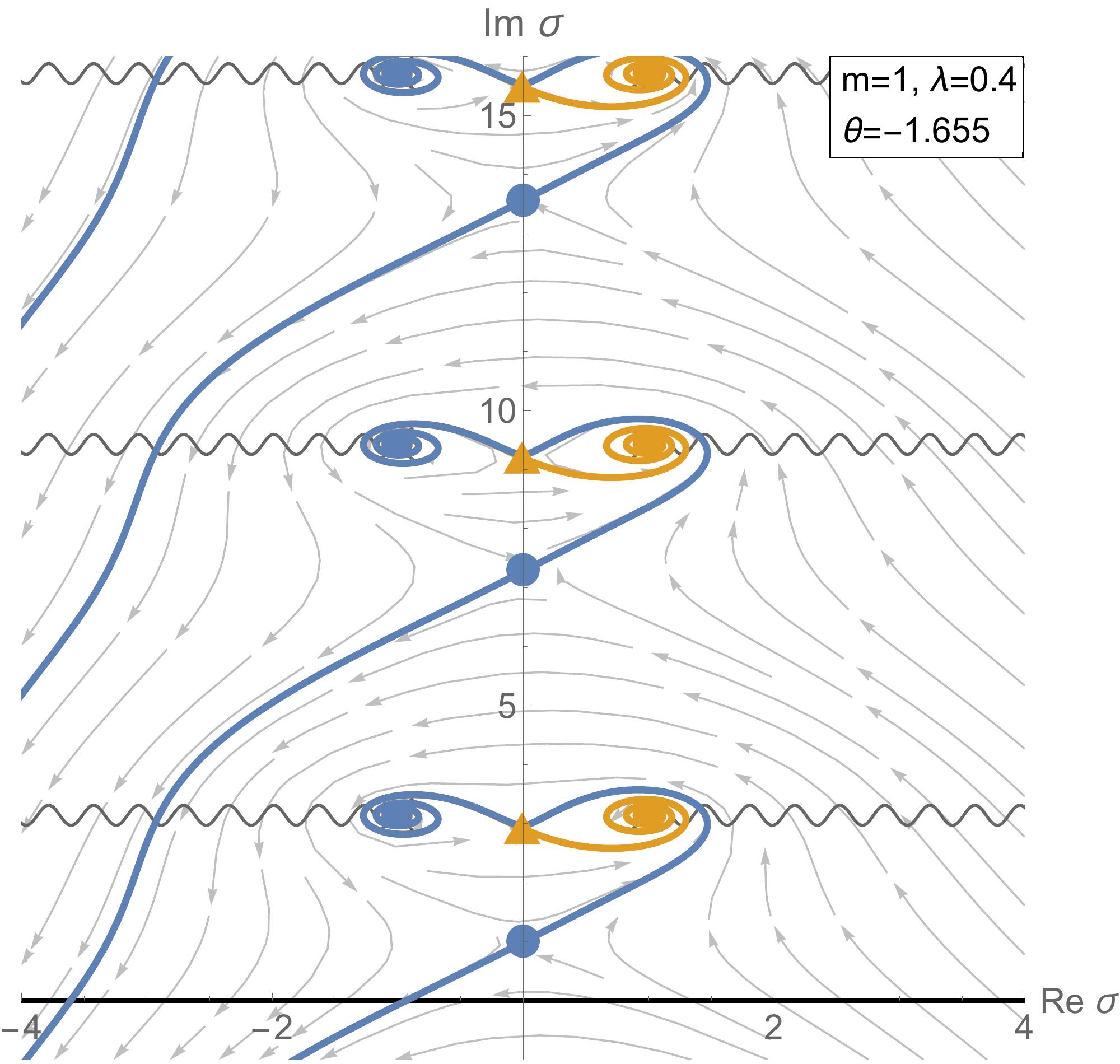} \hspace{8mm}
\includegraphics[width=75mm]{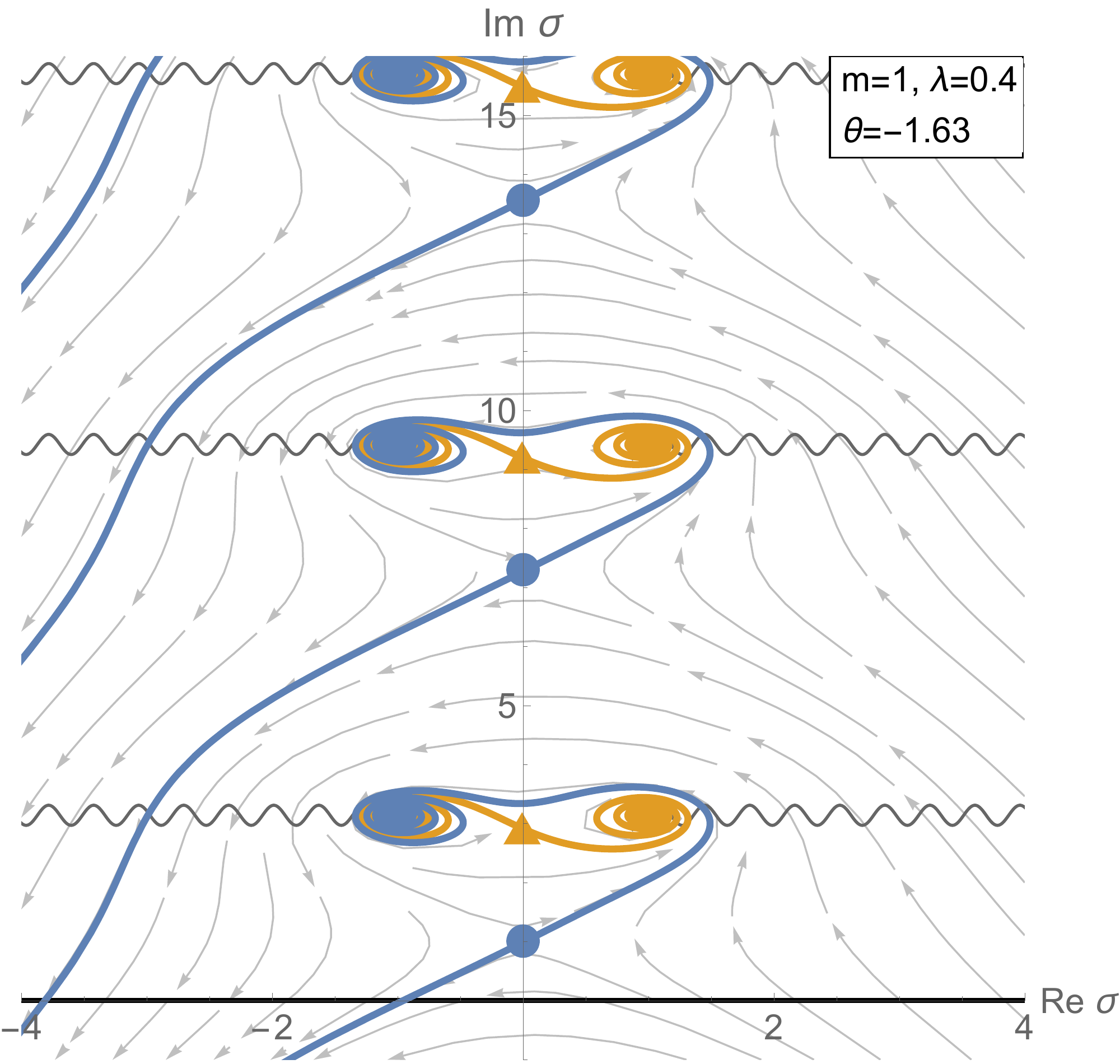} \\
\vspace{5mm}
\includegraphics[width=75mm]{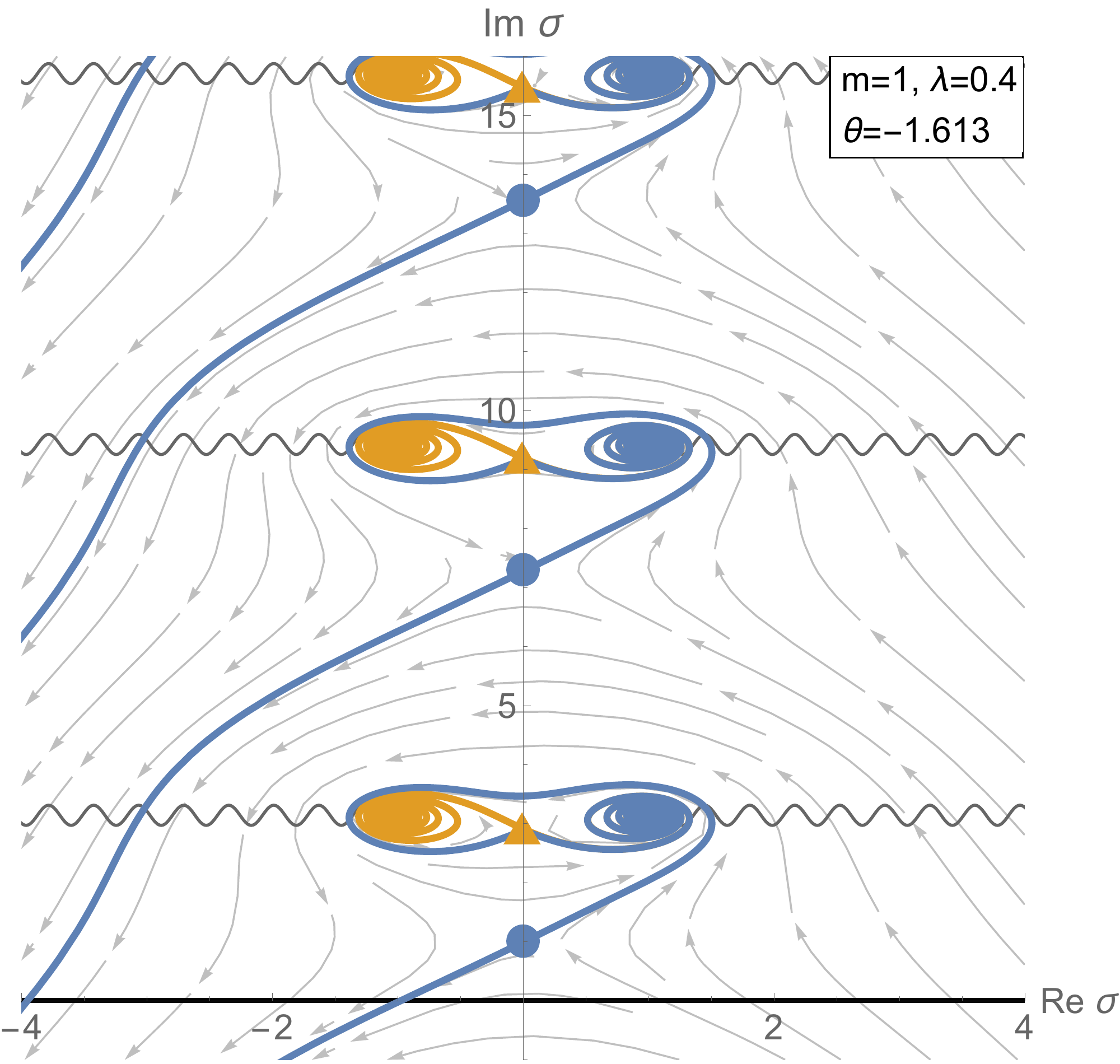} \hspace{8mm}
\includegraphics[width=75mm]{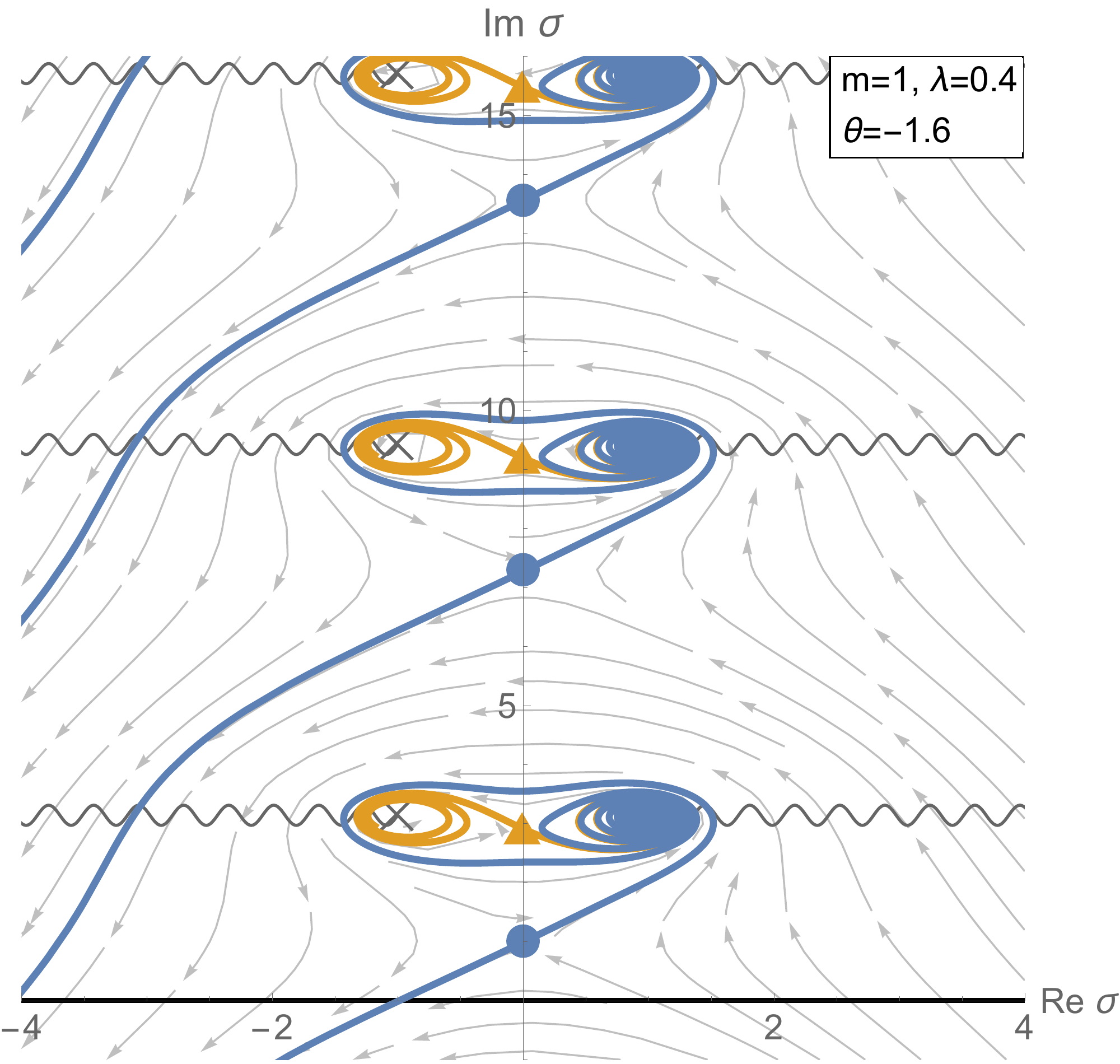} \\
\vspace{5mm}
\includegraphics[width=75mm]{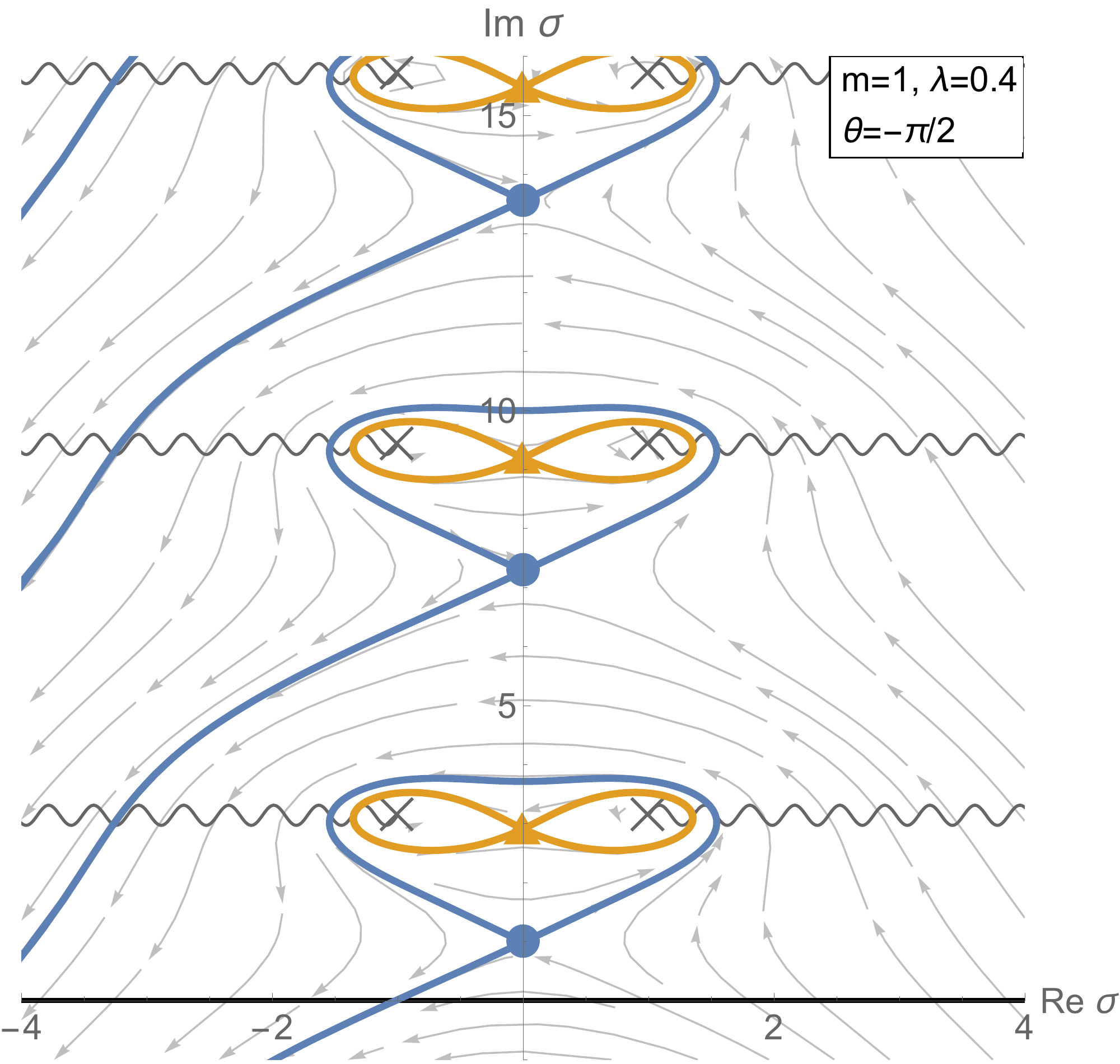} \hspace{8mm}
\includegraphics[width=75mm]{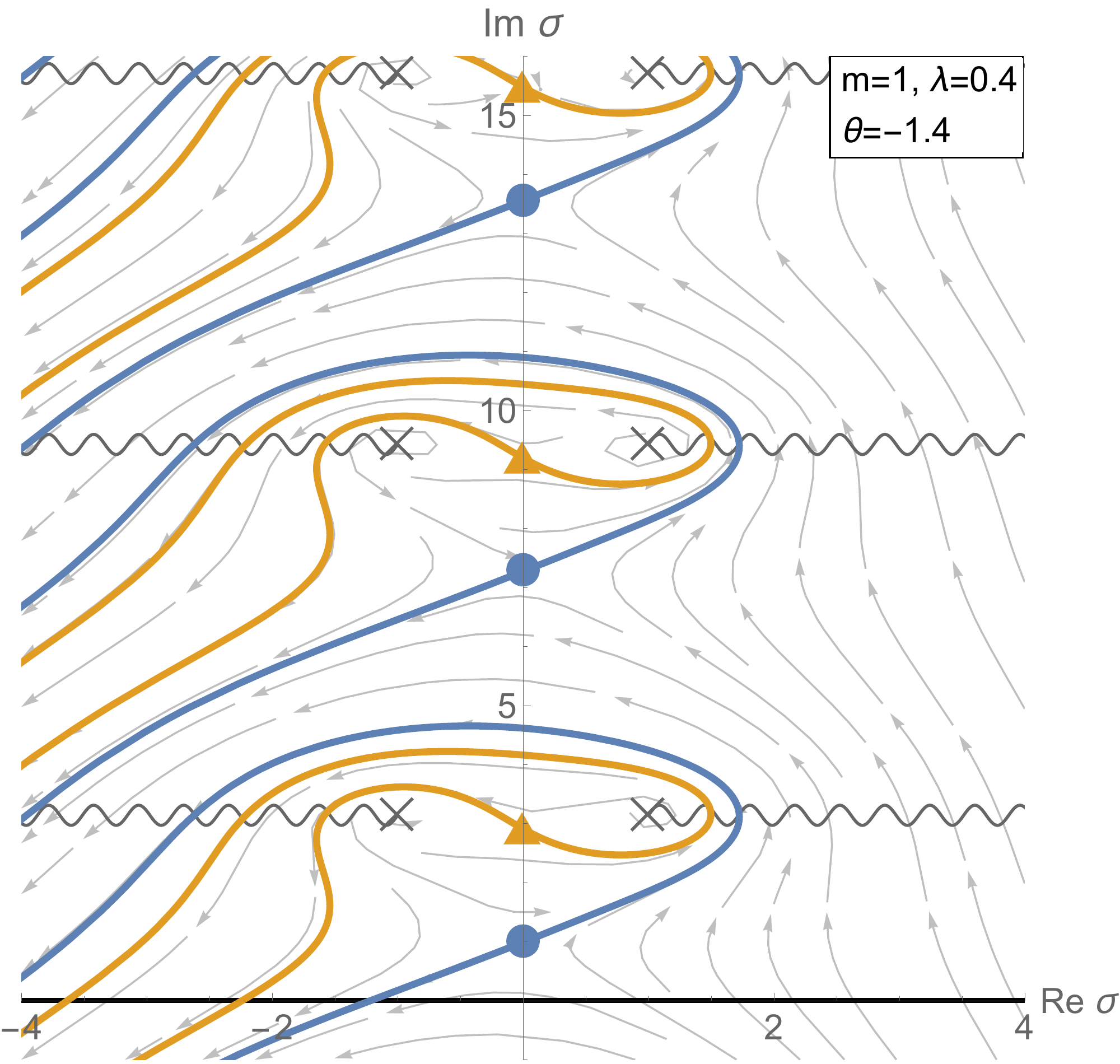}
\end{figure}

\begin{figure}
\centering
\includegraphics[width=75mm]{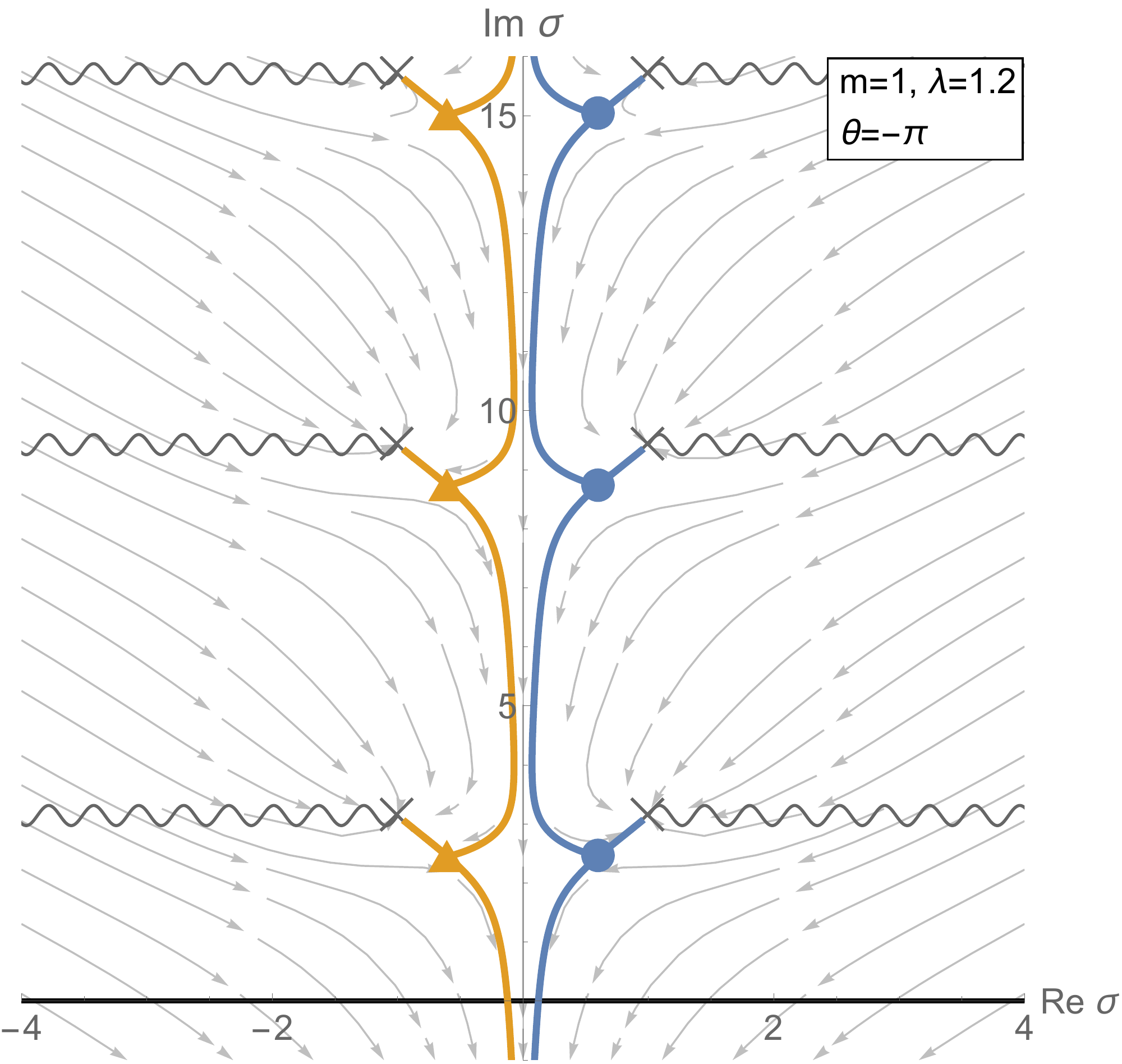} \hspace{8mm}
\includegraphics[width=75mm]{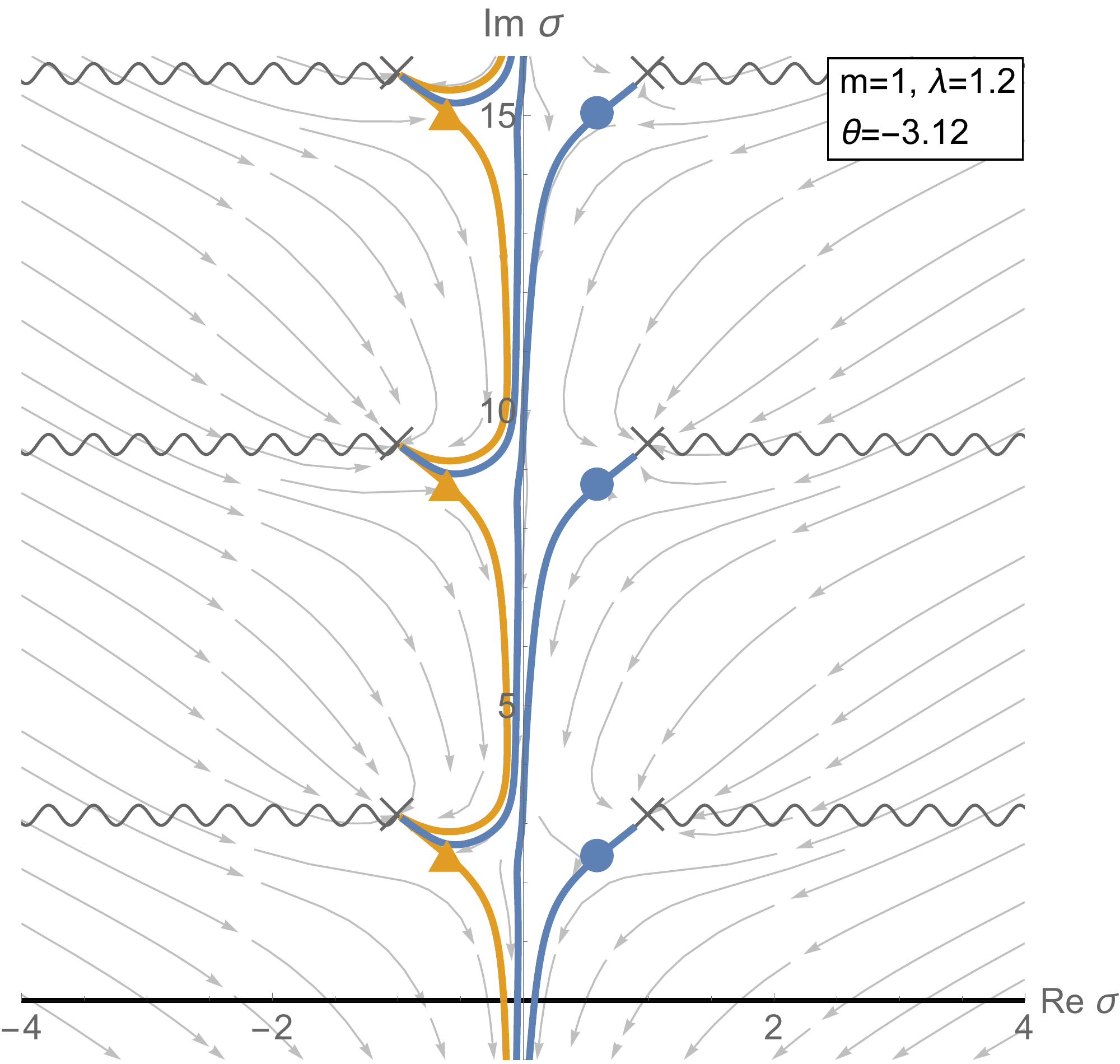} \\
\vspace{5mm}
\includegraphics[width=75mm]{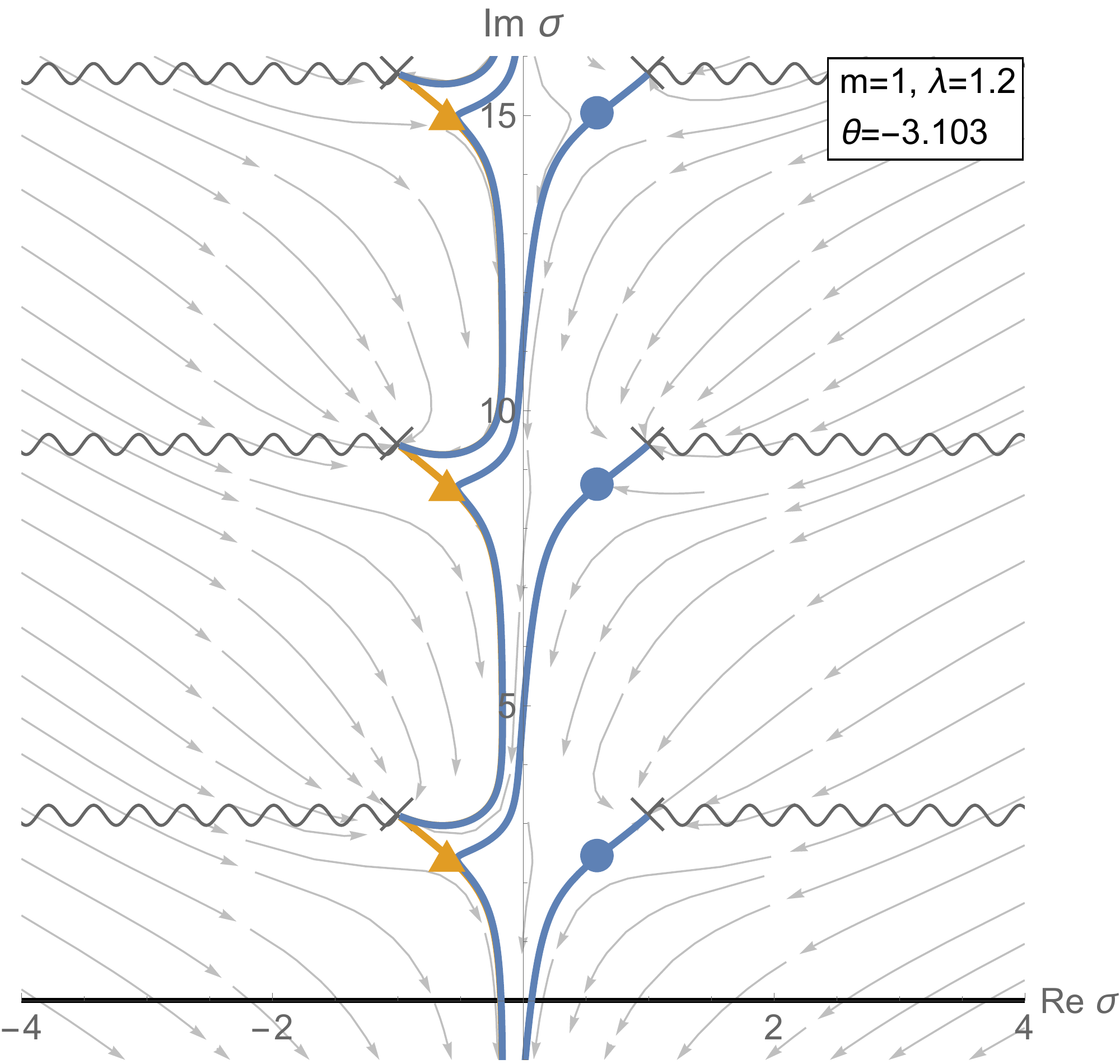} \hspace{8mm}
\includegraphics[width=75mm]{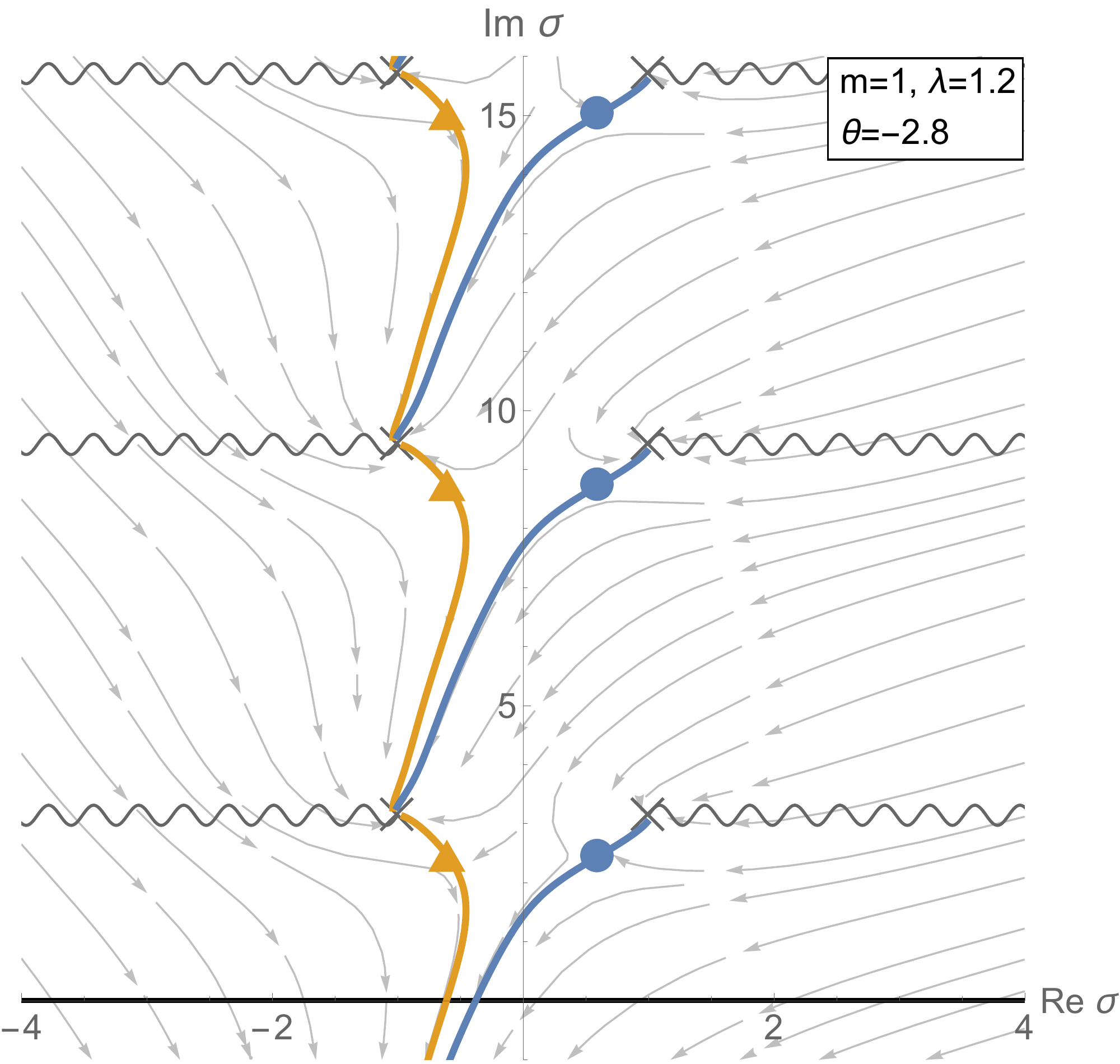} \\
\vspace{5mm}
\includegraphics[width=75mm]{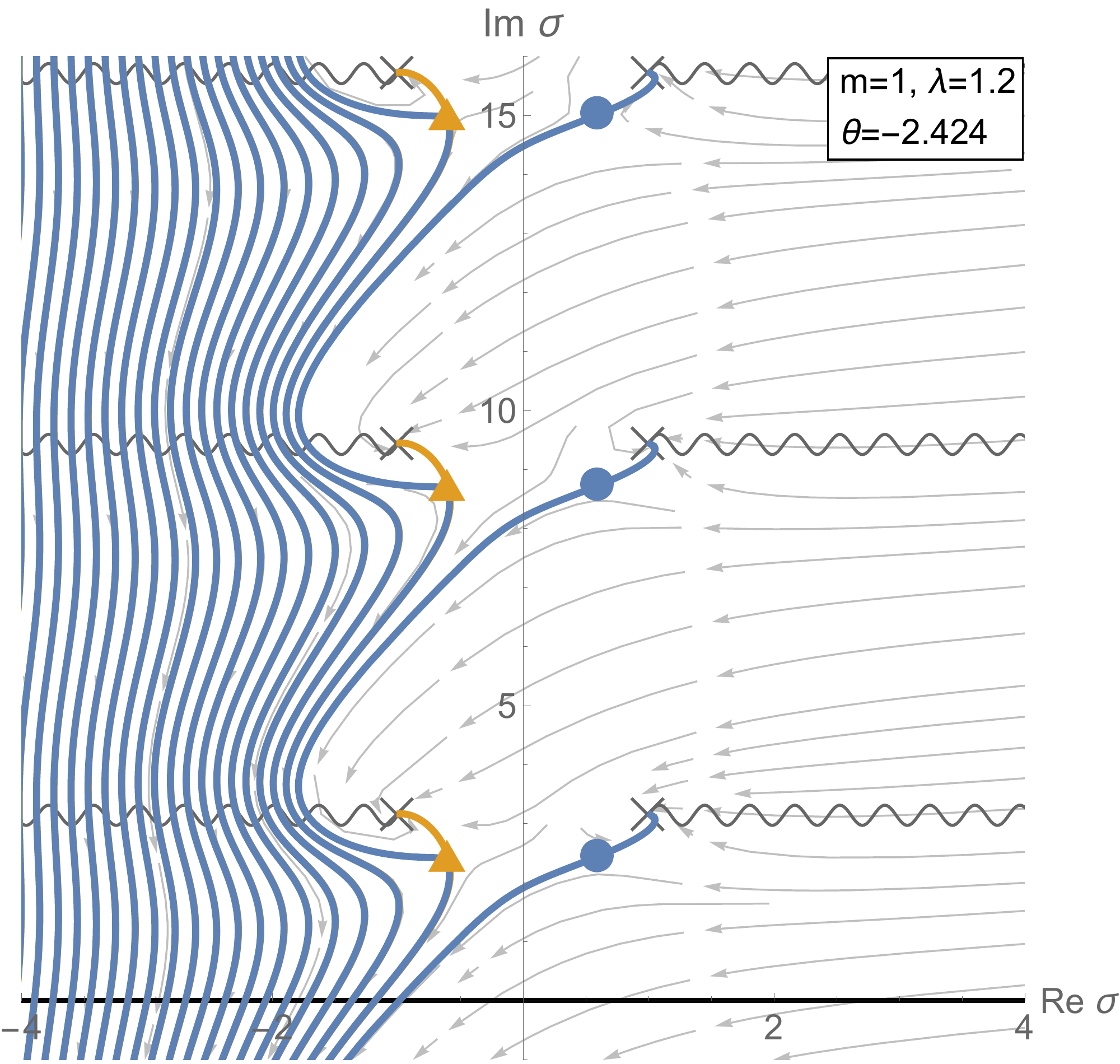} \hspace{8mm}
\includegraphics[width=75mm]{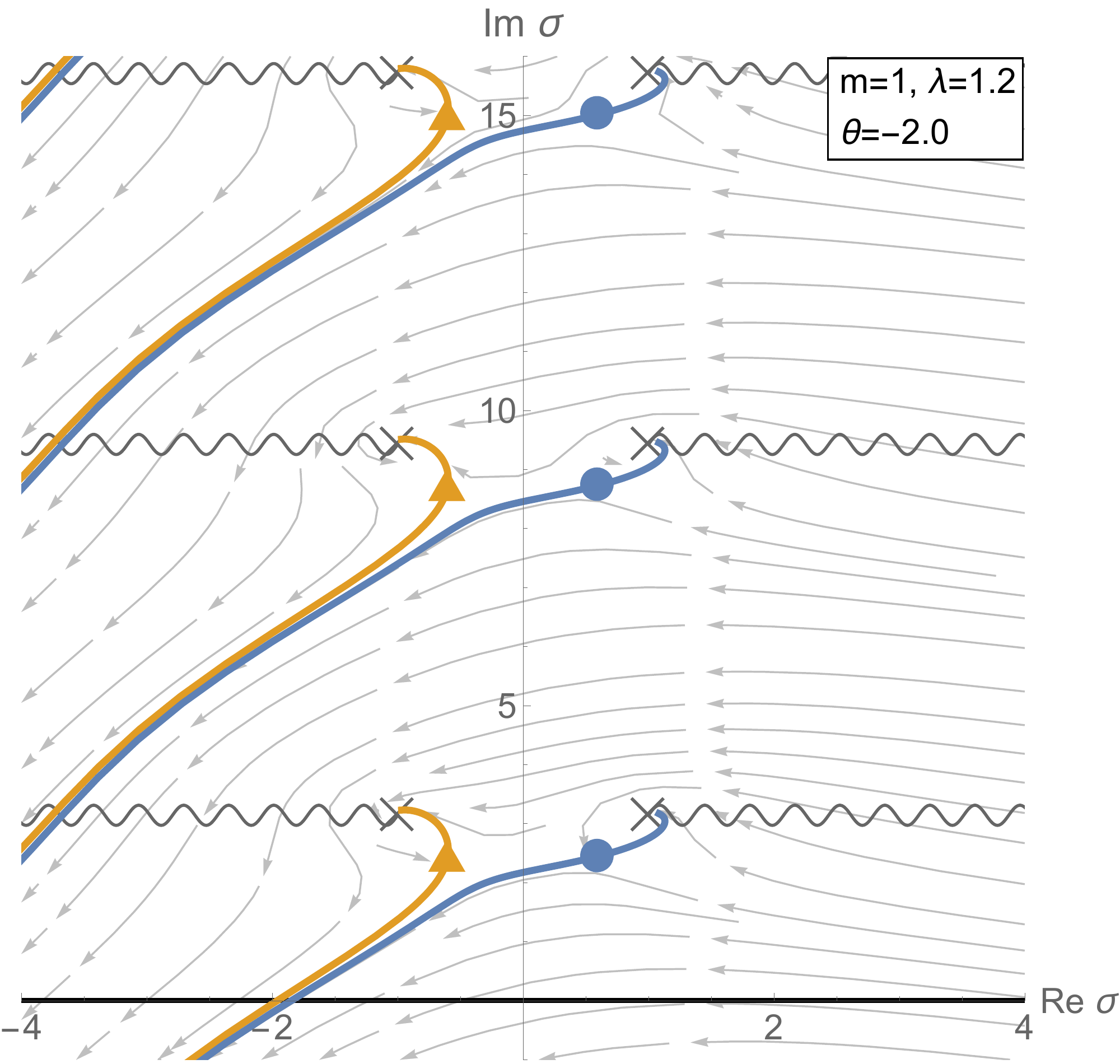}
\end{figure}
\begin{figure}
\centering
\includegraphics[width=75mm]{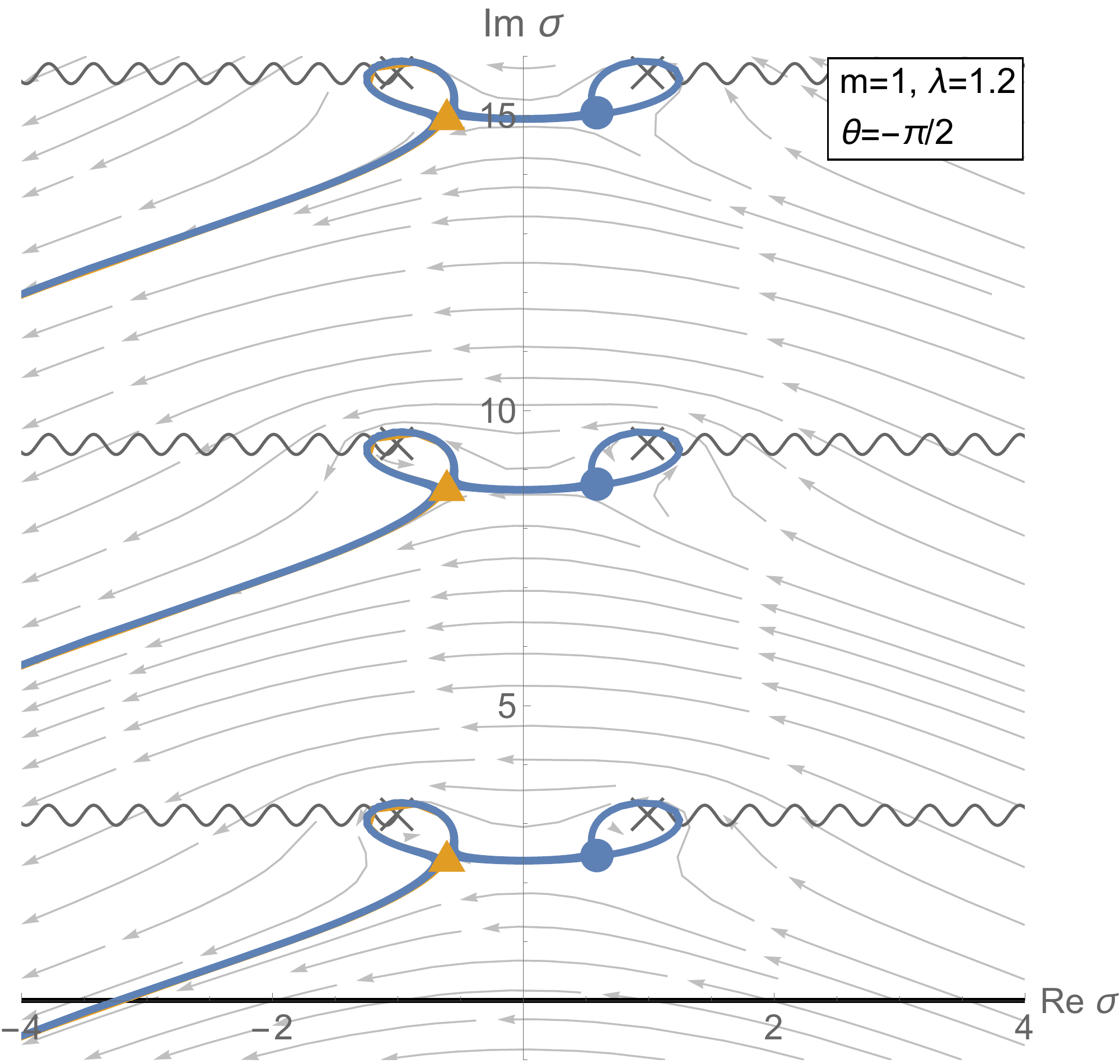} \hspace{8mm}
\includegraphics[width=75mm]{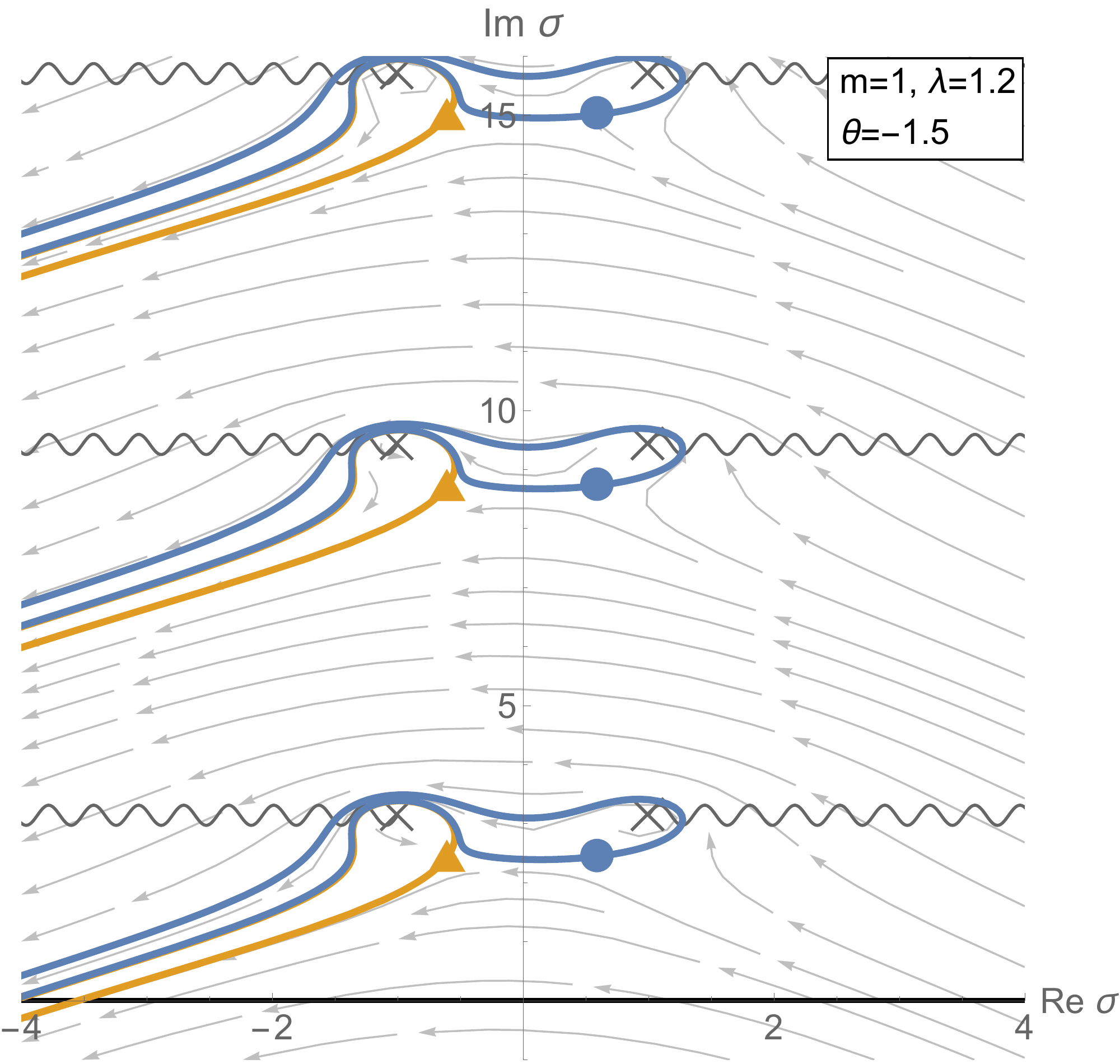} \\
\vspace{5mm}
\includegraphics[width=75mm]{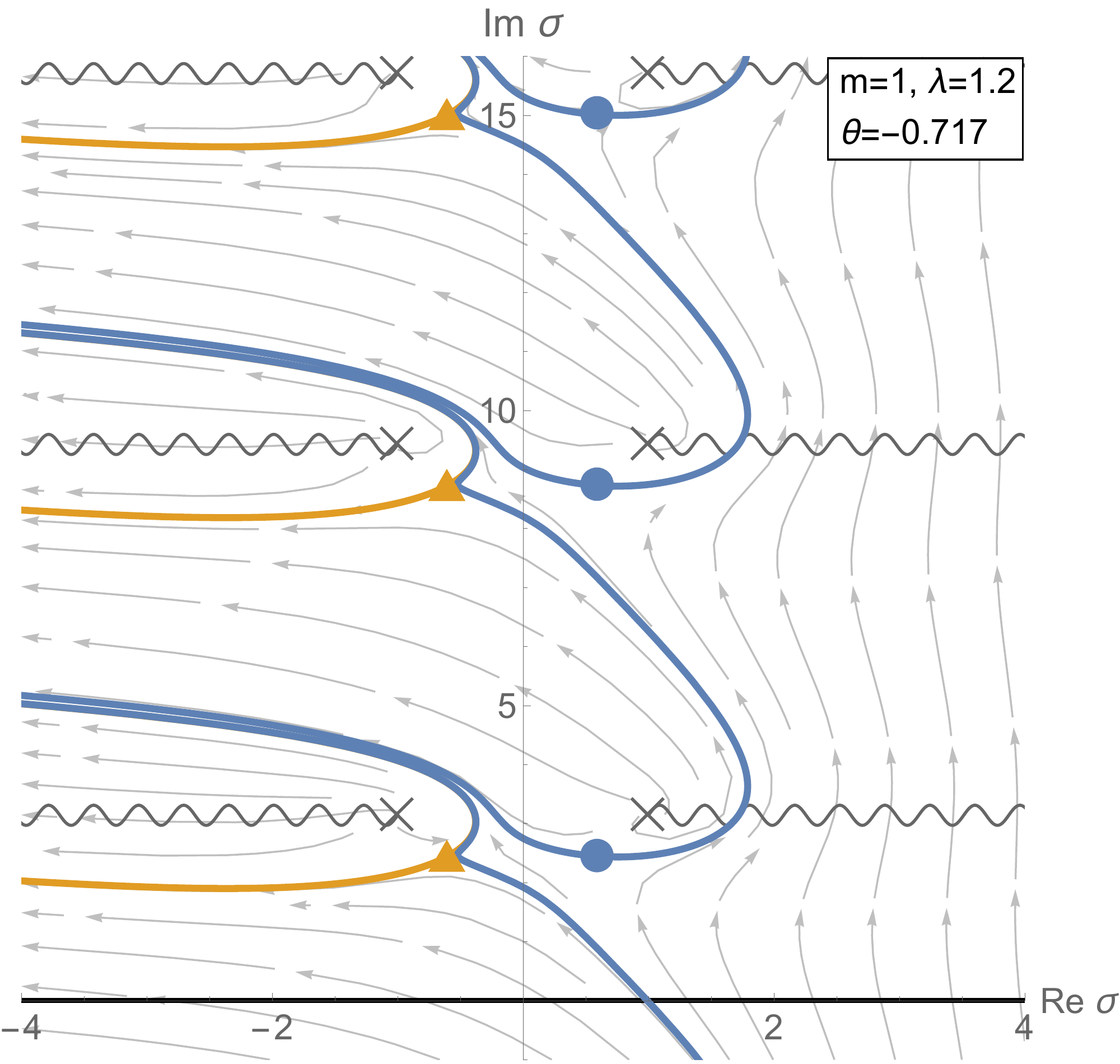} \hspace{8mm}
\includegraphics[width=75mm]{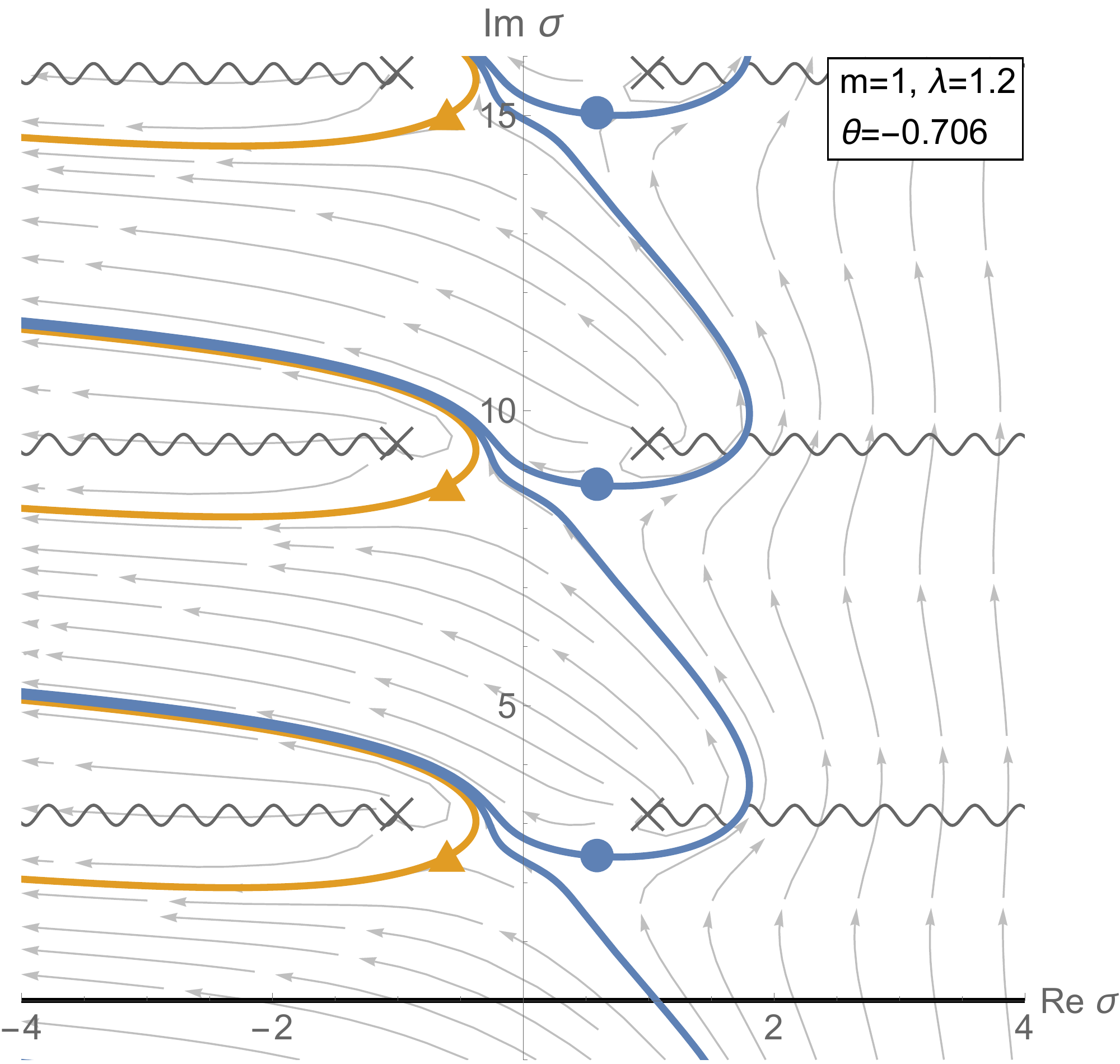} \\
\vspace{5mm}
\includegraphics[width=75mm]{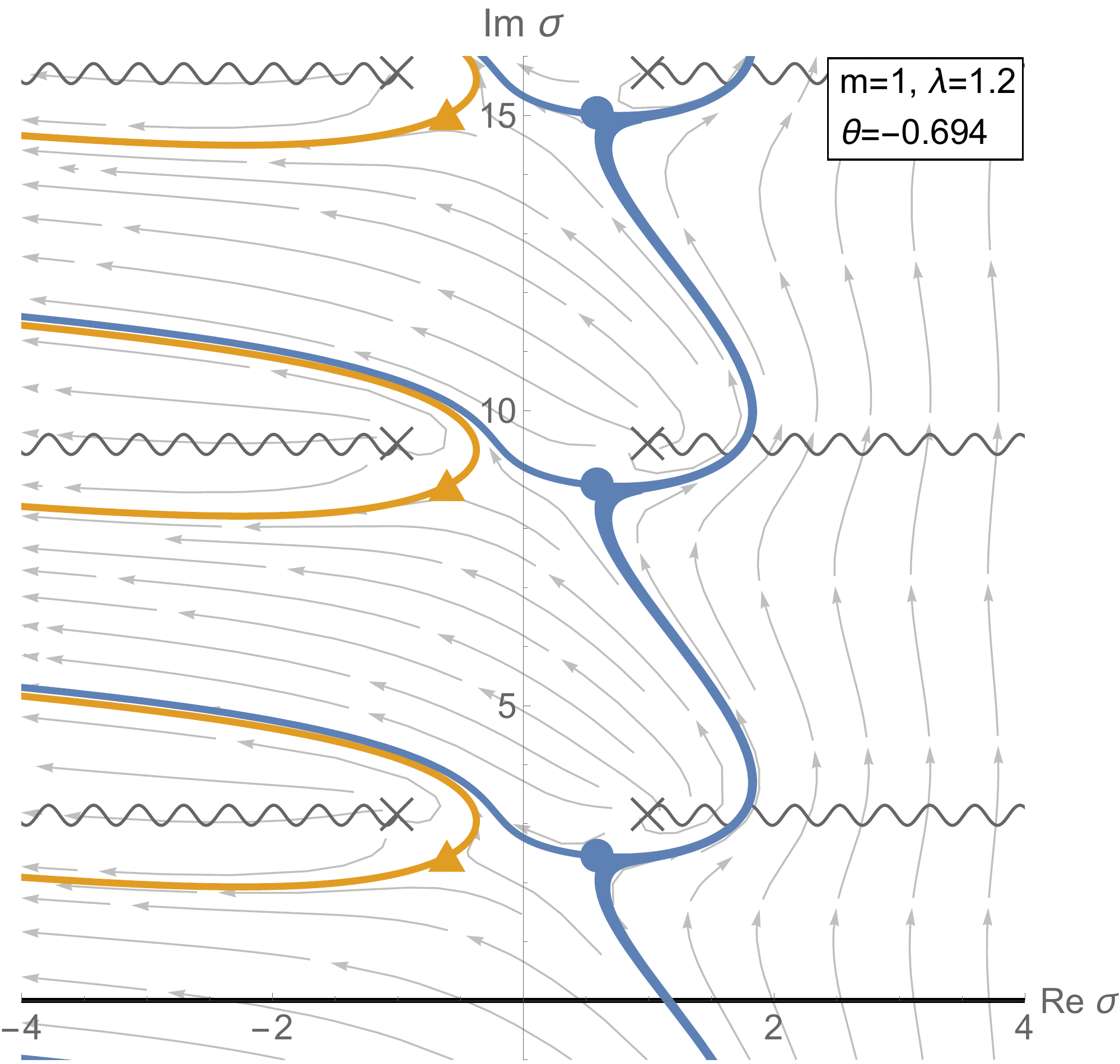} \hspace{8mm}
\includegraphics[width=75mm]{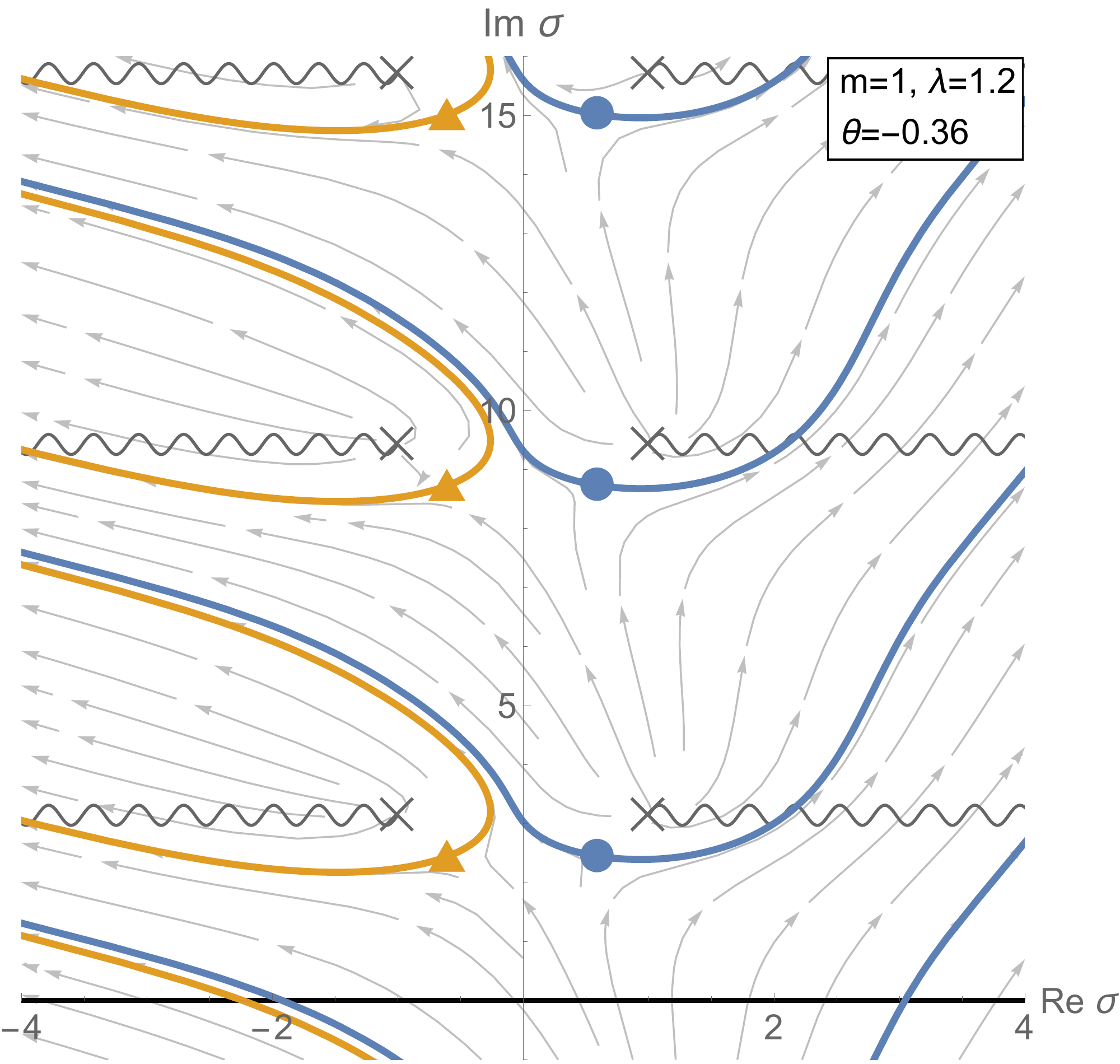}
\end{figure}
\begin{figure}
\centering
\includegraphics[width=75mm]{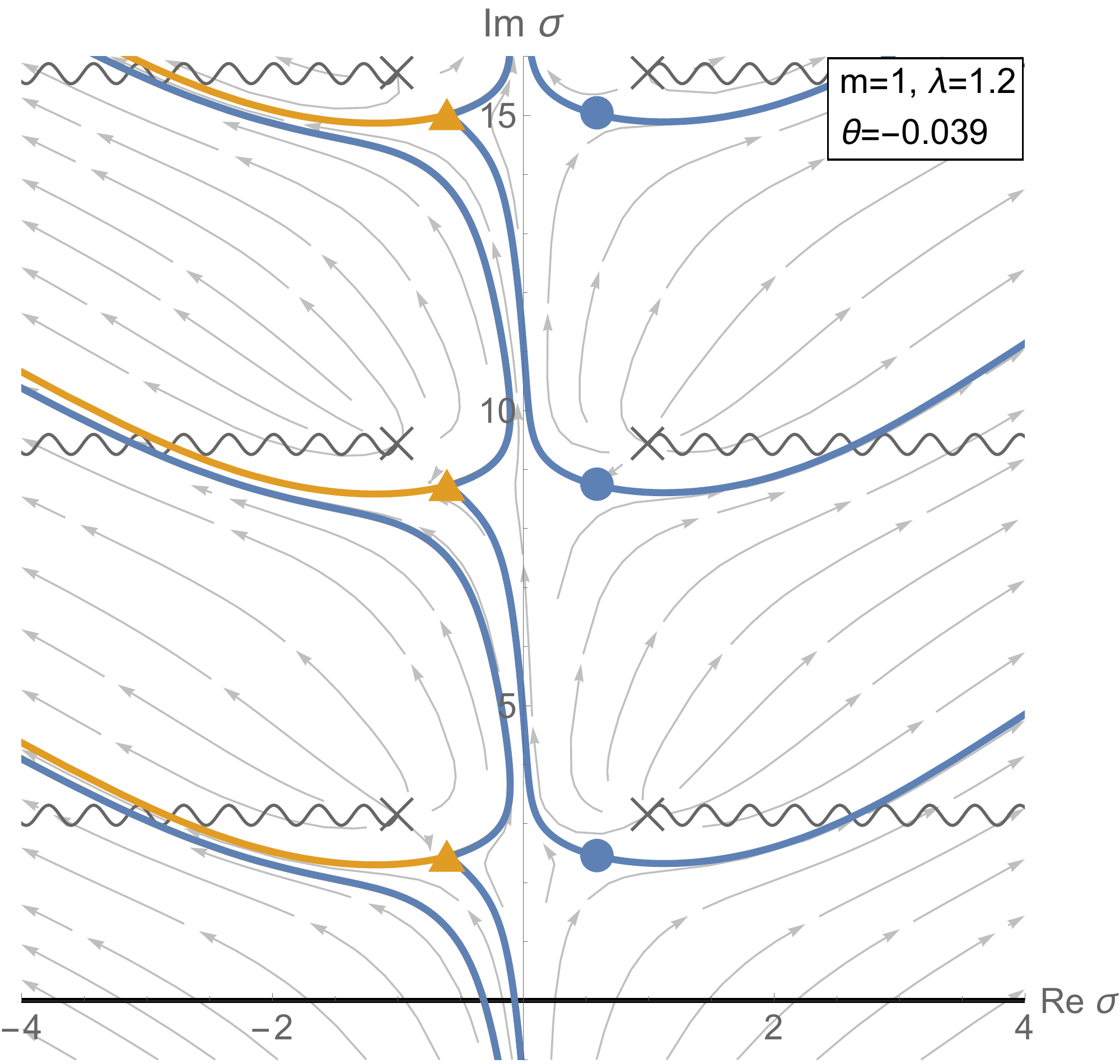} \hspace{8mm}
\includegraphics[width=75mm]{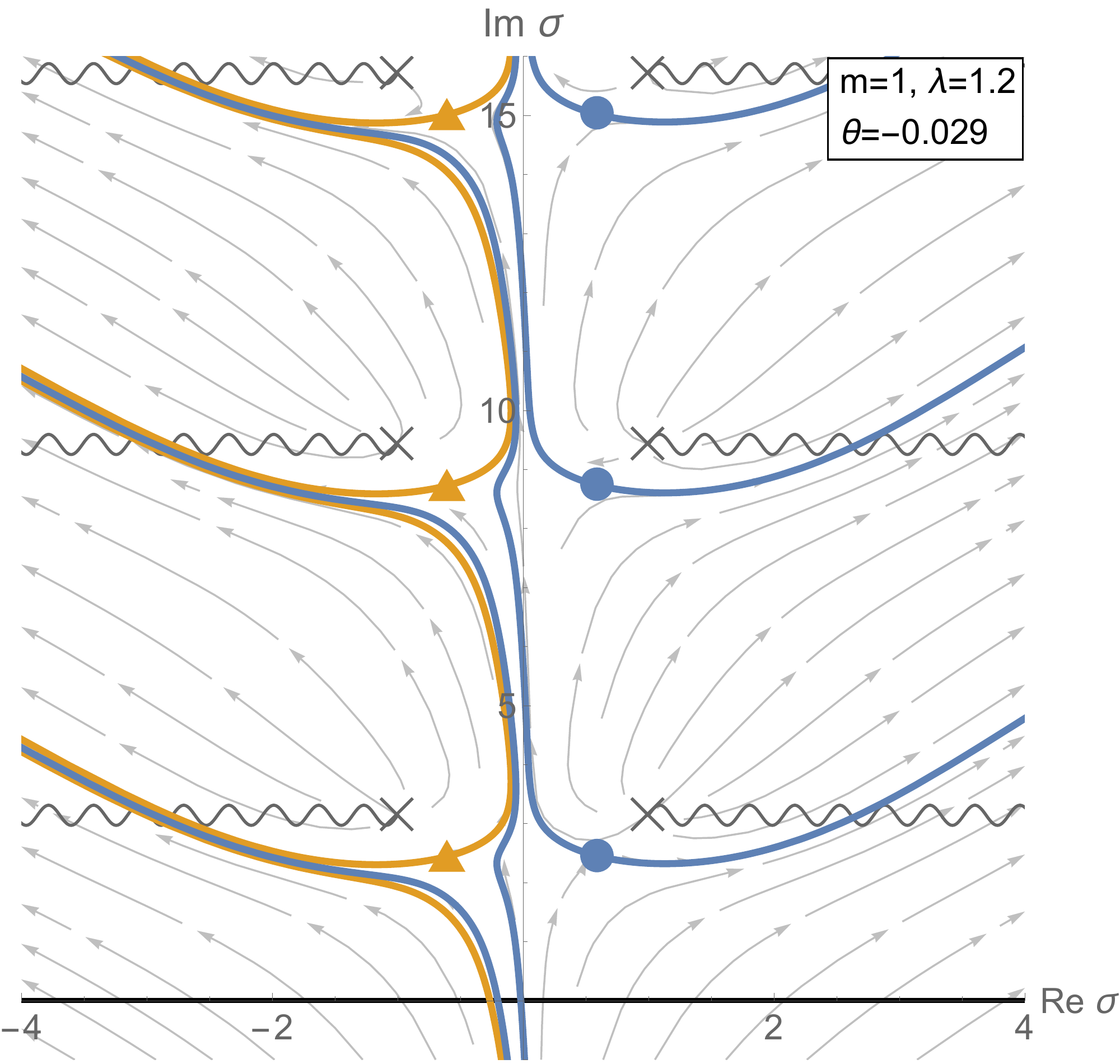} \\
\vspace{5mm}
\includegraphics[width=75mm]{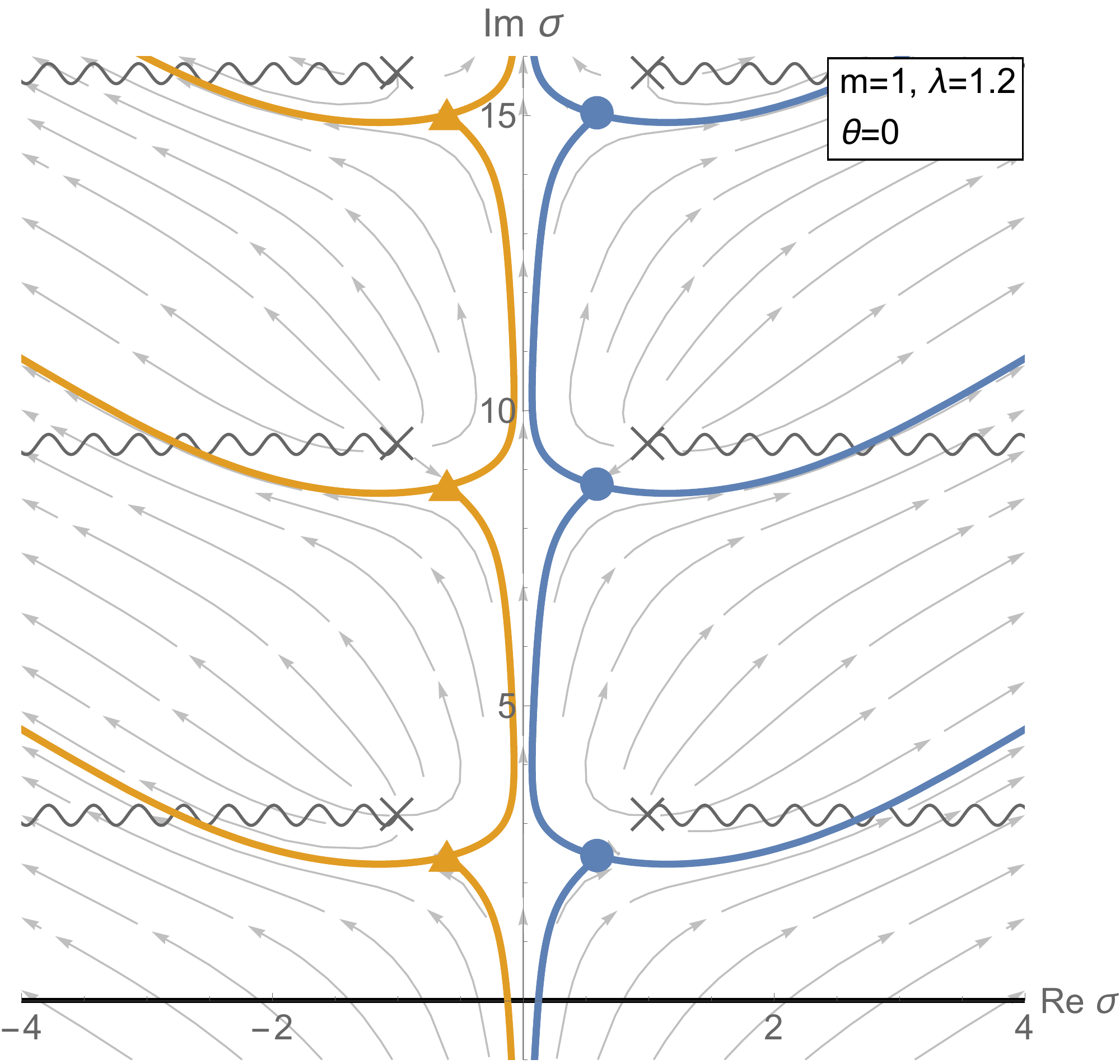} \hspace{8mm}
\includegraphics[width=75mm]{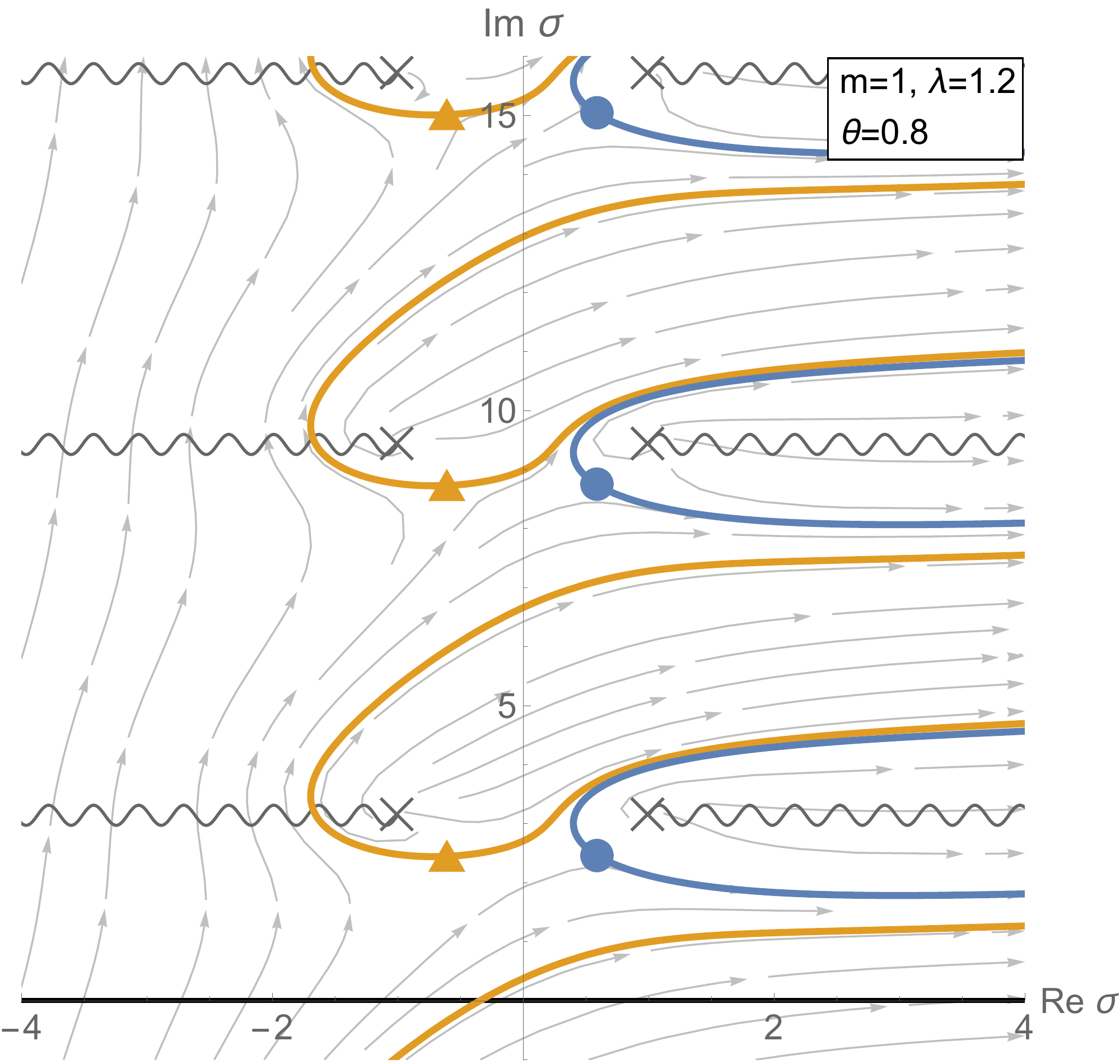} \\
\vspace{5mm}
\includegraphics[width=75mm]{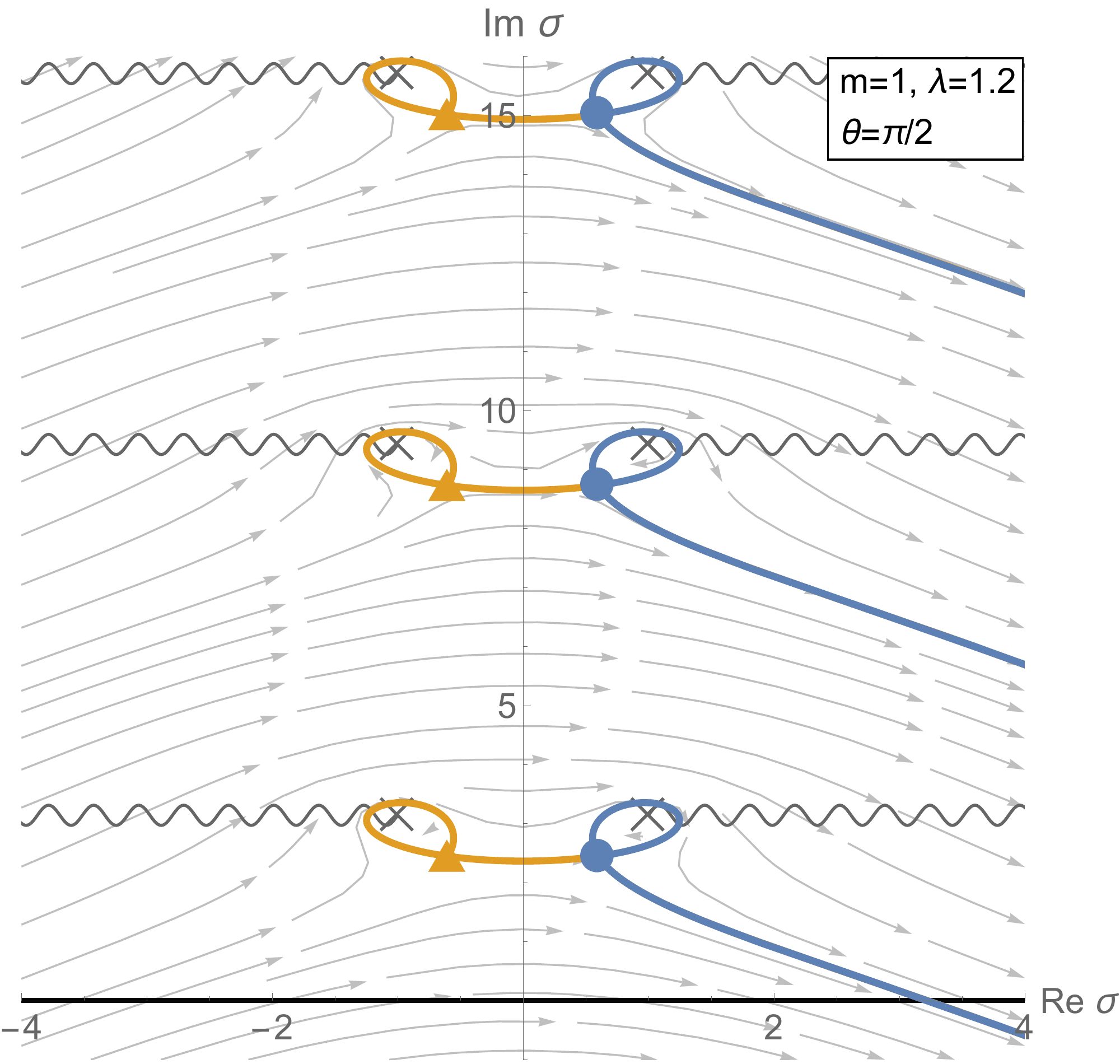} \hspace{8mm}
\includegraphics[width=75mm]{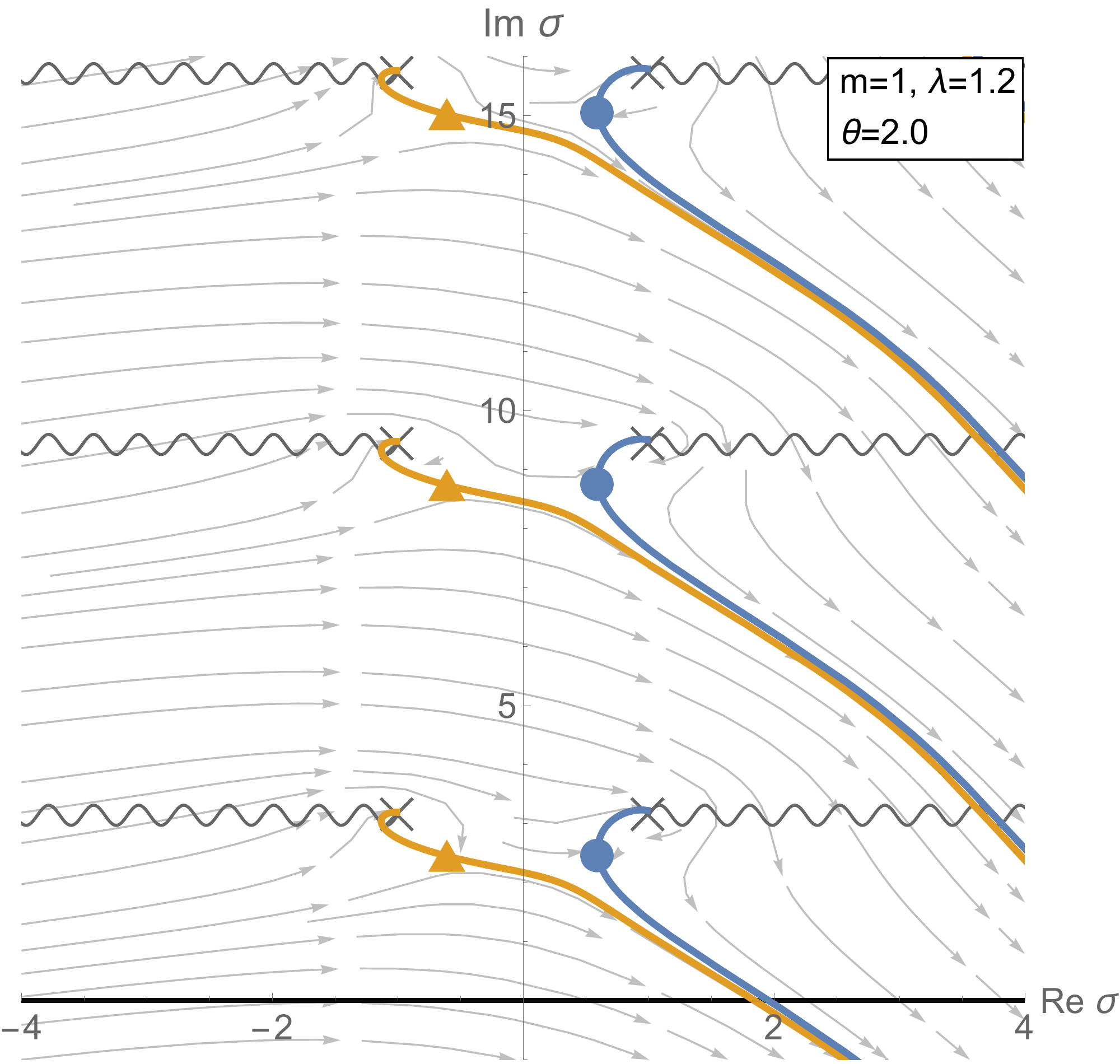}
\end{figure}
\clearpage

\section{Comments on the Pad\'{e}-Uniformized approximation}
\label{app:Borel_comments}
\begin{figure}[t]
	\centering
	\includegraphics[width=70mm]{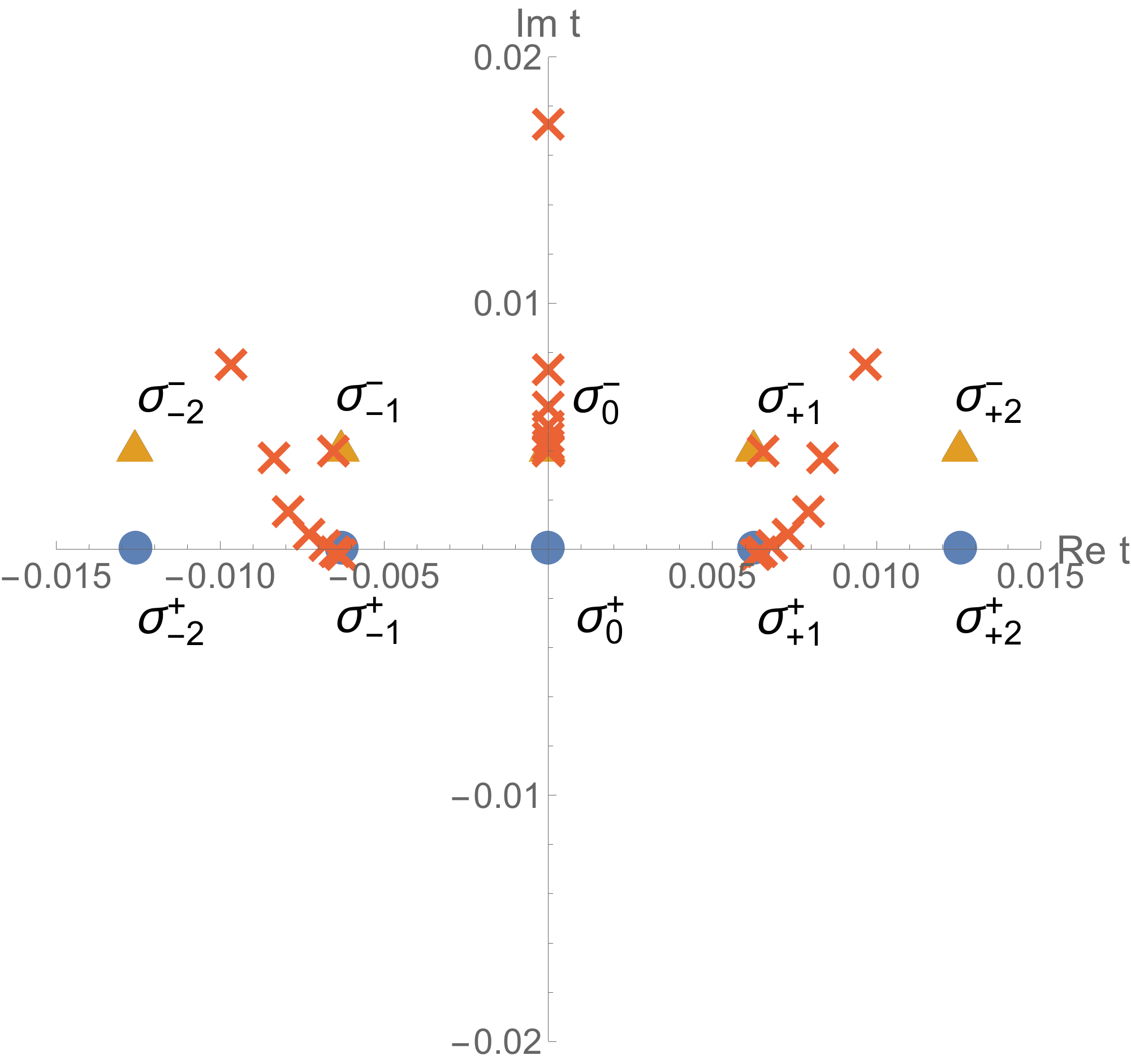}
	\hspace{8mm}
	\includegraphics[width=70mm]{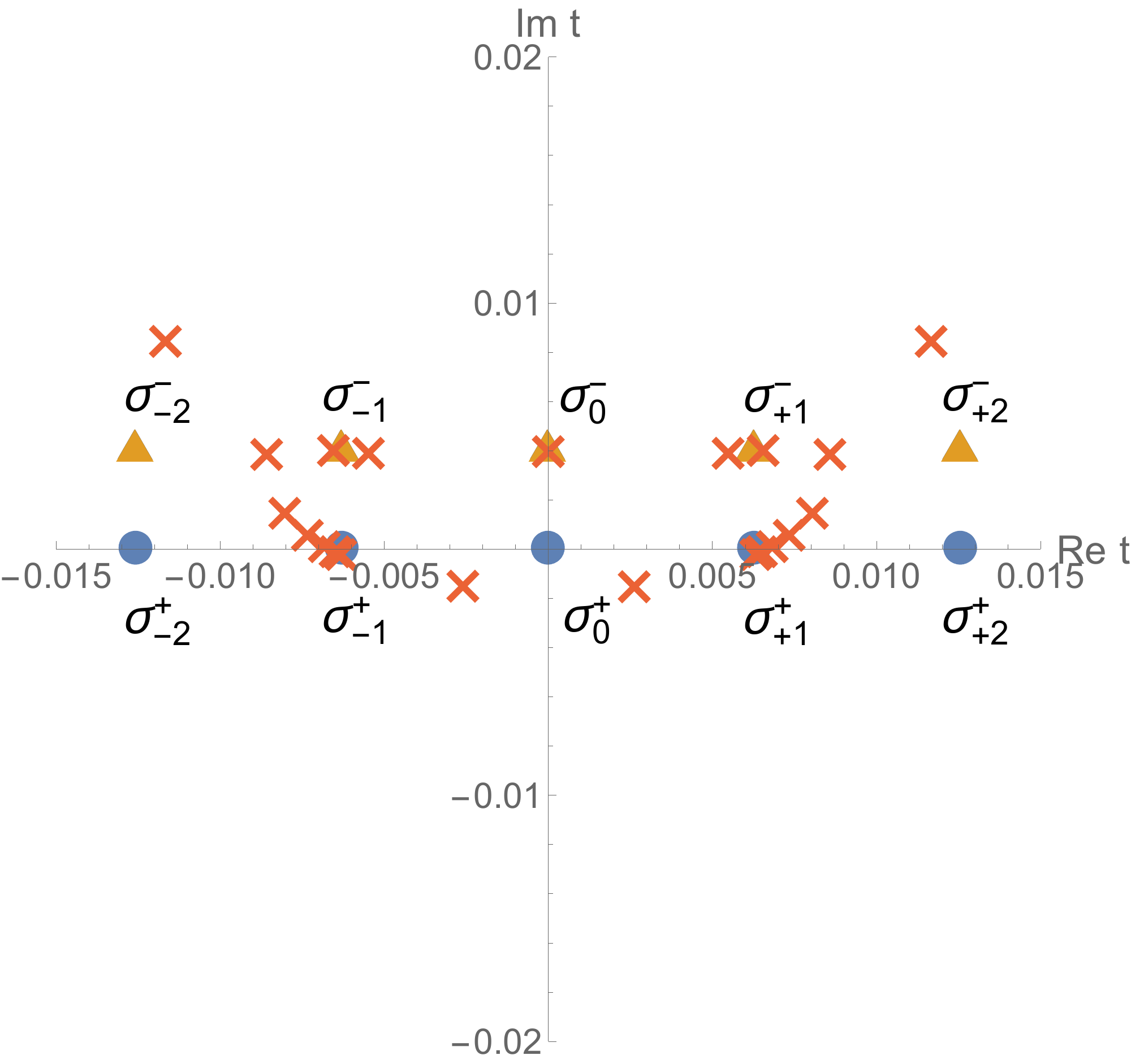}
	\caption{
		Illustrations of the Borel plane structure for the supercritical region.
		In this figure, we set $m=10$
		and set $\lambda=1.0\times10^{-3}\geq\lambda\crit=9.0\times10^{-5}$.
		The left panel is obtained by the Pad\'{e} approximation,
		and the right panel is by the Pad\'{e}-Uniformized approximation.
	}
	\label{fig:pade_m10_lam0p001}
\end{figure}
In the main text, 
we have seen that the Pad\'{e} approximation becomes worse due to branch cuts.
The branch cut singularities were associated with a saddle $\sigma_0^-$
	both in the subcritical and supercritical regions.
To make matters worse,
there are other signs of branch cuts on the Borel $t$-plane.
Fig.~\ref{fig:pade_m10_lam0p001} is an example.
We set $m=10$ (for which $\lambda\crit = 9.0\times 10^{-5}$)
so that singularities gather around the origin and the Pad\'{e} approximation works better.
We can see signs of branch cuts
associated not only with a saddle $\sigma_0^-$
but also with four other saddles $\sigma_{\pm1}^{\pm}$.
We claim that
even the Pad\'{e}-Uniformized approximation is obstructed by these branch cuts.
In this appendix,
we discuss this point by studying some simple examples of applications of the Pad\'{e}-Uniformized approximation.

\subsection{Single branch cut}
\begin{figure}[t]
	\centering
	\includegraphics[width=70mm]{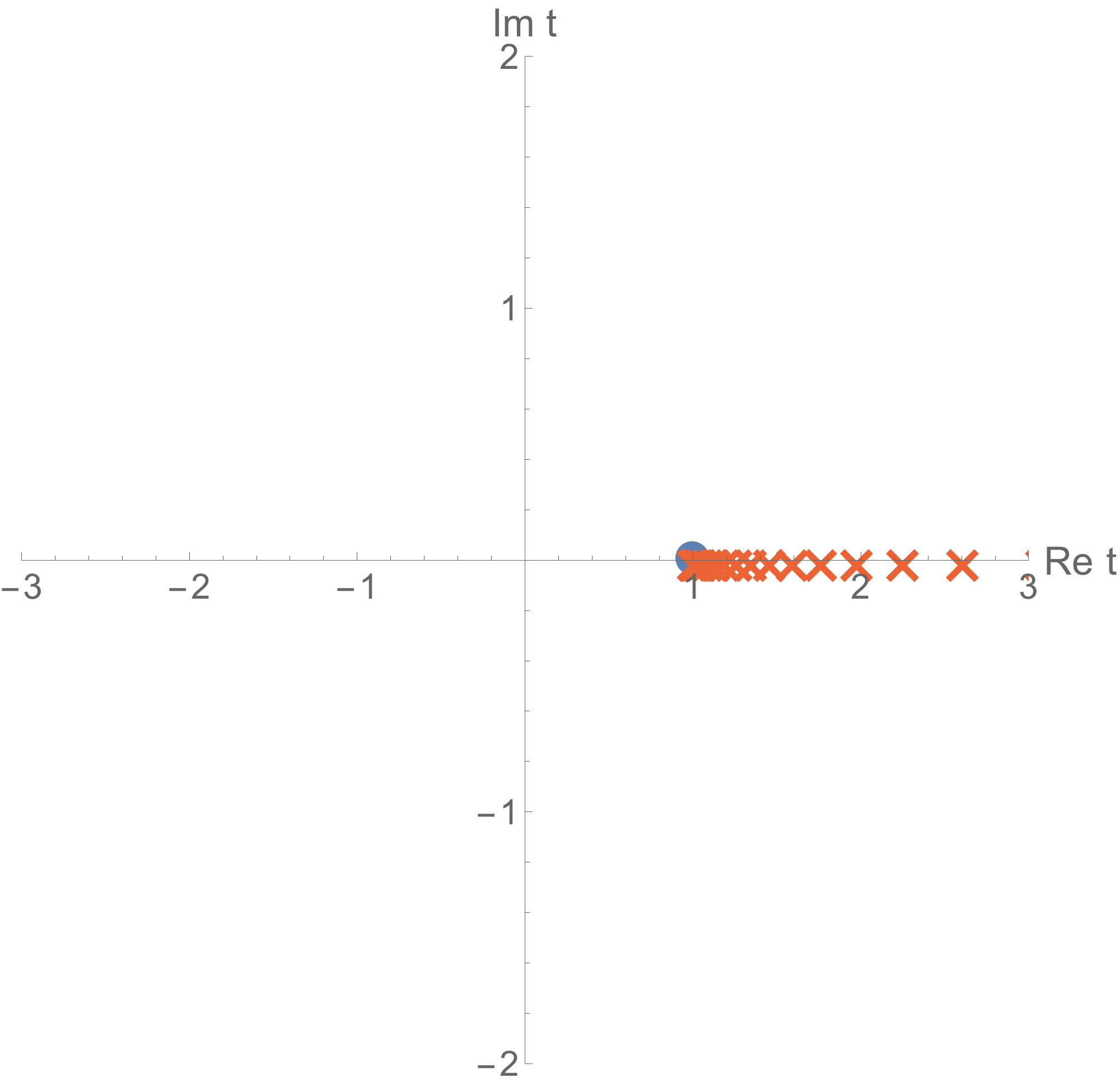}
	\hspace{8mm}
	\includegraphics[width=70mm]{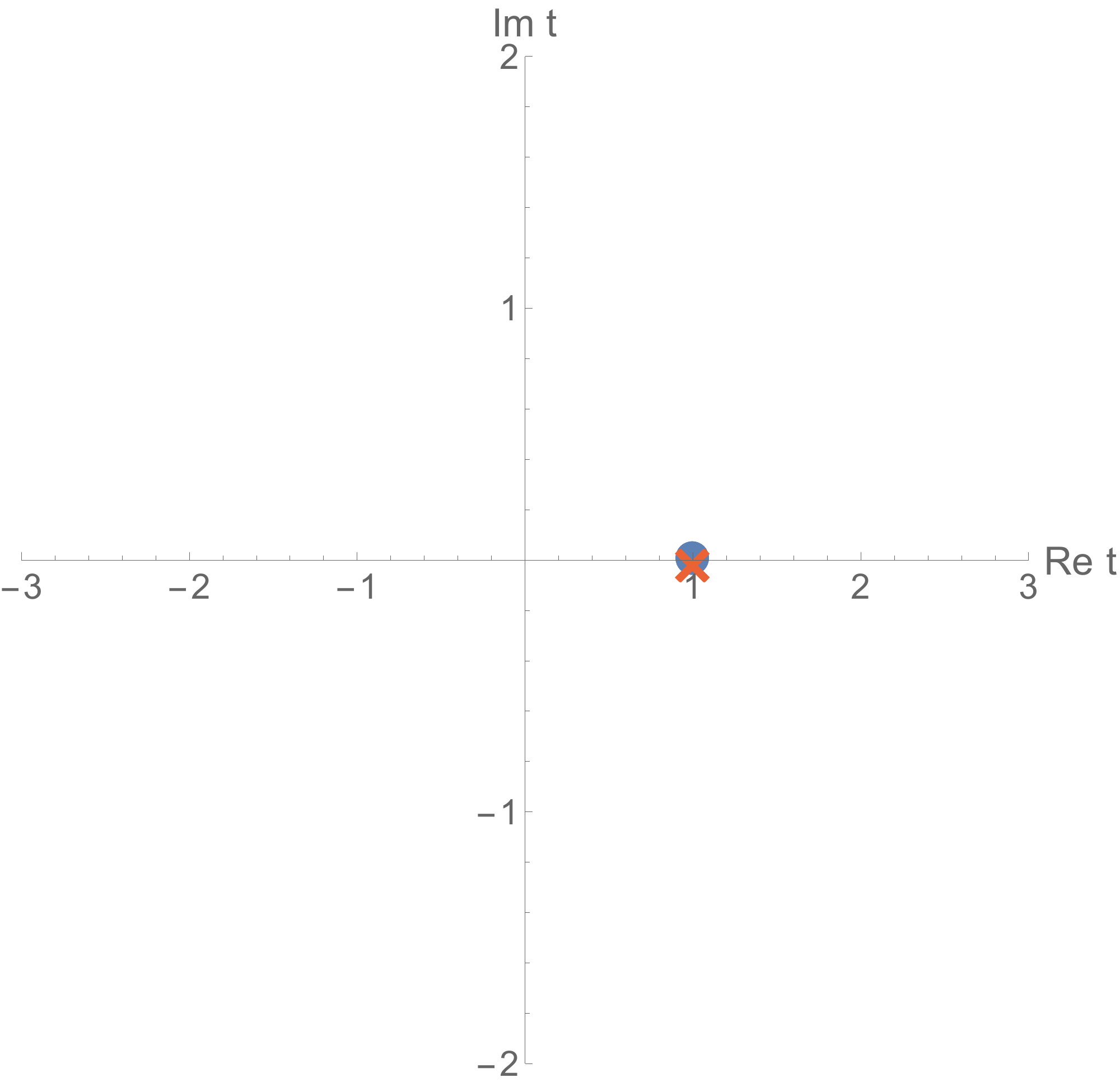}
	\caption{
		Comparison of the Pad\'{e} and the Pad\'{e}-Uniformized approximations (25,25)
		in the case of a single branch cut.
[Left] The Pad\'{e} approximation.
[Right] The Pad\'{e}-Uniformized approximation,
where the branch cut is removed by the uniformization map.
	}
	\label{fig:pade_improve_case1}
\end{figure}

Let us start with the simplest case.
Consider, for example,
a function which has a single branch cut
\begin{align}
	\mathcal{B}F(t)
	= \frac{1}{(1-t)^{1/5}},
\end{align}
and a uniformization map
\begin{align}
	u = \psi(t) = -\ln(1-t).
\end{align}
In Fig.~\ref{fig:pade_improve_case1},
we compare the standard Pad\'{e} approximation with the Pad\'{e}-Uniformized approximation for this example.
We see that the Pad\'{e}-Uniformized approximation works well in this case.

\subsection{Multiple branch cuts}
\begin{figure}[t]
	\centering
	\includegraphics[width=70mm]{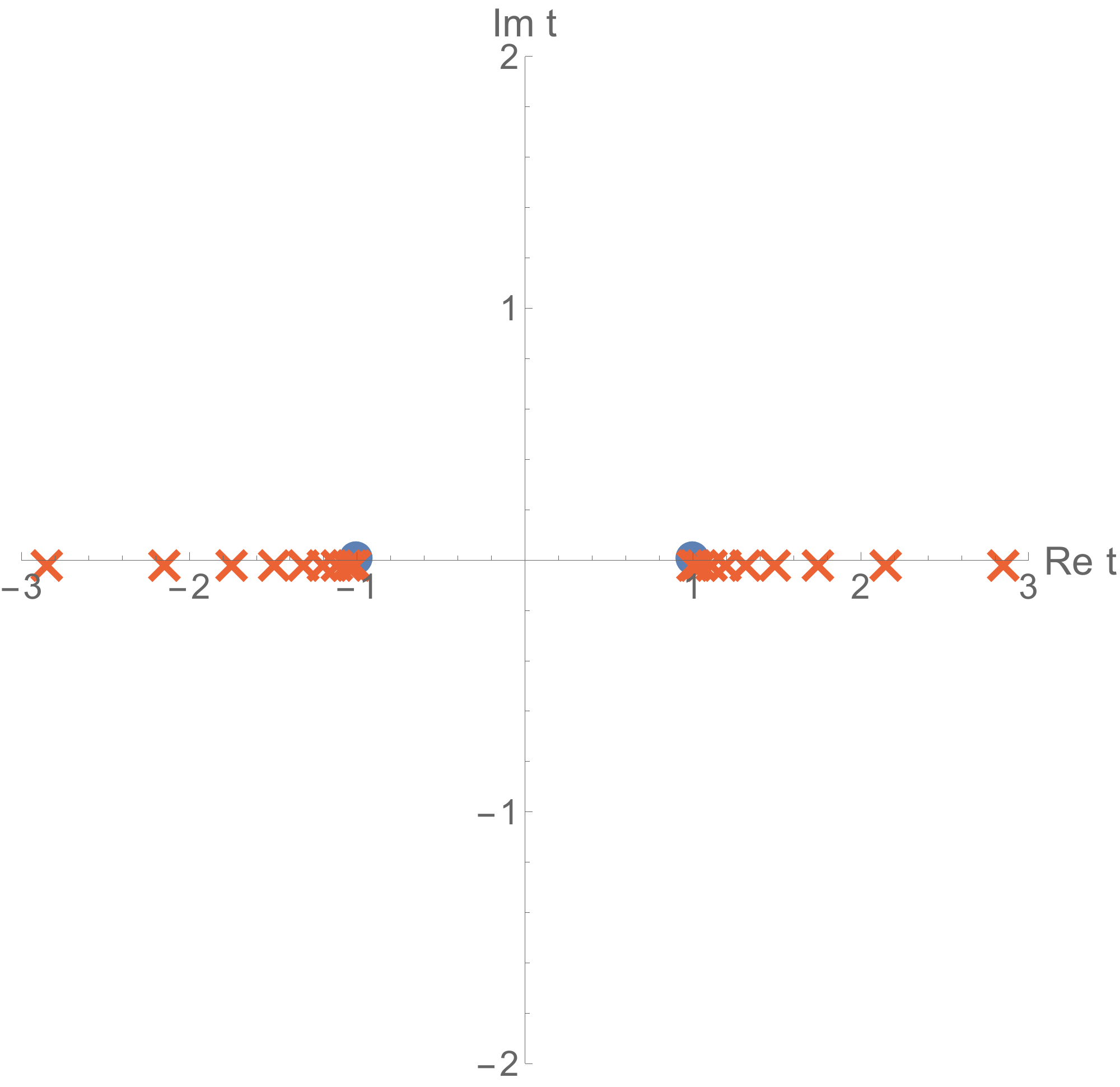}
	\hspace{8mm}
	\includegraphics[width=70mm]{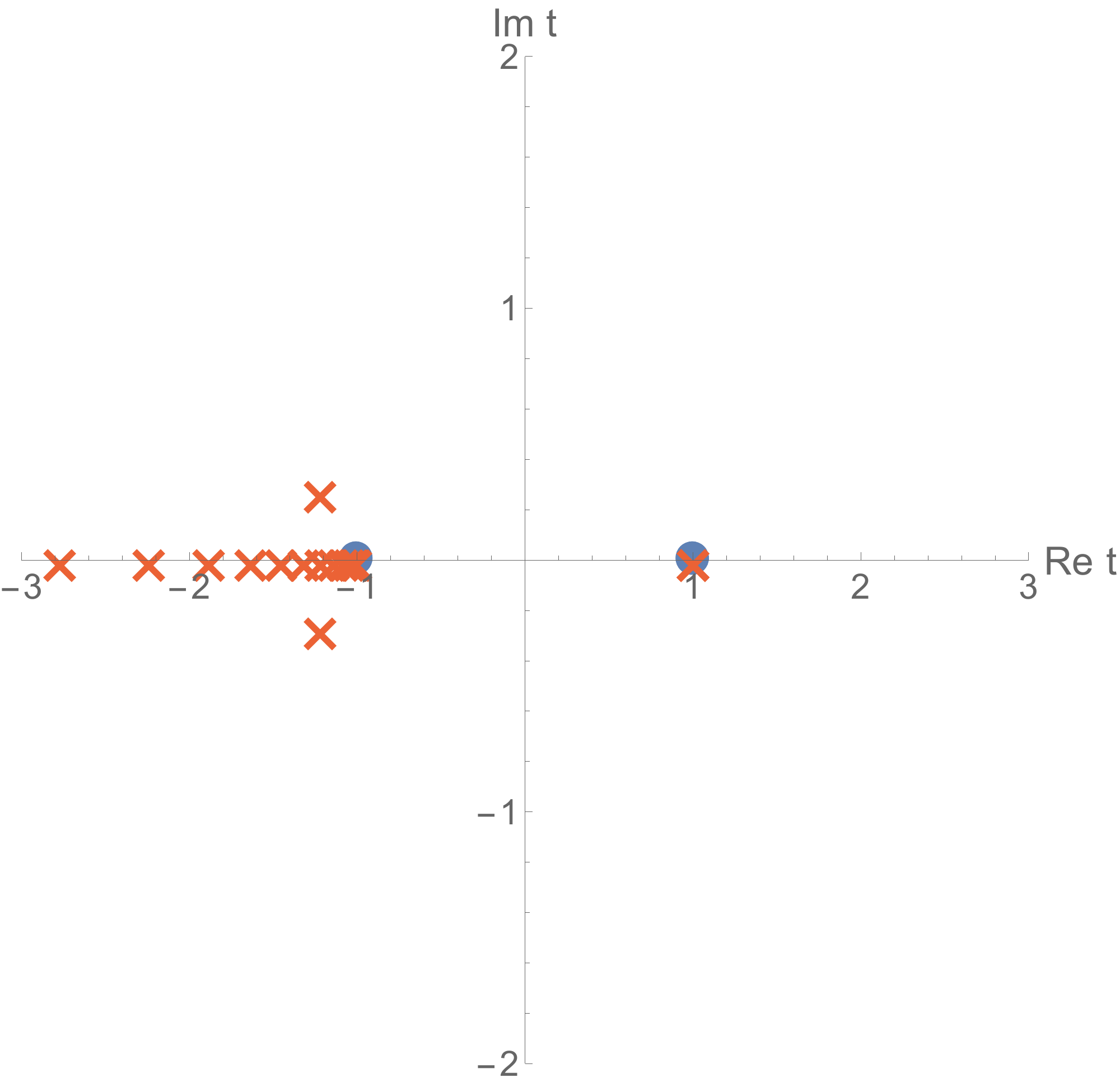}
	\caption{
		Comparison of the Pad\'{e} and the Pad\'{e}-Uniformized approximation (25,25)
		in the case with the two branch cuts.
		One of the branch cuts is removed by the uniformization map.
	}
	\label{fig:pade_improve_case2}
\end{figure}

Next let us consider the following example with two branch cuts
\begin{align}
	\mathcal{B}F(t)
	= \frac{1}{(1-t)^{1/5}(1+t)^{1/5}}.
\end{align}
The result is shown in Fig.~\ref{fig:pade_improve_case2}.
We see that the approximation becomes worse.
In the Pad\'{e}-Uniformized approximation (right panel),
we can see that there is a pair of singularities above and below the negative real axis.
This is an artifact due to the uniformization map
as explained below.
The $n$-th Riemann sheet of the Borel $t$-plane is sent to a region
\begin{align}
	-n\pi < \Im u < +n\pi
\end{align}
by the uniformization map.
The branch cut which starts from the singularity $t=-1$ is sent to
\begin{align}
	\Re u < -\ln 2, \quad
	\Im u = 2\pi n.
\end{align}
Then, the Pad\'{e}-Uniformized approximation on the $t$-plane
	(or the standard Pad\'{e} approximation on the $u$-plane)
	resembles singularities at
\begin{align}
	u = -\ln2+2\pi i n.
\end{align}
Ideally, all of which are sent back to the same point $t=-1$ by the inverse map
\begin{align}
	t=\phi(u) = 1-e^{-u}.
\end{align}
However, since the approximation becomes worse away from the origin $u=0$,
found singularities (particularly with $\abs{n}>0$) are not sent back exactly to $t=-1$.
As a result,
the Pad\'{e}-Uniformized approximation returns multiple singularities around $t=-1$.
The pair of singularities in the right panel of Fig.~\ref{fig:pade_improve_case2} corresponds to $n=\pm1$.
Other pairs of singularities which correspond to larger $\abs{n}$ are missing
	simply because they are too far away from the origin $u=0$.

%
\begin{figure}[t]
	\centering
	\includegraphics[width=70mm]{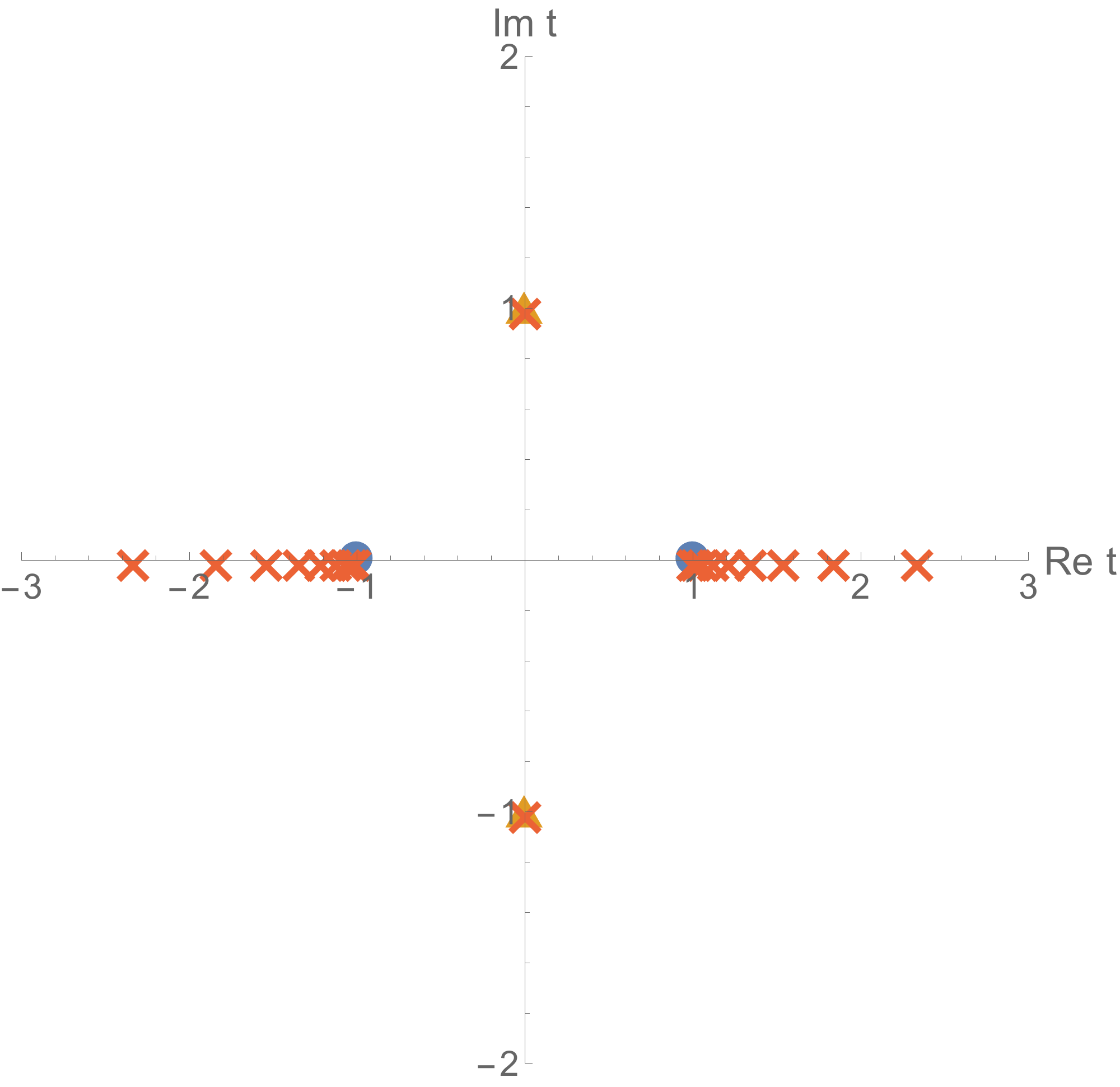}
	\hspace{8mm}
	\includegraphics[width=70mm]{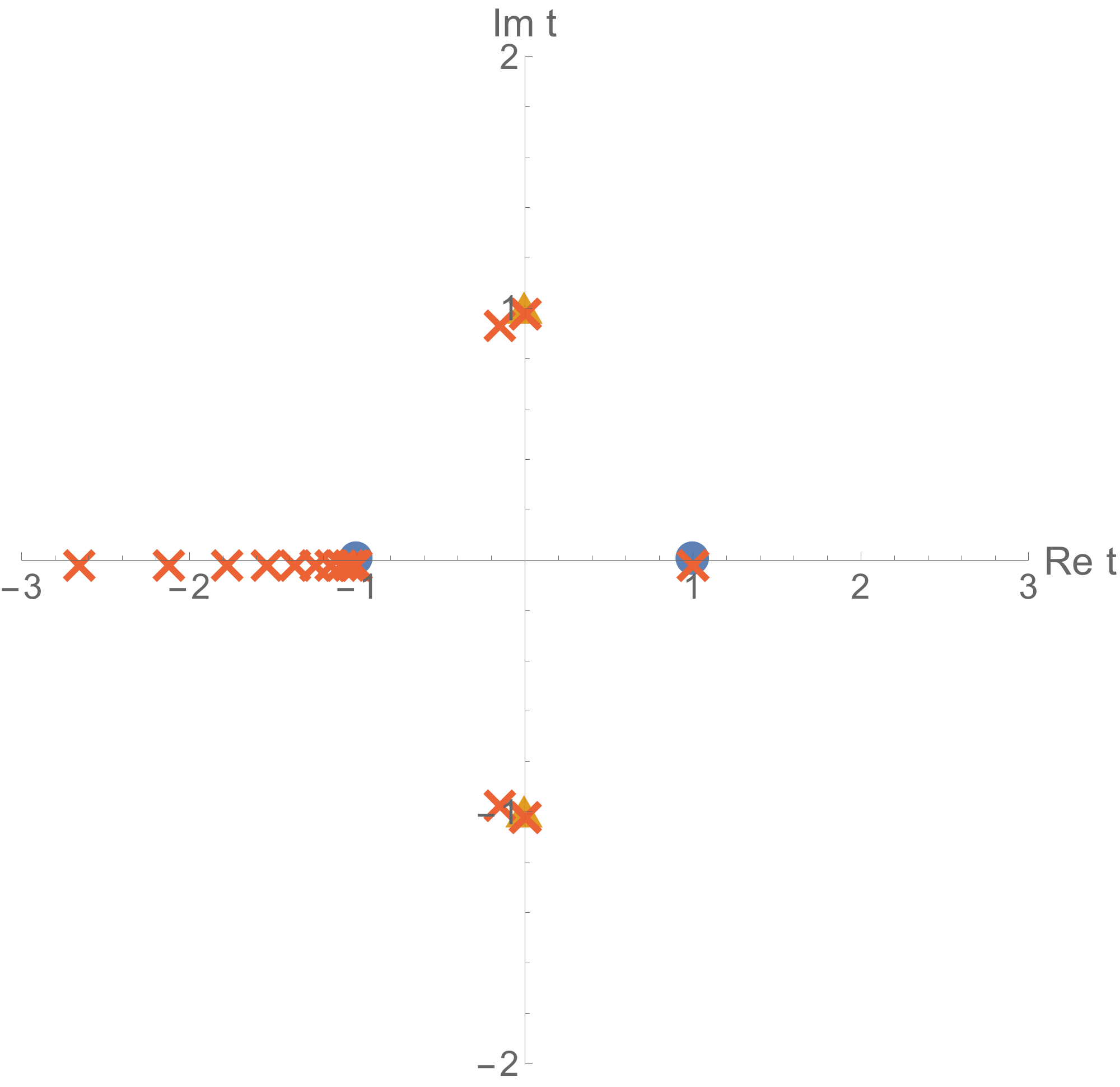}
	\caption{
		Comparison of the Pad\'{e} and the Pad\'{e}-Uniformized approximation (25,25)
		in the case of two branch cuts and two poles.
		One of the branch cuts is removed by the uniformization map.
	}
	\label{fig:pade_improve_case3}
\end{figure}
Such artifacts cause trouble
since they are indistinguishable from other genuine singularities.
Let us consider, for example, the following function
\begin{align}
	\mathcal{B}F(t)
	= \frac{1}{(1-t)^{1/5}(1+t)^{1/5}}\frac{1}{(i-t)(i+t)}.
\end{align}
The result of the Pad\'{e}(-Uniformized) approximation is shown in Fig.~\ref{fig:pade_improve_case3}.
The two poles $t=\pm i$ are sent to
\begin{align}
	u = -\ln(1\mp i)	
\end{align}
while the branch cut singularities $t=\pm 1$ are sent to
\begin{align}
	u = -\ln2+2\pi in.
\end{align}
These two singularities $u = -\ln2+2\pi in$ are further away from the poles $u = -\ln(1\mp i)$.
Then, the approximation for the two singularities is disturbed by the poles.
As a result,
the Pad\'{e}-Uniformized approximation returns the poles which originate in $t=\pm i$,
but with much worse artifacts which originate in $t=\pm 1$.
Indeed, 
in the right panel of Fig.~\ref{fig:pade_improve_case3},
the pair of artifacts is indistinguishable from the genuine poles.
We claim that some of singularities found in Sec.~\ref{sec:Borel_pade} are these types of artifacts.

\subsection{On elimination of multiple branch cuts}
\begin{figure}[th]
	\centering
	\includegraphics[width=70mm]{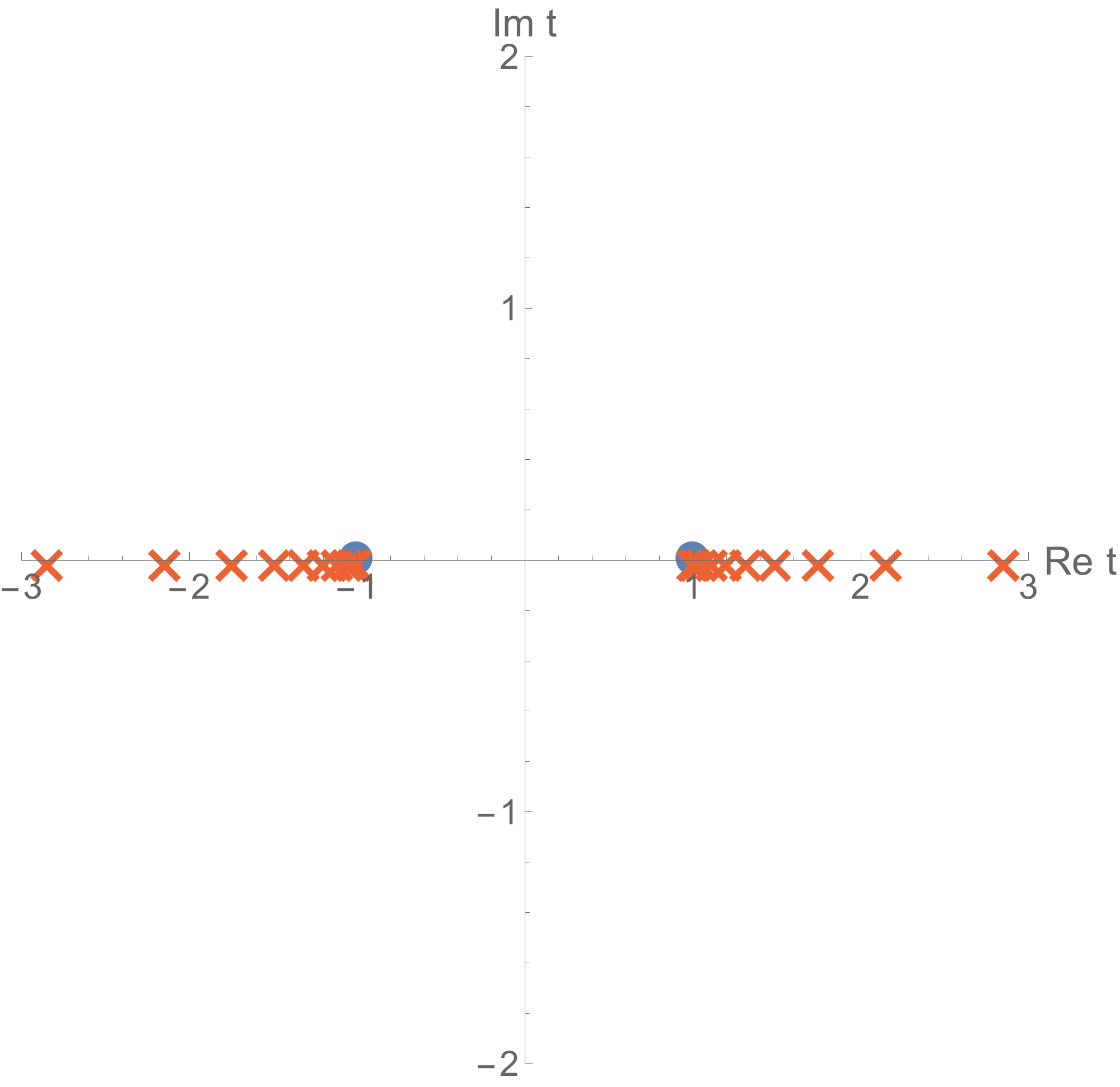}
	\hspace{8mm}
	\includegraphics[width=70mm]{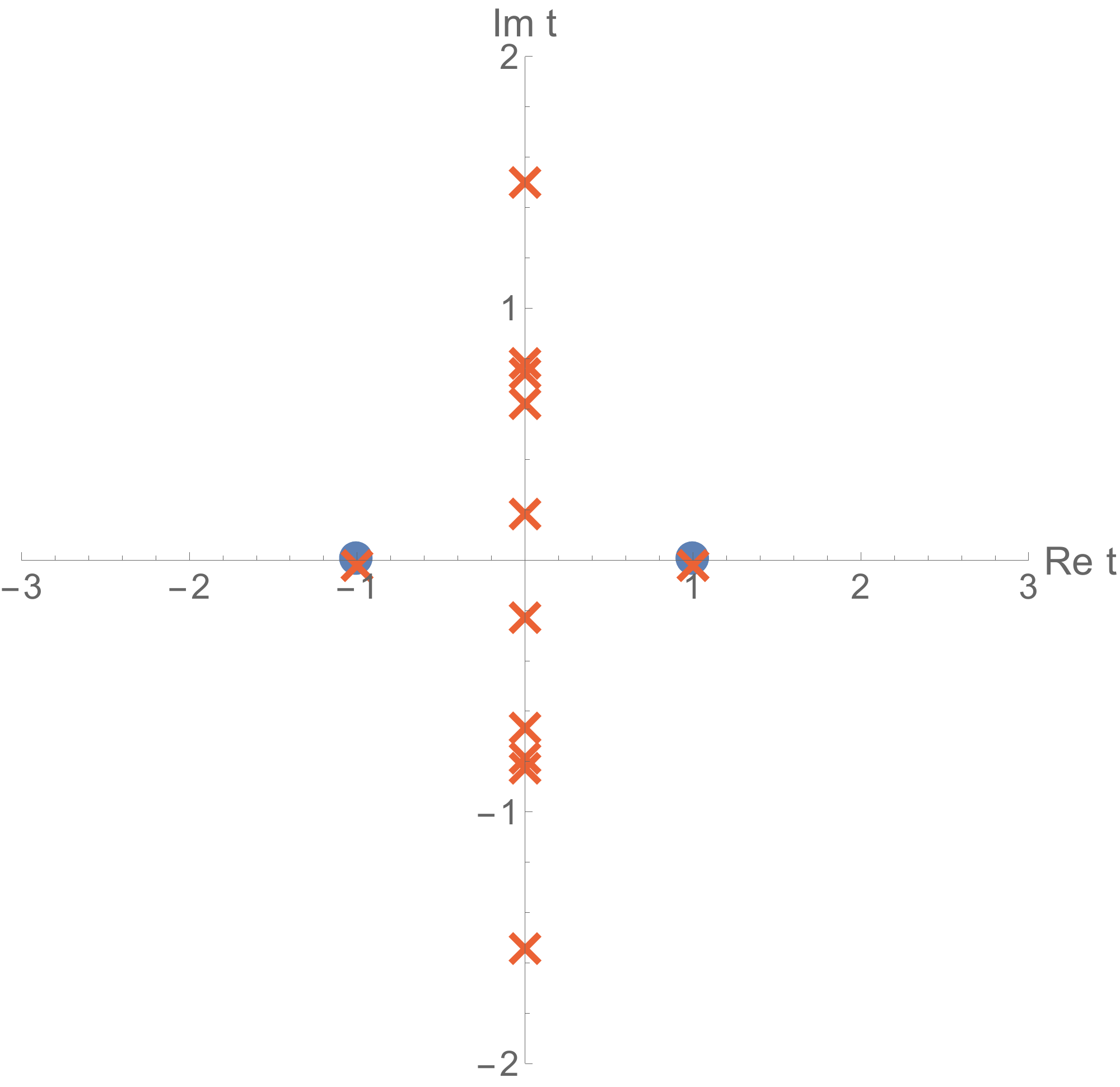}
	\caption{
		Comparison of the Pad\'{e} and the Pad\'{e}-Uniformized approximation (25,25)
		in the case of two branch cuts.
		The two singularities $t=\pm1$ are mapped to $u=\pm\infty$.
		The multiple branch cuts are eliminated by the map.
	}
	\label{fig:pade_improve_case4}
\end{figure}
One possible way to avoid such multiple branch cuts is to consider a map
\begin{align}
	u = \psi(t) = -\ln(1-t)+\ln(1+t),
\end{align}
which sends the singularities at $t=\pm1$ to $u=\pm\infty$.
However, this map causes other branch cuts on the $u$-plane.
As a result,
the Pad\'{e}-Uniformized approximation returns a lot of artifacts along the imaginary axis
as shown in Fig.~\ref{fig:pade_improve_case4}.
It seems that the problem resides in
the non-trivial topology of the Borel $t$-plane
due to multiple branch cuts.

For the above reasons,
the multiple branch cuts worsen the Pad\'{e}-Uniformized approximation.
Some artifacts are indistinguishable from other genuine singularities.
Also, multiple branch cuts are not eliminated simultaneously
at least by naive maps.
Further improvements of the Pad\'{e}-Uniformized approximation are left for future works.

\section{Transseries for a finite $\eta$}
\label{sec:trans_finite_eta}
In this appendix
we derive the transseries for the finite $\eta$ from the viewpoint of difference equation.
See \cite{Kuik:2003,Costin:1999798} for technical details.
We consider the formal transseries satisfying the difference equation ($x=N_f$).
\begin{\eqa}
&& Z(x+1) = f(x) Z(x) + g(x) Z(x-1), \label{eq:diff-z}
\end{\eqa}
where
\begin{\eqa}
&& f(x)= \frac{(2x-1) \cosh (m)}{2x \sinh^2(m)},\qquad g(x)= -\frac{(x-1)^2+ \eta^2}{4x(x-1) \sinh^2(m)}.
\end{\eqa}
We introduce $P(x) =  Z(x-1)$, and eq.(\ref{eq:diff-z}) can be written as a vectorial expression, given by
\begin{\eqa}
&& {\bf Z}(x+1) = M(x) {\bf Z}(x),
\end{\eqa}
where ${\bf Z}(x)  = (Z(x),P(x))^{\top}$, and the $2$-by-$2$ matrix $M(x)$ is defined as
\begin{\eqa}
&& M(x) =
\begin{pmatrix}
 f(x) & g(x) \\
 1 & 0
\end{pmatrix} \sim
\Lambda \left( {\mathbb I} + x^{-1} A \right) + O(x^{-2}),\\ \nonumber \\
&& \Lambda =
\begin{pmatrix}
 \frac{\cosh (m)}{\sinh^2(m) }  & - \frac{1}{4\sinh^2(m) } \\
 1 & 0
\end{pmatrix}, \qquad
A =
\begin{pmatrix}
  0 & 0 \\
  2\cosh (m) & -1
\end{pmatrix}.
\end{\eqa}

The transseries structure generally depends on the form of a difference equation and is uniquely determined from the change of asymptotic series by acting the shift operator. See Appendix  \ref{sec:prop_T} in detail.
In order to obtain the transseries based on the difference equation, one needs to diagonalize $\Lambda$ and $A$.
By using an invertible matrix $U$ to diagonalize $\Lambda$, one finds that
\begin{\eqa}
&& \hat{\Lambda} = U \Lambda U^{-1} \,=: \, {\rm diag} (\lambda_{-},\lambda_{+}), \\
&& \lambda_{\pm} = \frac{\cosh(m) \pm 1}{2 \sinh^2(m)}
= \frac{1}{2 (\cosh(m) \mp 1)}, \\
&& \tilde{A} := UAU^{-1} =
\begin{pmatrix}
-\frac{1}{2} & - \frac{1}{2} + \frac{\cosh(m)}{\cosh(m) -1} \\
- \frac{1}{2} + \frac{\cosh(m)}{\cosh(m)+1} & -\frac{1}{2}
\end{pmatrix}, \\ 
&& U =
\begin{pmatrix}
  -\sinh^2(m) & \frac{1}{2} + \frac{\cosh(m)}{2} \\
 \sinh^2(m) & \frac{1}{2} -\frac{\cosh(m)}{2}
\end{pmatrix}, \\ \nonumber \\
&\Rightarrow \quad& \tilde{\bf Z}(x+1) = \tilde{M}(x) \tilde{\bf Z}(x),
\end{\eqa}
where $\tilde{M}(x):=UM(x)U^{-1}$ and $\tilde{\bf Z}(x):=U{\bf Z}(x)$.
Next, we consider the diagonalization for $\tilde{A}$ by employing the technique in Appendix \ref{sec:diag_M}.
In order to perform it, we act $W(x)$ from the left as
\begin{\eqa}
\check{\bf Z}(x+1) &=& \check{M}(x) \check{\bf Z}(x), \qquad \check{\bf Z}(x) := W(x) \tilde{\bf Z}(x),
\end{\eqa}
where
\begin{\eqa}
&& \check{M}(x) = W(x+1) \tilde{M}(x)W^{-1}(x), \qquad W(x) := {\mathbb I} + x^{-1} V,
\end{\eqa}
and
\begin{\eqa}
&& {\rm diag}_{M}[\tilde{A}] =: {\rm diag}(a_{-},a_{+}), \qquad a_{\pm} = -\frac{1}{2}.
\end{\eqa}
Hence,
\begin{\eqa}
Z(x) = \sum_{s=\{+,-\}} \sum_{n=0}^{\infty} \sigma_{s} c_{s,n} e^{\log \lambda_{s} \cdot x}x^{a_{s}-n},
\end{\eqa}
where the transseries in terms of $x$ is completely determined but it is not for other parameters such as $m$ and $\eta$.
Hence, $\sigma_s=\sigma(m,\eta),c_{s,n}=c_{s,n}(m,\eta)$ in general.
Since $c_{\pm,0}$ is relevant only to the normalization, one can take $c_{\pm,0}=1$ without loss of generality.
$\sigma_{\pm}$ can be determined from the partition function,  and they are given by
\be
\sigma_{+}(m,\eta) = 0, \qquad \sigma_{-}(m,\eta) = \sqrt{2 \pi (1 + \cosh m)}.
\ee
Notice that $c_{\pm,n>0}$ are recursively determined from the difference equation.

\subsection{Properties of the shift operator $T$} 
\label{sec:prop_T}
We define the shift operator $T$ as $T[f(x)]=f(x+1)$.
It satisfies the following homomorphic properties for summation and multiplication:
\begin{\eqa}
&& T[f(x) + g(x)] = T[f(x)] + T[g(x)], \\
&& T[f(x) \cdot g(x)] = T[f(x)] \cdot T[g(x)].
\end{\eqa}
The action of $T$ to transmonomials gives
\begin{\eqa}
&& T[x^{-1}] = \frac{1}{x+1} = \sum^{\infty}_{n=0}  (-1)^n x^{-1-n}, \\
&& T[x^{a}] = (x+1)^a = x^{a}(1+x^{-1})^a = x^{a} \sum^{\infty}_{n=0} \frac{a\cdot(a-1) \cdots (a-n+1)}{n !} \, x^{-n}, \\
&& T[e^{-\mu x}] = e^{-\mu (x+1)} =e^{-\mu} \cdot  e^{-\mu x}, \\
&& T[\log (x)] = \log (x+1) = \log(x) + \log(1+x^{-1}) = \log(x) + \sum_{n=1}^{\infty} \frac{(-1)^{n+1}x^{-n}}{n}.
\end{\eqa}
Notice that when $g(x+1)=g(x)$, the action of $T$ to $g(x)$ gives the identity map.
As one can see easily, for example, the below type of transseress is closed under action of the shift operator $T$:
\begin{\eqa}
\sum_{n,k=0}^{\infty} e^{-n \log \mu} c_{n,k}x^{a-k}  
\ \xrightarrow[]{\ \ T \ \ } \
\sum_{n,k=0}^{\infty} e^{-n \log \mu} c^\prime_{n,k}x^{a-k}.
\end{\eqa}
The type of transseries and the propagation of integration constants are determined by the form of difference equation such as (non)linearity, (non)autonomous, and so on.

\subsection{Diagonalization of $M(x)$} \label{sec:diag_M}
Suppose $f(x)$ is a transseries determined by a difference equation having a form $f(x+1)=M(x) f(x)$, where $M(x)$ is a function satisfying $M(x) \sim \Lambda ( 1 + x^{-1} A) + O(x^{-2})$ with a constant $\Lambda$ and $A$.
Assume that $\Lambda$ is positive.
If the transseries includes $e^{-\mu x} x^{a}$, since the action of $T$ gives
\begin{\eqa}
T[e^{-\mu x} x^{a}] = e^{-\mu} \cdot e^{-\mu x} x^{a} \left( 1 + a x^{-1} + O(x^{-2})\right),
\end{\eqa}
hence, $\mu$ and $a$ are determined from $\Lambda$ and $A$, respectively,  as
\begin{\eqa}
\mu = - \log (\Lambda), \qquad a = A.
\end{\eqa}

We would extend $f(x)$ to multi-variables, ${\bf f}(x)=(f_1(x),\cdots,f_N(x))^{\top}$,
\begin{\eqa}
{\bf f}(x+1) = M(x) {\bf f}(x),
\end{\eqa}
where $M(x)$ is an $N$-by-$N$ matrix given by
\begin{\eqa}
M(x) = \Lambda \left( {\mathbb I} + x^{-1} A \right) + O(x^{-2}),
\end{\eqa}
with constants matrix $\Lambda$ and $A$.
Assume that $\Lambda$ is diagonalizable by an invertible matrix $U$ and that the all of eigenvalues are positive.
By acting $U$ from the left in the both sides, one can obtain that
\begin{\eqa}
\tilde{\bf f}(x+1) = \tilde{M}(x) \tilde{\bf f}(x),
\end{\eqa}
where
\begin{\eqa}
&& \tilde{M}(x) = \hat{\Lambda} \left( {\mathbb I} + x^{-1} \tilde{A} \right) + O(x^{-2}), \\
&& \hat{\Lambda} = U \Lambda U^{-1} =: {\rm diag} (\lambda_{1},\cdots,\lambda_{N}), \\
&& \tilde{A} = U A U^{-1}.
\end{\eqa}
In order to diagonalize $\hat{\Lambda} \tilde{A}$, we redefine $\tilde{\bf f}(x)$ and the difference equation as
\begin{\eqa}
&& \check{\bf f}(x+1) = \check{M}(x) \check{\bf f}(x), \\
&& \check{\bf f}(x) = W(x) \tilde{\bf f}(x), \\
&& W(x) = {\mathbb I} + x^{-1} V,
\end{\eqa}
with an $N$-by-$N$ constant matrix $V$ and $\check{M}(x)$ is given by
\begin{\eqa}
\check{M}(x) &=& W(x+1) \tilde{M}(x) W^{-1}(x)  \nonumber \\
&=& \left( {\mathbb I} + (x+1)^{-1} V \right) \tilde{M}(x) \left( {\mathbb I} + x^{-1}V \right)^{-1} \nonumber \\
&\sim&  \hat{\Lambda} + x^{-1} \left( [V,\hat{\Lambda}] + \hat{\Lambda} \tilde{A} \right) + O(x^{-2}),
\end{\eqa}
where $[A,B]:= AB-BA$.
Notice that all the diagonal parts of $V$ can be taken as zero and one can determine other $N(N-1)$-components of $V$ such that 
\begin{\eqa}
&&  [V,\hat{\Lambda}] + \hat{\Lambda} \tilde{A}  = {\rm diag}_{M}[\hat{\Lambda} \tilde{A}] = \hat{\Lambda} {\rm diag}_{M}[\tilde{A}], \\
&& {\rm diag}_{M}[\tilde{A}] =: {\rm diag}(a_1, \cdots, a_{N}),
\end{\eqa}
where ${\rm diag}_{M}[\tilde{A}]$ is a diagonal matrix having diagonal parts of $\tilde{A}$.

\section{Transseries for a finite $\lambda=\eta /N_f$
}
\label{sec:trans_finite_lam}
Let us rewrite the recursion relation (\ref{eq:diff-z})  in terms of a fixed $\lambda=\eta /N_f$
\begin{\eqa}
&& Z(x+1) = f(x) Z(x) + g(x) Z(x-1), \label{eq:diff-z-lambda}
\end{\eqa}
where
\begin{\eqa}
&& f(x)= \frac{(2x-1) \cosh (m)}{2x \sinh^2(m)},\qquad g(x)= -\frac{(x-1)^2+ x^2\lambda^2}{4x(x-1) \sinh^2(m)}.
\end{\eqa}
If we ignore the extra dependence of $Z(x,\eta)$ on $x$ through $\eta=\lambda/x$, we find that the vectorial recursion relation for  ${\bf Z}(x)  = (Z(x),P(x))^{\top}$ is modified to   
\begin{\eqa}
  && {\bf Z}(x+1) = M(x) {\bf Z}(x),
\end{\eqa}
\begin{\eqa}
&&  M(x) =
\begin{pmatrix}
 f(x) & g(x) \\
 1 & 0
\end{pmatrix} \sim
\Gamma \left( {\mathbb I} + x^{-1} {\cal A} \right) + O(x^{-2}),\\ \nonumber \\
&& \Gamma =
\begin{pmatrix}
 \frac{\cosh (m)}{\sinh^2(m) }  & - \frac{\lambda^2+1}{4\sinh^2(m) } \\
 1 & 0
\end{pmatrix}, \qquad
{\cal A}=
\begin{pmatrix}
  0 & 0 \\
  \frac{2\cosh (m)}{\lambda^2+1} &\frac{\lambda^2 -1}{\lambda^2+1}
\end{pmatrix}.
\end{\eqa}
By diagonalizing the matrix $\Gamma$, we find two exponents $\gamma_1, \gamma_2$  
\begin{\eqa}
&& \tilde{M}(x) = \hat{\Gamma} \left( {\mathbb I} + x^{-1} \tilde{\cal A} \right) + O(x^{-2}), \\
&& \hat{\Gamma} = U \Gamma U^{-1} =: {\rm diag} (\gamma_{+},\gamma_{-}), \\
&& \tilde{\cal A} = U {\cal A} U^{-1}.
\end{\eqa}
\begin{\eqa}
 \gamma_{\pm}= \frac {\cosh m \pm \sqrt{1-\lambda^2\sinh^2 m}}{2\sinh^2 m} .
\end{\eqa}
These two exponents are degenerate when 
\begin{\eqa}
 \lambda=\frac{1}{\sinh m} ,
\end{\eqa}
which agrees with the critical point $\lambda_c$ in (\ref{eq:criticalPoint}) for the phase transition.

\section{A possible relation between the Borel singularities and complex SUSY solutions}
\label{app:CSS}
In this appendix,
we point out that
a path integral interpretation of the Borel singularities appearing the main text may be complex supersymmetric solutions (CSS) found in \cite{Honda:2017qdb}.
It was proposed that Borel transformation of large Chern-Simons level expansion includes the following factor
\begin{\eq}
\mathcal{B}Z(t)
\supset \prod_{\rm CSS} \frac{1}{(t-S_c )^{n_B -n_F}} ,
\end{\eq}
where $n_B (n_F )$ is the number of bosonic (fermionic) solutions.
In the SQED studied in this paper, 
there are two types of CSS with $n_B -n_F \neq 0$,
more precisely $n_B -n_F =N_f$.
Actions of the solutions are
\begin{\eq}
S_{\rm CSS} = N_f (2\pi n \lambda \mp i\lambda m ) ,
\label{eq:CSS}
\end{\eq} 
while the action at the saddle points $\sigma_n^\pm$ in the localization formula is
\begin{\eqa}
S_n^\pm 
&=& N_f \Biggl[ 2\pi n \lambda 
-\frac{i \lambda}{2} \log{ \frac{( -\lambda\cosh{m} \mp i\Delta )(-i+\lambda )} 
                                  {( -\lambda\cosh{m} \pm i\Delta )(i+\lambda )} }
 +\log{\frac{\cosh{m} \pm \Delta } {1+\lambda^2  } } 
 \Biggr] .
\end{\eqa}
Comparing this with \eqref{eq:CSS}, 
we find that these actions agree when $\lambda$ is large $\lambda \gg 1$.
Thus it seems plausible that
the Borel singularities correspond to the CSS in the original path integral
at least for large $\lambda$. 
It would be interesting to extend the analysis in this appendix to finite $\lambda$.

\bibliographystyle{utphys}
\bibliography{thimble_phase.bib}

\end{document}